\documentclass[longauth, letter]{aa}  

\usepackage{graphicx}
\usepackage[utf8]{inputenc}

\usepackage{txfonts}
\usepackage[]{hyperref}
\usepackage{url,natbib}

\bibpunct{(}{)}{;}{a}{}{,}    

\usepackage{booktabs,array}

\hypersetup{
  colorlinks=true,   
  urlcolor=blue,     
  linkcolor=blue,     
  citecolor = blue
}
\usepackage{siunitx}
\usepackage{comment}
\usepackage[]{xcolor} 
\usepackage{mathabx}

\usepackage{float}
\usepackage{caption}
\usepackage{subcaption}
\usepackage{multicol}
\usepackage{multirow}
\usepackage{stfloats}
\usepackage{placeins}

\newcommand{\ms}{\mbox{m\,s$^{-1}$}}
\newcommand{\kms}{\mbox{km\,s$^{-1}$}}

\newif\ifshowcomOD
\showcomODtrue

\newif\ifshowcomTS
\showcomTStrue

\usepackage[normalem]{ulem}

\begin{document}

   \title{Detection of barium in the atmospheres of the ultra-hot gas giants WASP-76b and WASP-121b\thanks{Based in part on Guaranteed Time Observations collected at the European Southern Observatory under ESO programmes 1102.C-0744, 1102.C-0958, and 1104.C-0350 by the ESPRESSO Consortium.}}

   \subtitle{Together with new detections of Co and Sr+ on WASP-121b }

   \author{T. Azevedo Silva\inst{\ref{IA-Porto}, \ref{UPorto}\thanks{\email{Tomas.Silva@astro.up.pt}}}
           \and O. D. S. Demangeon\inst{\ref{IA-Porto}, \ref{UPorto},\thanks{\email{Olivier.Demangeon@astro.up.pt}}} 
           \and N. C. Santos \inst{\ref{IA-Porto},\ref{UPorto}}
           \and R. Allart \inst{\ref{Montreal},\ref{Geneve}}
           \and F. Borsa \inst{\ref{INAF - Brera}}
           \and E. Cristo \inst{\ref{IA-Porto}, \ref{UPorto}}
           \and E. Esparza-Borges \inst{\ref{IAC}, \ref{ULaguna}}
           \and J. V. Seidel \inst{\ref{ESO}}
           \and E. Palle \inst{\ref{IAC-VL}}
           \and S. G. Sousa \inst{\ref{IA-Porto}}
           \and H. M. Tabernero \inst{\ref{INTA}}
           \and M. R. Zapatero Osorio \inst{\ref{INTA}}
           \and S. Cristiani \inst{\ref{INAF - Trieste}}
           \and F. Pepe \inst{\ref{Geneve}}
           \and R. Rebolo \inst{\ref{IAC},\ref{ULaguna}}
           \and V. Adibekyan \inst{\ref{IA-Porto}, \ref{UPorto}}
           \and Y. Alibert \inst{\ref{Bern}}
           \and S. C. C. Barros \inst{\ref{IA-Porto},\ref{UPorto}}
           \and F. Bouchy \inst{\ref{Geneve}}
           \and V. Bourrier \inst{\ref{Geneve}}
           \and G. Lo Curto \inst{\ref{ESO}}
           \and P. Di Marcantonio \inst{\ref{INAF - Trieste}}
           \and V. D'Odorico \inst{\ref{INAF - Trieste},\ref{Pisa},\ref{Inst-Trieste}}
           \and D. Ehrenreich \inst{\ref{Geneve}, \ref{Centre-Geneva}}
           \and P. Figueira \inst{\ref{Geneve}, \ref{IA-Porto}}
           \and J. I. Gonz\'alez Hern\'andez\inst{\ref{IAC}, \ref{ULL}}
           \and C. Lovis \inst{\ref{Geneve}}
           \and C. J. A. P. Martins \inst{\ref{IA-Porto},\ref{CAUP}}
           \and A. Mehner \inst{\ref{ESO}}
           \and G. Micela \inst{\ref{INAF - Palermo}}
           \and P. Molaro \inst{\ref{INAF - Trieste}, \ref{IFPU}}
           \and D. Mounzer \inst{\ref{Geneve}}
           \and N. J. Nunes \inst{\ref{IA-Lisboa},\ref{ULisboa}}
           \and A. Sozzetti \inst{\ref{INAF - Torino}}
           \and A. Su\'arez Mascareño \inst{\ref{IAC},\ref{ULaguna}}
           \and S. Udry \inst{\ref{Geneve}}
           }

    \institute{Instituto de Astrof\'isica e Ci\^{e}ncias do Espa\c co, Universidade do Porto, CAUP, Rua das Estrelas, 4150-762           Porto, Portugal \label{IA-Porto}
            \and
            Departamento de F\'isica e Astronomia, Faculdade de Ciências, Universidade do Porto, Rua do Campo Alegre, 4169-007 Porto, Portugal. \label{UPorto}
            \and
            Department of Physics, and Institute for Research on Exoplanets, Universit\'e de Montr\'eal, Montr\'eal, H3T 1J4, Canada. \label{Montreal}
            \and
            Observatoire astronomique de l’Universit\'e de Gen\`{e}ve, Chemin Pegasi 51, 1290 Versoix, Switzerland. \label{Geneve}
            \and
             INAF – Osservatorio Astronomico di Brera, Via Bianchi 46, 23807 Merate, Italy. \label{INAF - Brera}
            \and
             Instituto de Astrof\'isica de Canarias (IAC), 38205 La Laguna, Tenerife, Spain. \label{IAC}
            \and
            Departamento de Astrof\'isica, Universidad de La Laguna, E-38206, La Laguna, Tenerife, Spain. \label{ULaguna}
            \and
             European Southern Observatory, Alonso de C\'ordova 3107, Vitacura, Regi\'on Metropolitana, Chile. \label{ESO}
            \and
             Instituto de Astrof\'isica de Canarias, C. V\'ia L\'actea, s/n, E-38205, Spain. \label{IAC-VL}
            \and
            Centro de Astrobiolog\'ia (CSIC-INTA), Crta. Ajalvir km 4, E-28850 Torrej\'on de Ardoz, Madrid, Spain. \label{INTA}
            \and
             INAF – Osservatorio Astronomico di Trieste, via G. B. Tiepolo 11, I-34143, Trieste, Italy. \label{INAF - Trieste}
            \and
            Physikalisches Institut, University of Bern, Sidlerstrasse 5, 3012 Bern, Switzerland. \label{Bern}
            \and
            Scuola Normale Superiore, Piazza dei Cavalieri 7, I-56126 Pisa, Italy. \label{Pisa}
            \and
            Institute for Fundamental Physics of the Universe, Via Beirut 2, I-34151 Miramare, Trieste, Italy. \label{Inst-Trieste}     
            \and
            Centre Vie dans l'Univers, Facult\'e des sciences de l'Universit\'e de Gen\`eve, Quai Ernest-Ansermet 30, 1205 Geneva, Switzerland. \label{Centre-Geneva}
            \and
            Universidad de La Laguna (ULL), Departamento de Astrof\'isica,
            38206 La Laguna, Tenerife, Spain. \label{ULL}  
            \and
            Centro de Astrof\'isica da Universidade do Porto, Rua das Estrelas, 4150-762 Porto, Portugal. \label{CAUP}     
            \and
            INAF – Osservatorio Astronomico di Palermo, Piazza del Parlamento 1, Palermo, Italy. \label{INAF - Palermo}  
            \and
             Institute for Fundamental Physics of the Universe, Via Beirut 2, I-34151 Miramare, Trieste, Italy. \label{IFPU}
            \and
            Instituto de Astrof\'isica e Ci\^encias do Espa\c{c}o, Faculdade de Ci\^encias da Universidade de Lisboa \label{IA-Lisboa}
            \and
            Departamento de F\'isica, Faculdade de Ciências da Universidade de Lisboa, Campo Grande, Edificio C8, P-1749-016, Lisboa, Portugal. \label{ULisboa}
            \and
            INAF - Osservatorio Astrofisico di Torino, Via Osservatorio 20, I-10025 Pino Torinese (TO), Italy. \label{INAF - Torino}
            }

   \date{Received September 15, 1996; accepted March 16, 1997}

 
  \abstract
   {High-resolution spectroscopy studies of ultra-hot Jupiters have been key in our understanding of exoplanet atmospheres. Observing into the atmospheres of these giant planets allows for direct constraints on their atmospheric compositions and dynamics while laying the groundwork for new research regarding their formation and evolution environments.}
   {Two of the most well-studied ultra-hot Jupiters are WASP-76b and WASP-121b, with multiple detected chemical species and strong signatures of their atmospheric dynamics. We take a new look at these two exceptional ultra-hot Jupiters by reanalyzing the transit observations taken with ESPRESSO at the Very Large Telescope and attempt to detect additional species.}
   {To extract the planetary spectra of the two targets, we corrected for the telluric absorption and removed the stellar spectrum contributions. We then exploited new synthetic templates that were specifically designed for ultra-hot Jupiters in combination with the cross-correlation technique to unveil species that remained undetected by previous analyses.}
   {We add a novel detection of Ba+ to the known atmospheric compositions of WASP-76b and WASP-121b, the heaviest species detected to date in any exoplanetary atmosphere, with additional new detections of Co and Sr+ and a tentative detection of Ti+ for WASP-121b. We also confirm the presence of Ca+, Cr, Fe, H, Li, Mg, Mn, Na, and V on both WASP-76b and WASP-121b, with the addition of Ca, Fe+, and Ni for the latter. Finally, we also confirm the clear asymmetric absorption feature of Ca+ on WASP-121b, with an excess absorption at the bluer wavelengths and an effective planet radius beyond the Roche lobe. This indicates that the signal may arise from the escape of planetary atmosphere. 
   }
   {}


   \maketitle
%

\section{Introduction}

Ultra-hot Jupiters are currently the most readily accessible laboratories for the study of exoplanet atmospheres. Their size and large atmospheric scale heights, combined with the proximity to the host stars, make them appealing targets for the study of light that is transmitted through planetary atmospheres. With the recent developments of instruments for high-resolution spectroscopy \citep[e.g., ESPRESSO;][]{2010SPIE.7735E..0FP, 2021A&A...645A..96P}, the enhanced ability to retrieve high-resolution planetary spectra from the transit observations of ultra-hot Jupiters provided unique glimpses into the atmosphere of these extreme worlds. From the detection of several chemical species \citep[e.g.,][]{2019A&A...628A...9C, 2021A&A...646A.158T} to evaporating atmospheres \citep[e.g.,][]{2018NatAs...2..714Y} and the study of winds \citep[e.g.,][]{2020Natur.580..597E, 2021A&A...653A..73S}, resolving line features over short exposures has proven to be key to unraveling these distant alien atmospheres. 

Two of the best examples are WASP-76b and WASP-121b \citep{2016A&A...585A.126W, 2016MNRAS.458.4025D}. Both planets are inflated ultra-hot Jupiters on orbits with periods shorter than two days and equilibrium temperatures close to 2500 K (for more information, see Appendix \ref{table:system-params}). They were observed with ESPRESSO, and on one of the observing nights, the four UTs of the Very Large Telescope (VLT) were used. These targets are currently benchmarks in the study of atmospheric composition and dynamics \citep{2019A&A...623A.166S, 2020Natur.580..597E, 2021A&A...646A.158T, 2021JGRE..12606629F, 2022A&A...661A..78S}. Transmission spectroscopy studies report the detection of species from 
H\footnote{\label{footnote:tentative} For these species, only tentative detections are given. We confirm H in WASP-76b and Mn in WASP-121b as detected.}, Li, Na, Mg, K, Ca+, V, Cr, Mn, Fe, Co\footnotemark[\value{footnote}], Ni, and Sr+ 
for WASP-76b \citep{2021A&A...646A.158T, 2022AJ....163..107K} and 
H, Li, Na, Mg, K, Ca, Ca+, Sc+, V, Cr, Mn\footnotemark[\value{footnote}], Fe, Fe+, and Ni
for WASP-121b \citep{2020A&A...641A.123H, 2021A&A...645A..24B, 2021MNRAS.506.3853M} using ESPRESSO high-resolution observations for two transits each. The detection of these species, together with their respective velocities, broadening, and depths, provides important insights into the composition and dynamics of these extreme atmospheres.


In this study we revisit these datasets and extend the list of currently detected species. This paper is structured as follows. In Section \ref{section:Observational} we describe the observations, and in Section \ref{section:Planetary} we show how we reduced the data, extracted the planetary spectra, and cross-correlated these spectra with synthetic spectra that were specifically designed for ultra-hot Jupiters. The results and discussion are then given in Section \ref{section:Results}.

\section{Observational data}
\label{section:Observational}

We analyzed data from two transits of WASP-76b and WASP-121b that were observed with ESPRESSO. These same datasets have recently been studied and led to the detection of multiple species for each of the planets (WASP-76b: \citealt{2021A&A...646A.158T, 2022AJ....163..107K}; WASP-121b: \citealt{2021A&A...645A..24B}). The 1UT observations we used were obtained as part of the ESPRESSO Guaranteed Time Observations (program 1102.C-0744, PI: F. Pepe) that cover a wavelength range from 3800 \r{A} to 7880 \r{A}. The HR21 mode \citep[high resolution with 2$\times$1 binning, ][]{2021A&A...645A..96P} was used, and a resolution of $R \sim 140\,000$ was achieved. The 4UT observations were taken during the commissioning of ESPRESSO and used the MR42 mode (medium resolution with 4$\times$ and a 2$\times$ binning in the spatial and dispersion directions, $R \sim 70\,000$). The observations are summarized in Table \ref{table:nights}. For more information, we refer to \cite{2021A&A...646A.158T} and \cite{2021A&A...645A..24B}.

The data were reduced using version 2.2.8 of the Reduction Software (DRS) pipeline\footnote{Available at \url{http://www.eso.org/sci/software/pipelines/index.html}}. From the obtained data products, we used the sky-subtracted 1D (orders merged) spectra \texttt{S1D\_SKYSUB\_A}.

{\renewcommand{\arraystretch}{1.}
\begin{table*}
\caption{Summary of the transit observations of WASP-76b and WASP-121b.}
\label{table:nights}
\centering
\begin{tabular}{cccccccc}
\hline
Target & Night & Exp. time (s) & $\text{N}_{\text{obs}}$ & $\text{N}_{\text{obs, in-transit}}$ $^{a}$ & Airmass change$^{b}$ & S/N@550nm & Mode  \\ \hline
WASP-76b  & 2018-09-03 (N1) & 600 & 35 & 16 (20) & 1.99 - 1.13 - 1.37 & $\sim$101 & 1-UT HR21 \\
WASP-76b  & 2018-10-31 (N2) & 300 & 70 & 30 (37) & 1.42 - 1.13 - 2.64 & $\sim$79 & 1-UT HR21 \\
WASP-121b & 2018-11-30 (N1) & 300 & 29 & 17 (20) & 1.11 - 1.03 - 1.07 & $\sim$151 & 4-UT MR42 \\
WASP-121b & 2019-01-06 (N2) & 400 & 52 & 17 (23) & 1.20 - 1.03 - 1.61 & $\sim$47 & 1-UT HR21 \\

\hline
\\\end{tabular}
\tablefoot{$^{a}$ Number of spectra obtained between second and third contact. In parentheses, we list the number for the duration of the first to forth contact.
$^{b}$Airmass change during the night (start - minimum - end). 
}
\end{table*}}

\section{Planetary transmission spectrum and cross-correlation analysis}
\label{section:Planetary}

\subsection{Telluric correction}
\begin{figure*}[!h]
        \centering
        \includegraphics[width=\linewidth]{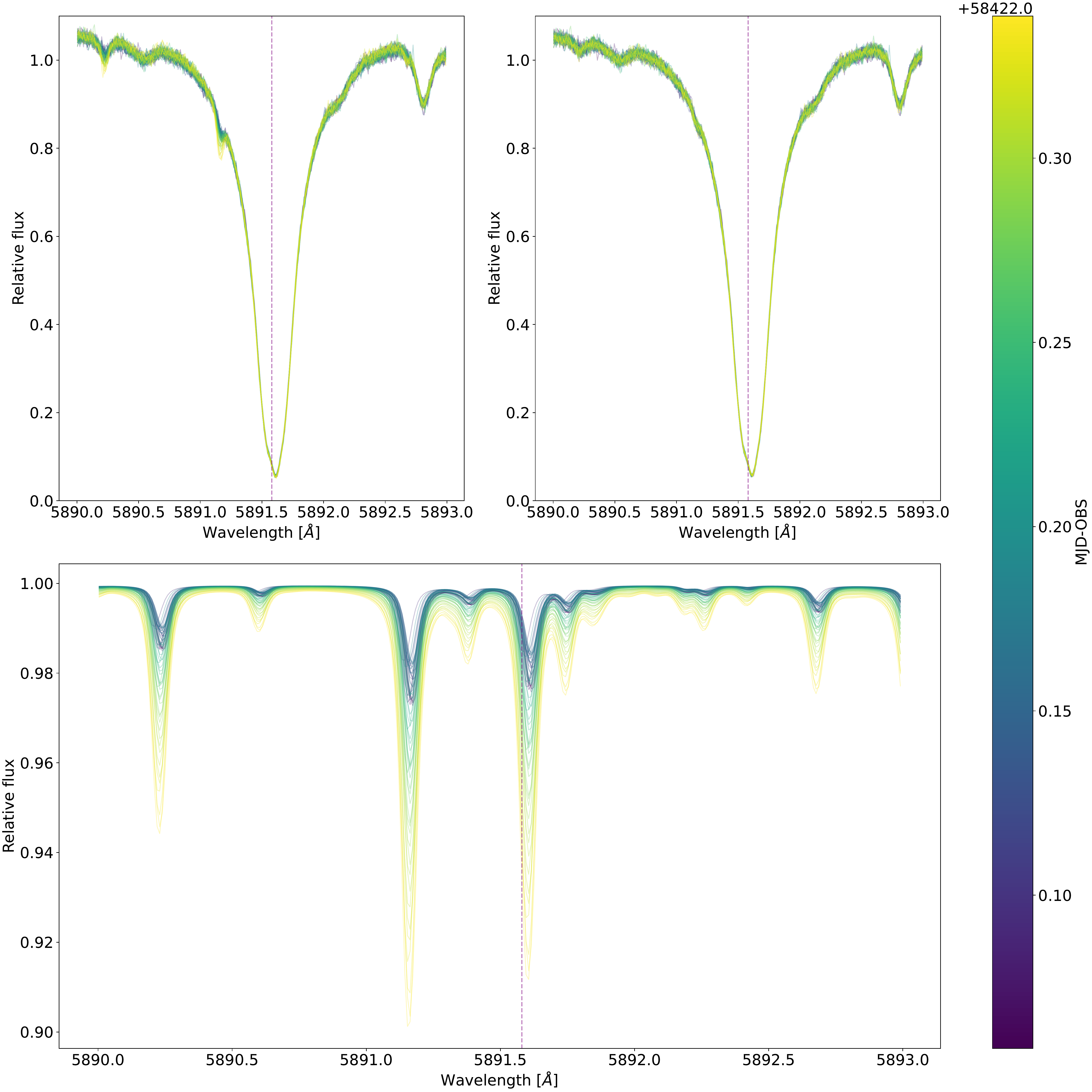}
        \put(-260,30){\frame{\includegraphics[width=6.5cm]{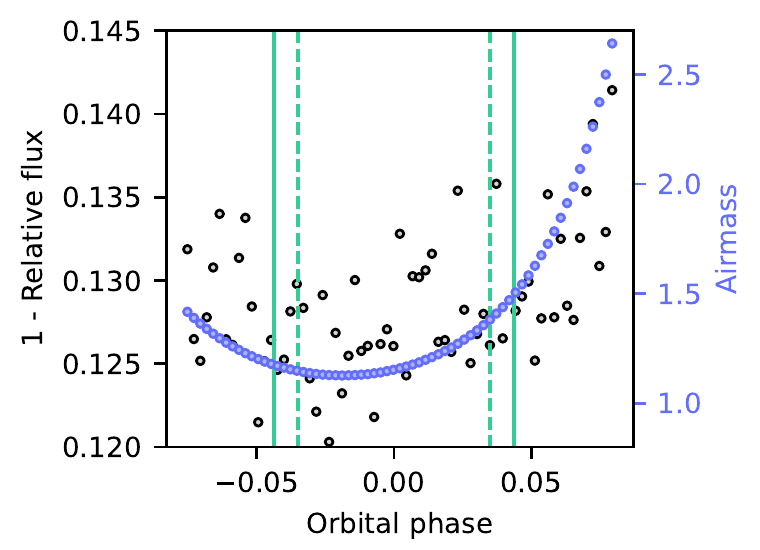}}}

        \caption{Illustration of the \texttt{Molecfit} telluric correction on a single line from the sodium D-line doublet in the observed spectrum of WASP-76b on the night of 2018 October 31 in all exposures. The vertical dashed purple line represents the expected line position.
        \textbf{Top left}: Sodium line prior to the telluric correction. \textbf{Top right}: Sodium line after the \texttt{Molecfit} telluric correction. \textbf{Bottom}: Ratio of the nontelluric-corrected over telluric-corrected spectral lines. \textbf{Inset}: Flux and air-mass variation over the transit observation. As expected for observations later in the night, with higher airmass, the removed telluric features are stronger.
}
    \label{fig:telluric_comp}
\end{figure*}

Before light reaches the VLT mirrors, it crosses Earth's atmosphere, which leaves its imprints on the observed spectra in the form of terrestrial absorption features. To correct for Earth absorption lines, we use the \texttt{Molecfit}\footnote{Available at \url{https://www.eso.org/sci/software/pipelines/skytools/ molecfit}} pipeline \citep{2015A&A...576A..77S, 2015A&A...576A..78K} within the \texttt{ExoReflex} environment  \citep{2013A&A...559A..96F} as in \cite{2017A&A...606A.144A}. With this same tool, we simultaneously accounted for the barycentric Earth radial velocity (BERV) in each exposure \citep{2017A&A...606A.144A}. The wavelength regions we used to fit the terrestrial features are 6890-6900 \r{A}, 7160-7340 \r{A}, and 7590-7770 \r{A} \citep[as in ][]{2021A&A...646A.158T}. These regions were selected because they are rich in telluric lines and poor in stellar features. We alternatively attempted to optimize these regions by defining more smaller regions and were very careful to exclude any stellar features. However, the two reductions produced similar results (illustrated in Fig. \ref{Appendix:Telluric}). In Appendix \ref{Appendix:Telluric} we provide our \texttt{Molecfit} input parameters together with an illustrative plot of \textit{\textup{before}} and  \textit{\textup{after}} telluric correction around a single line of the sodium doublet (5891.58\,\r{A}, Fig. \ref{fig:telluric_comp}).

\subsection{Planetary spectrum extraction}

To extract the planetary transmission spectrum, we followed a reasoning similar to the techniques outlined by \cite{2008ApJ...673L..87R} and \cite{2015A&A...577A..62W}. These techniques have been successfully applied and improved in many recent studies using high-resolution transmission spectroscopy \citep[e.g.,][]{2017A&A...603A..73Y, 2020A&A...644A.155A, 2021A&A...646A.158T, 2021A&A...645A..24B, 2021A&A...647A..26C, 2022MNRAS.513L..15S}.

We started by shifting each individual spectrum into the reference frame of the star. By applying Keplerian models, we corrected for the Doppler shifts arising from the orbital pull of the planet. We then created a template of the star without any planetary absorption. To do this, we combined the out-of-transit exposures into a stellar spectrum with a high signal-to-noise ratio, called the master-out. To build this master-out, we first corrected the shape of the retrieved continuum changes for flux variations across the exposures that are due to different atmospheric conditions (airmass, seeing, etc.). We started by selecting the spectrum with the lowest airmass as the reference, and divided each of the individual spectra by this reference. We then masked the region of the spectrum in which the signal-to-noise ratio was low\footnote{Regions in which the signal-to-noise ratio is significantly below the baseline (> $5\sigma$), e.g., <3900 \r{A}.} or that included strong telluric features\footnote{Regions in which the telluric correction was inadequate because of low flux from strong telluric absorption,  e.g., 6850-7700 \r{A}.} and fit a third-degree polynomial function to the unmasked data. We divided out the trend for each of the individual spectra and combined the corrected out-of-transit spectra with a weighted average to create a preliminary master-out, using the inverse square of flux uncertainties as the weights. 

We then performed a similar procedure, in which we instead divided the individual spectra by the created preliminary master-out. The reason for this additional step was that all the individual spectra (in-transit and out-of-transit) were later divided by the same out-of-transit template. As a result of this division, a clear wiggle pattern becomes visible across the spectra. This is to be expected for the ESPRESSO data, as noted by \cite{2021A&A...646A.158T} and \cite{2021A&A...645A..24B}. In Appendix \ref{Appendix:Wiggles} we show the wiggles across the wavelength range for a single exposure of WASP-121b with the 4UTs. The wiggle amplitude and frequency vary across wavelengths and do not behave as perfect sinusoids. To correct for these wiggles 
together with 
residual features from the flux normalization, we fit each spectrum with splines\footnote{We fit the splines over spectral segments of $\sim$ 200 \r{A}. As a first step, we also tried to fit the wiggles in each exposure using varying sinusoids on limited wavelength intervals, but we found better results with the spline method.}. We then divided our spectra over the fit splines. An example of this fitting is illustrated in Fig. \ref{fig:wiggles}.

Finally, we shifted these individual spectra according to the orbital velocity of the planet and computed the weighted averages of the spectral regions of interest. We thus created masters of the out-of-transit and in-transit data in the rest frames of the planet and star. 

\subsection{Cross-correlation analysis}

After the planetary transmission spectrum was retrieved for each of the exposures, we performed a cross-correlation analysis with all the templates from \texttt{The Mantis Network I}\footnote{We performed the CCF study with templates from the following species: 
Li, O, Na, Mg, Al, Si, P, S, K, Ca, Ca+, Sc, Sc+, Ti, Ti+, V, V+, Cr, Cr+, Mn, Mn+, Fe, Fe+, Co, Ni, Ni+, Cu, Zn, Ga, Ge, Rb, Sr, Sr+, Y, Y+, Zr, Zr+, Nb, Mo, Ru, Rh, Pd, Cd, In, Sn, Te, Cs, Ba, Ba+, La, La+, Ce, Ce+, Pr, Nd, Nd+, Sm, Sm+, Eu, Eu+, Gd, Gd+, Tb, Tb+, Dy, Dy+, Ho, Ho+, Er, Er+, Tm, Tm+, Yb, Lu, Lu+, Hf, Hf+, W, Re, Os, Ir, Pt, Tl, Pb, Bi, Th, Th+, U, and U+. Some other species were available but had no lines in our wavelength region of interest.} 
 \citep{2021arXiv211211380K}. These templates allow for the use of individual lines that would otherwise go unnoticed due to the high noise in the retrieved planetary spectrum. By using the cross-correlation function (CCF) approach, these individual lines can be combined into a detectable line profile. For the WASP-76b and WASP-121b line mask, we considered an isothermal atmosphere at 2500 K\footnote{For Fe+, given the empty line list at 2500 K, we used the mask corresponding to a temperature of 3000 K.} because of the equilibrium temperatures of the two planets \citep{2016A&A...585A.126W, 2016MNRAS.458.4025D} and in line with the application outlined in \cite{2021arXiv211211380K}. The number of lines used per species in our wavelength range is listed in Appendix \ref{Appendix:lines_mask}.

We computed the CCFs in the planet rest frame, with the previously mentioned binary masks, using the \texttt{SciPy} cross-correlation tool \citep{2020SciPy-NMeth}. For an easier visualization of the absorption signal, we also produced tomography plots and computed the absorption across a 2D cross-correlation grid in the Kp - planet velocity plane (hereafter \textit{\textup{Kp plots}}). For examples, see Fig \ref{fig:W76 three barium}.

The tomography plots show the signature of the Rossiter-McLaughlin effect (RM). 
Because of current limitations and uncertainties \citep{2021A&A...647A..26C} in the modeling of the RM and the associated center-to-limb variation (CLV), we chose not to attempt to model these effects. Instead of modeling, we masked the region that matched the stellar velocities (-15 km/s to +15 km/s in the stellar rest frame) and combined the remaining signal. In addition to masking the stellar component, this process removes part of the planetary signal, which at times can cause a diminished S/N. An illustration of this masking for the F9 mask is shown in Appendix \ref{Appendix: 15Masking}. 

\begin{figure*}[th]
     \centering
     \begin{subfigure}[b]{0.33\textwidth}
         \centering
         \includegraphics[width=\textwidth]{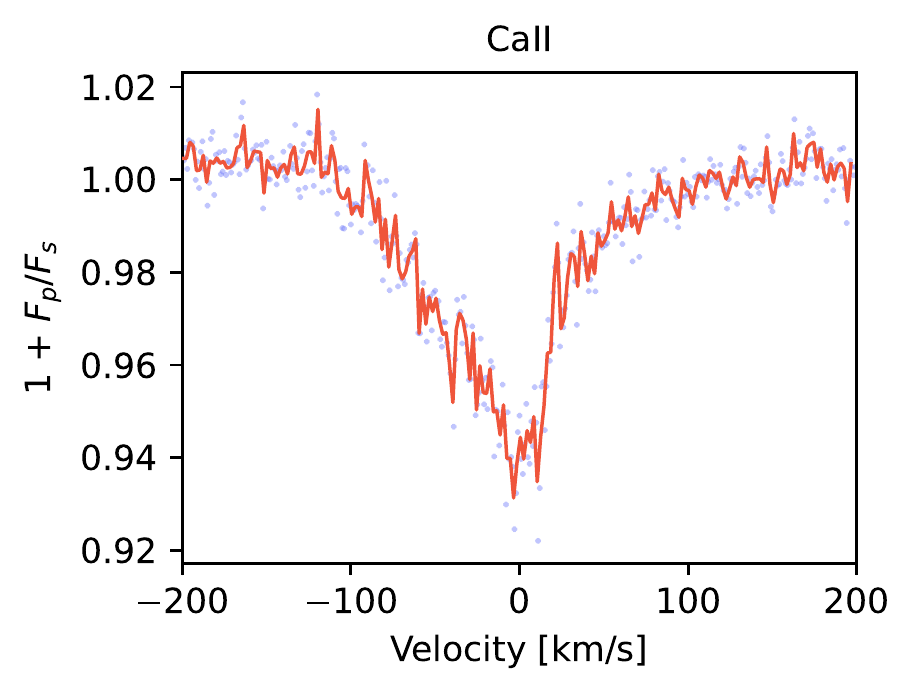}
         \label{fig:y equals x}
     \end{subfigure}
     \hfill
     \begin{subfigure}[b]{0.33\textwidth}
         \centering
         \includegraphics[width=\textwidth]{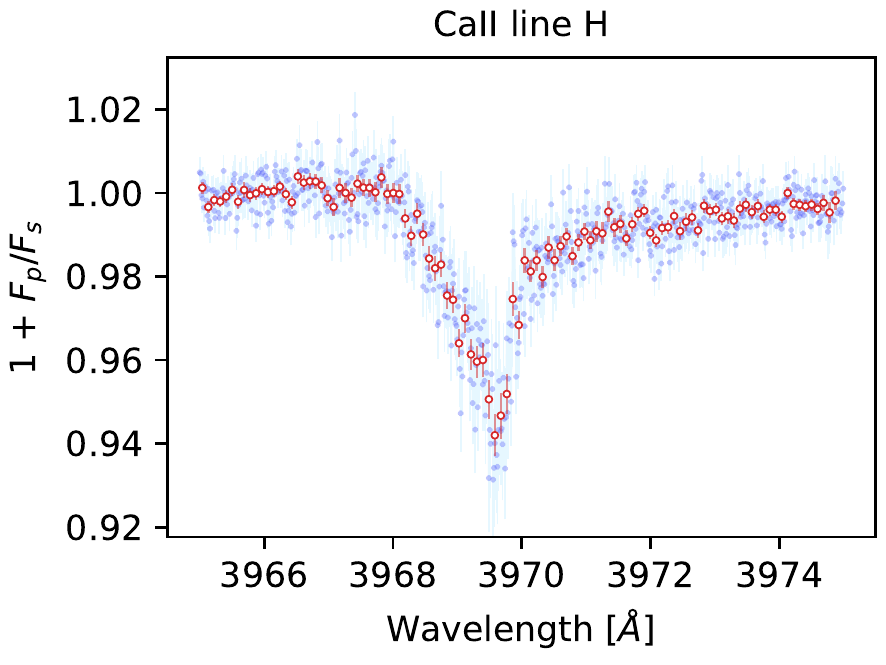}
         \label{fig:three sin x}
     \end{subfigure}
     \hfill
     \begin{subfigure}[b]{0.33\textwidth}
         \centering
         \includegraphics[width=\textwidth]{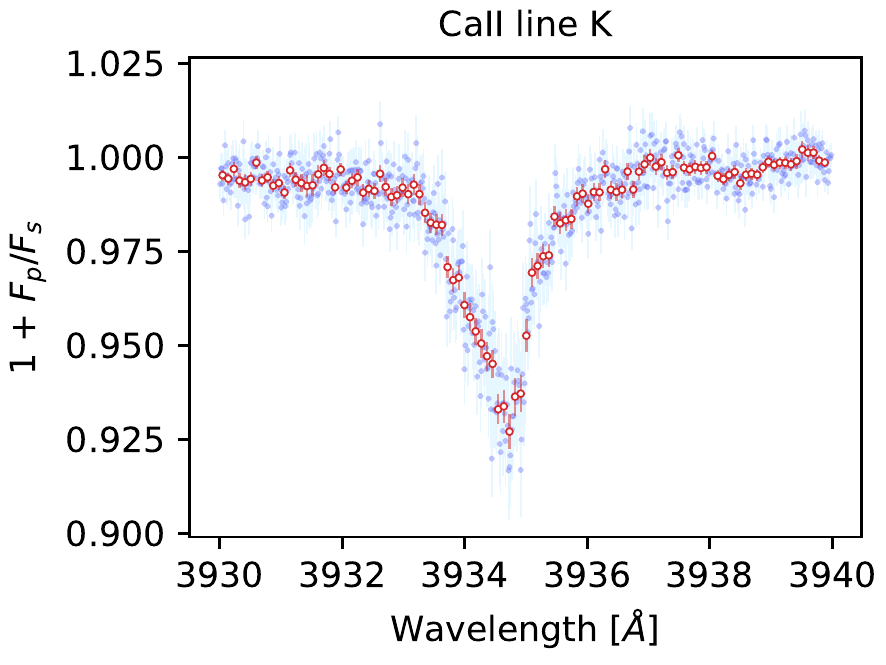}
         \label{fig:five over x}
     \end{subfigure}
        \caption{ \textbf{Left}: CCF of the Ca+ mask with the retrieved planetary spectrum for WASP-121b (night 2, 4UTs) at the planetary rest frame. \textbf{Center and right}: WASP-121b retrieved planetary spectrum at the Ca+ H\&K lines.}
        \label{fig:Calcium}
\end{figure*}

We fit Gaussian profiles to the CCFs of the species with visible planetary absorption features. We used the \texttt{lmfit} Python package \citep{2016ascl.soft06014N} to find the best-fit values and respective uncertainties. We did not perform a CCF for hydrogen\footnote{No \texttt{Mantis} masks were available for hydrogen.}, but instead directly fit the H$\alpha$ line on the retrieved planetary spectrum (master-in).
However, we do not expect the Ca+ absorption to follow a Gaussian profile.
As mentioned in \cite{2021A&A...645A..24B}, the Ca+ H\&K lines are significantly blueshifted and their profiles are wider and deeper than those of any other species, which suggests that it extends beyond the Roche lobe \footnote{We estimated the effective planet radius at the line center by assuming $R_\text{eff}^2 / R_\text{p}^2 = (\delta + h)/\delta$, where $\delta$ is the transit depth and $h$  is the transmitted line amplitude \citep{2020A&A...635A.171C}. For Ca+ in WASP-121b, we find effective radii of $\sim$1.7 $R_\text{p}$ (night 1) and $\sim$2.1 $R_\text{p}$ (night 2), in agreement with \cite{2021A&A...645A..24B}}. The authors also noticed slight asymmetries in these line profiles and pointed to planetary atmospheric escape as a possible explanation. In our analysis, we found similar results and a clear asymmetric profile for the 4UT partial transit. The individual H\&K lines and the combined CCF using the \texttt{Mantis} mask both show extended blueshifted absorption (see Fig. \ref{fig:Calcium}) beyond an underlying Gaussian profile. This indicates that calcium may be escaping the planetary atmosphere. A detailed modeling of this profile might confirm this hypothesis and allow an estimate of the atmospheric escape rate of calcium.


\section{Results and discussions}
\label{section:Results}

In Table \ref{table:species} we list the species for which a visual inspection showed tentative absorption features, together with the best-fit values and uncertainties from the Gaussian fit parameters (amplitude, center RV, and full width at half maximum, FWHM). The CCF line profile and Kp-plots for each of these are provided in Appendix \ref{Appendix:WASP-all}.

For WASP-76b, we find a new chemical species in the atmosphere of this planet, Ba+. It shows a strong absorption signal on both of the observing nights. We also confirm the detections made by \cite{2021A&A...646A.158T} and \cite{2022AJ....163..107K} of H, Li, Na, Mg, Ca+, V, Cr, Mn, and Fe. The strongest K lines lie in a region that we excluded because of high telluric contamination; therefore, K was not included in our study. We did not recover Co, Ni, and Sr+ either, which were claimed by \cite{2022AJ....163..107K}. The reason might be differences in the retrieval process. 

\begin{figure*}[h!]
     \centering
     \begin{subfigure}[b]{0.33\textwidth}
         \centering
         \includegraphics[width=\textwidth]{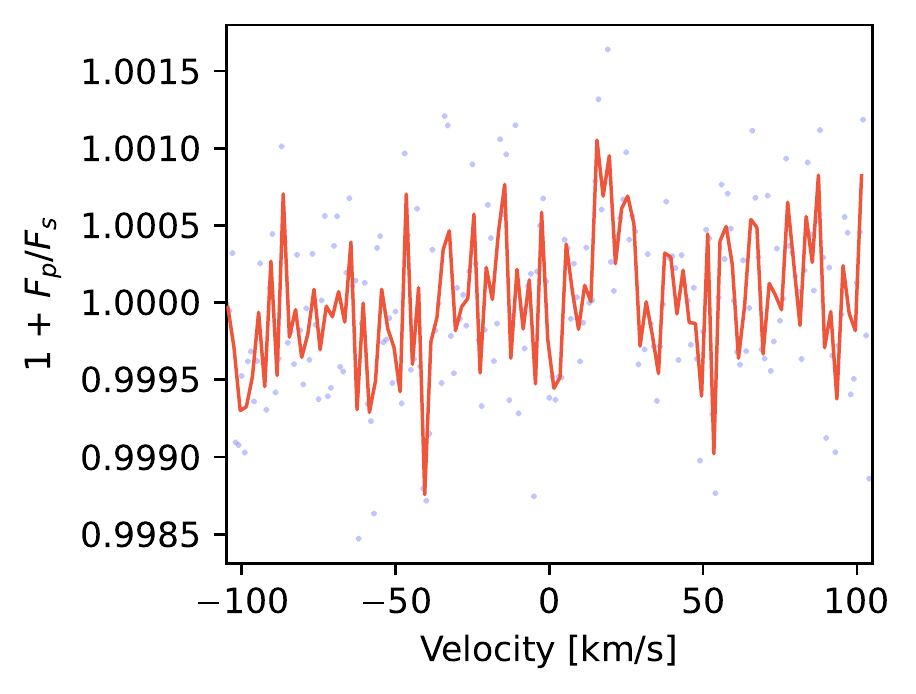}
         \label{fig:y equals x}
     \end{subfigure}
     \hfill
     \begin{subfigure}[b]{0.33\textwidth}
         \centering
         \includegraphics[width=\textwidth]{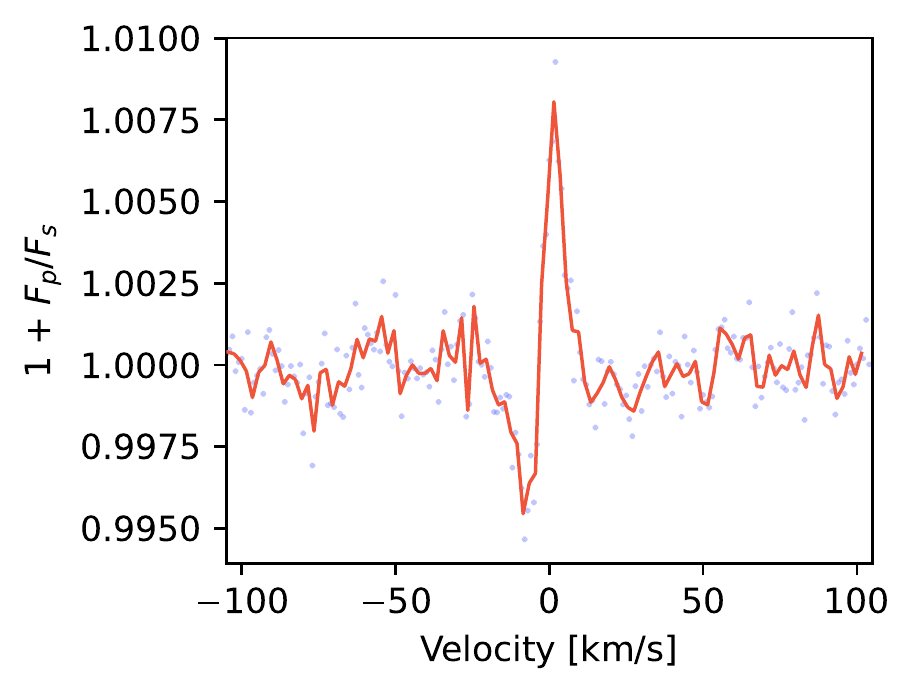}
         \label{fig:three sin x}
     \end{subfigure}
     \hfill
     \begin{subfigure}[b]{0.33\textwidth}
         \centering
         \includegraphics[width=\textwidth]{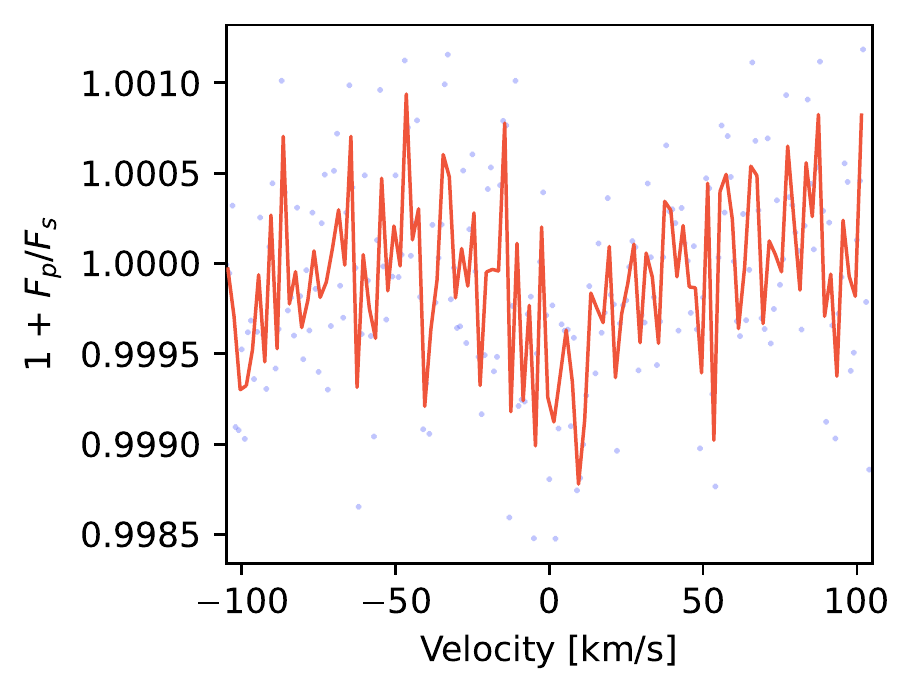}
         \label{fig:five over x}
     \end{subfigure}
        \caption{ \textbf{Left}: CCF of the Ti+ mask with the retrieved planetary spectrum for WASP-121b (night 2, 4UTs). \textbf{Center:} CCF built from considering the contributions from spectra between -15 km/s and 15 km/s in the stellar frame of reference alone. \textbf{Right}: Same as the initial CCF, but excluding the previous stellar contribution region. All the plots are in the planetary rest frame.}
        \label{fig:Ti}
\end{figure*}

We confirm the detections of H, Li, Na, Mg, Ca+, V, Cr, Mn, Fe, and Fe+ that were claimed by \cite{2021A&A...645A..24B} and of Ca and Ni by \cite{2021MNRAS.506.3853M} in WASP-121b. The new species we claim to be present in the atmosphere of WASP-121b are Ba+, Co, and Sr+. In addition to these, we also find tentative evidence of the presence of Ti+. It was visible when we removed the signal that matched the stellar velocities (see Fig. \ref{fig:Ti}). 

Overall, we see the trend that most of the detected species are blueshifted relative to their expected line positions. This trend was also observed in previous studies \citep[e.g.,][]{2021AJ....161..152C, 2022AJ....163..155P} and is expected to be due to the winds across the terminator, which move from the dayside to the nightside of the planet.

\begin{figure*}[h]
     \centering
     \begin{subfigure}[b]{0.3\textwidth}
         \centering
         \includegraphics[width=\textwidth]{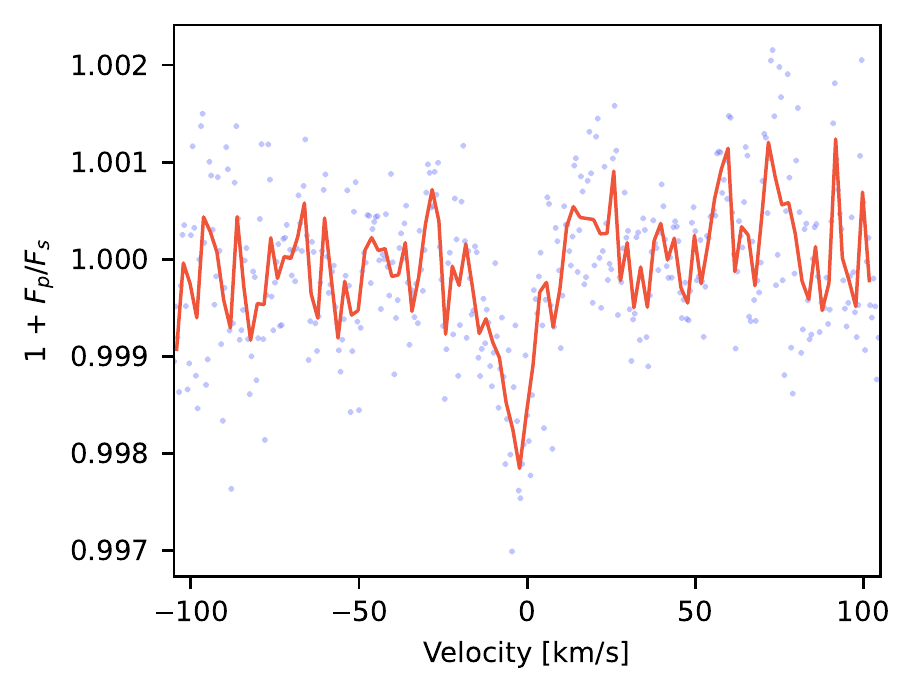}
         \label{fig:y equals x}
     \end{subfigure}
     \hfill
     \begin{subfigure}[b]{0.3\textwidth}
         \centering
         \includegraphics[width=\textwidth]{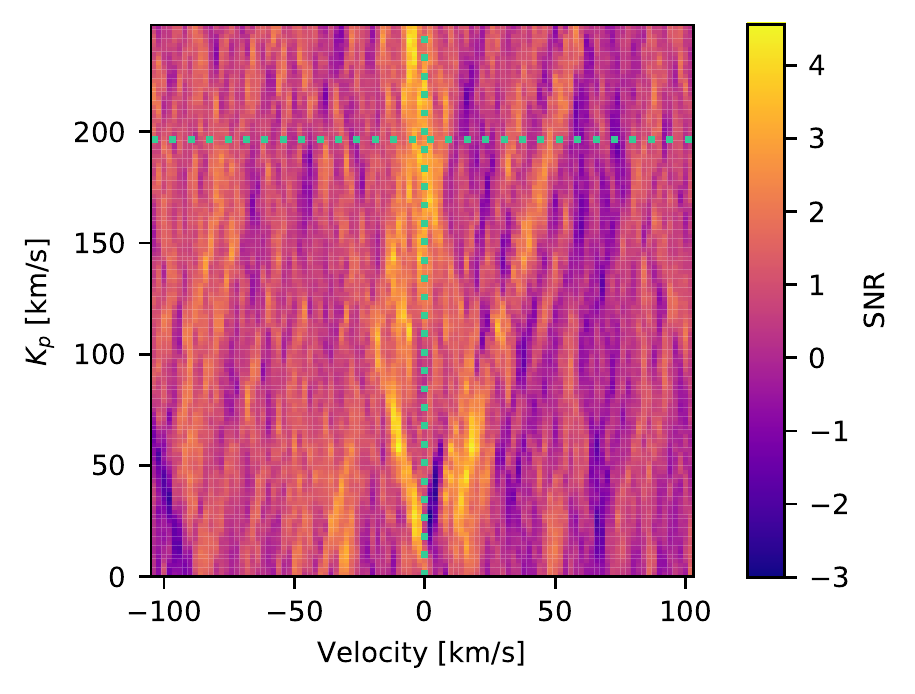}
         \label{fig:three sin x}
     \end{subfigure}
     \hfill
     \begin{subfigure}[b]{0.35\textwidth}
         \centering
         \includegraphics[width=\textwidth]{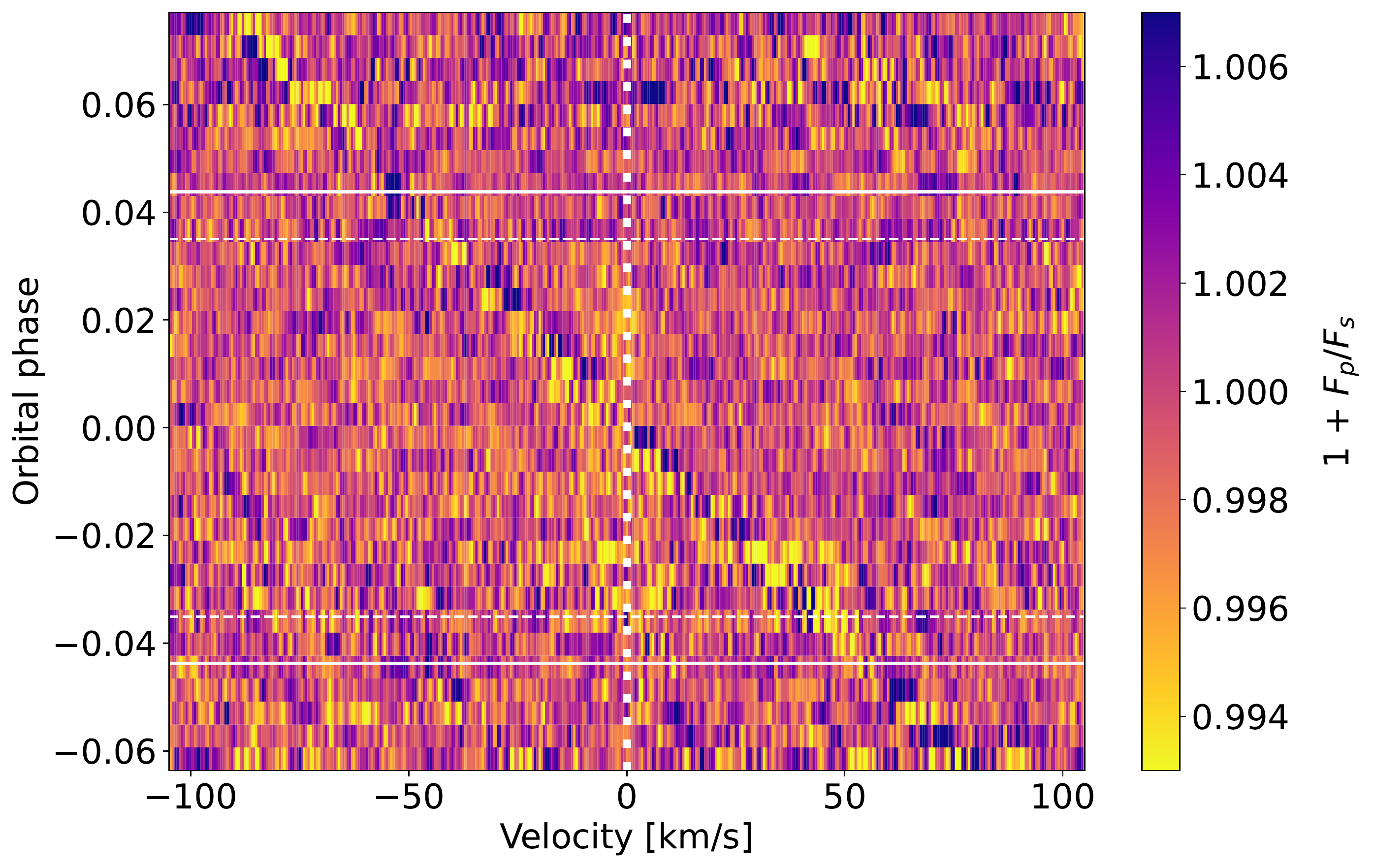}
         \label{fig:five over x}
     \end{subfigure}

     \begin{subfigure}[b]{0.3\textwidth}
         \centering
         \includegraphics[width=\textwidth]{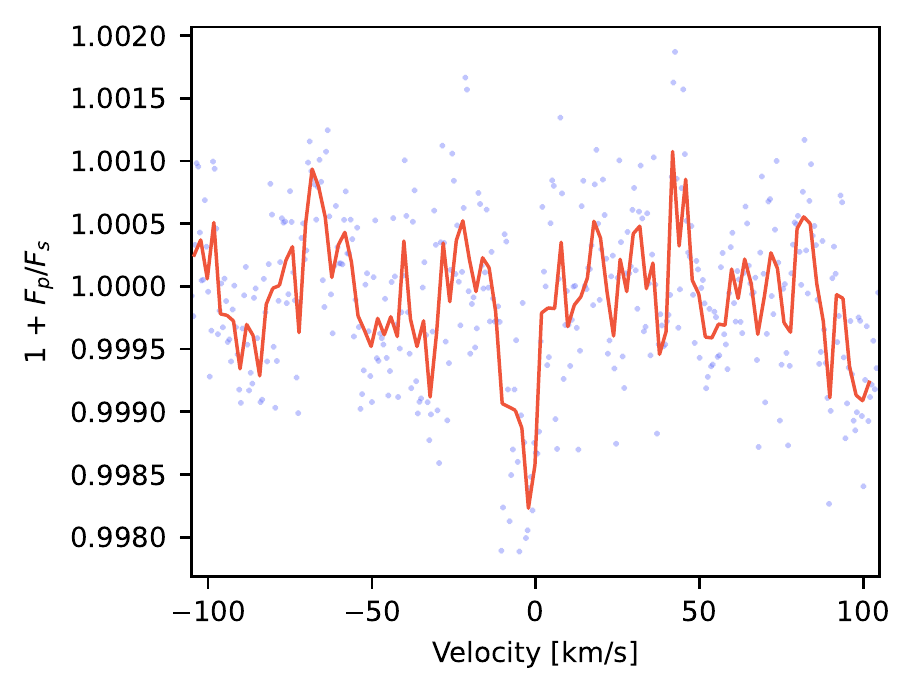}
         \label{fig:y equals x}
     \end{subfigure}
     \hfill
     \begin{subfigure}[b]{0.3\textwidth}
         \centering
         \includegraphics[width=\textwidth]{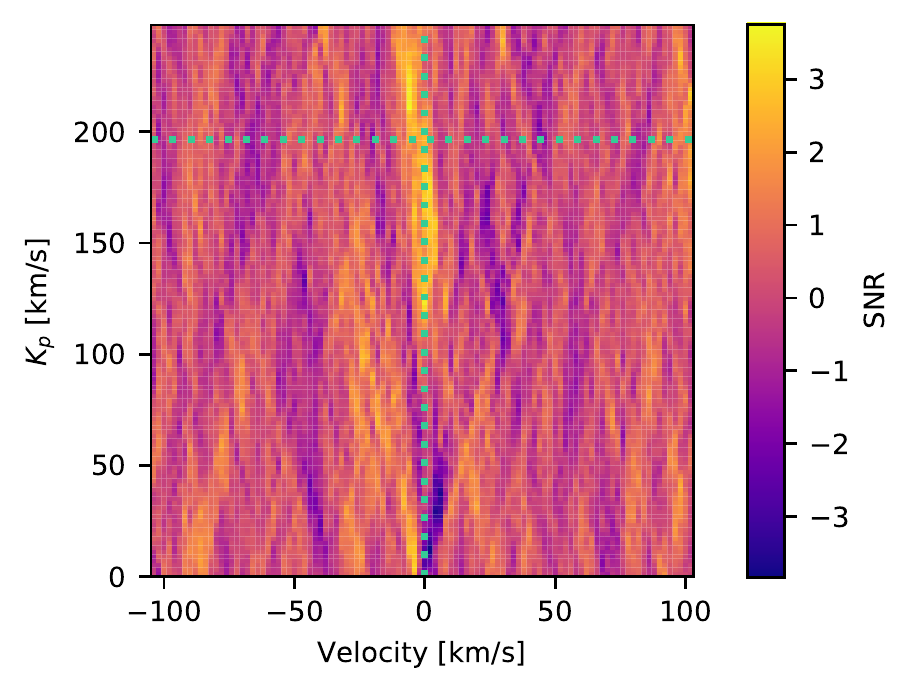}
         \label{fig:three sin x}
     \end{subfigure}
     \hfill
     \begin{subfigure}[b]{0.35\textwidth}
         \centering
         \includegraphics[width=\textwidth]{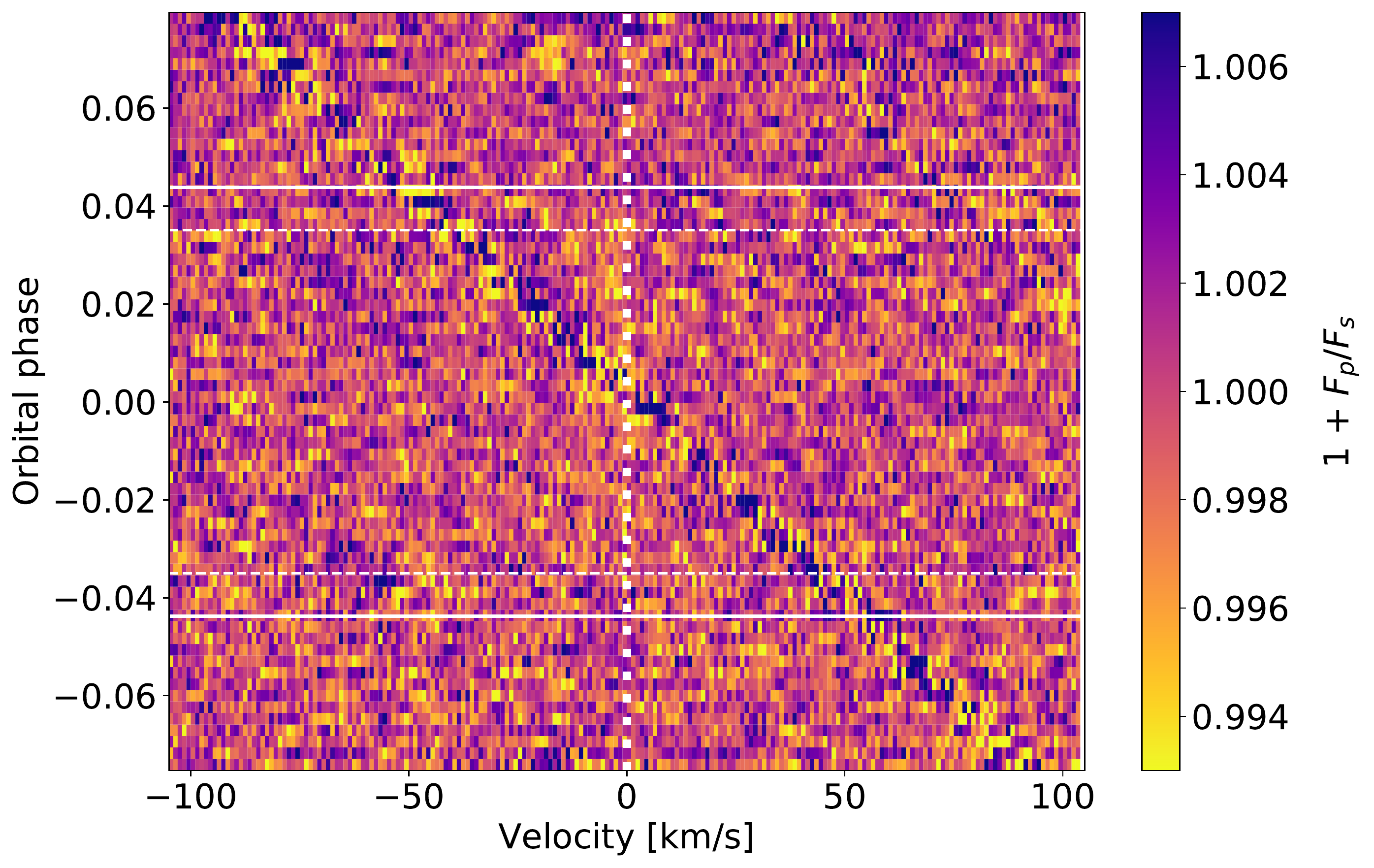}
         \label{fig:five over x}
     \end{subfigure}
        \caption{Cross-correlation of the Ba+ mask with the retrieved planetary spectra for the two observation nights of WASP-76b (\textbf{top} on 2018 September 3, and \textbf{bottom} on 2018 October 31). \textbf{Left:}  Cross-correlation functions of the averaged in-transit exposures on the planetary frame of reference. \textbf{Center:} \textit{\textup{Kp-plots.}} Map of the sum of all the individual exposures in the planet rest frame across different values in the \textit{\textup{Kp - planet velocity}} plane. The dashed green lines represent the expected position of the planetary signal on this map. \textbf{Right:} Tomographic plots in the planetary rest frame. The white lines represent the transit contacts.  }
        \label{fig:W76 three barium}
    
    \vspace{1cm}
    
     \centering
     \begin{subfigure}[b]{0.3\textwidth}
         \centering
         \includegraphics[width=\textwidth]{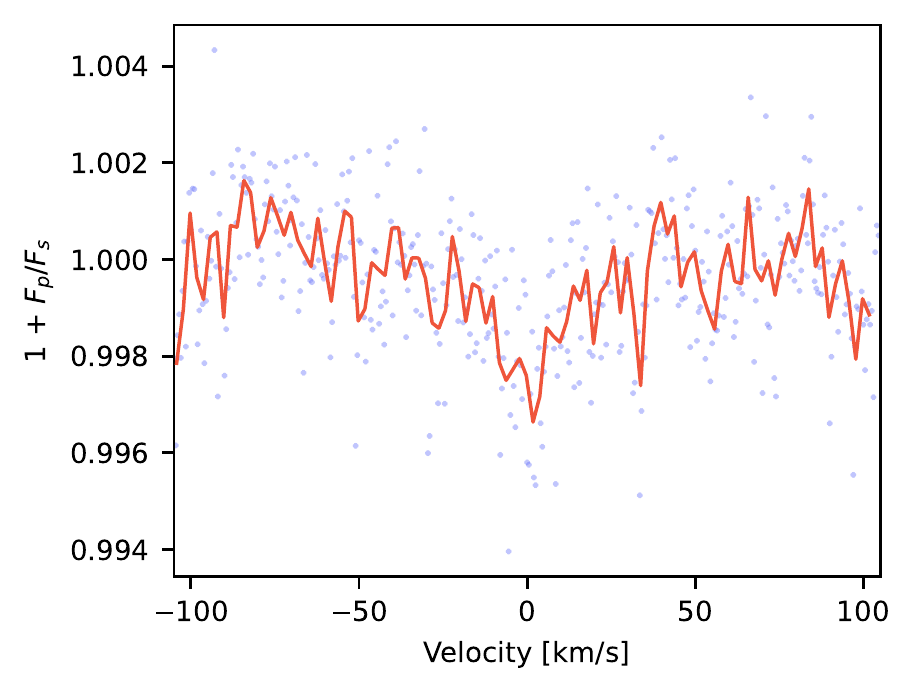}
         \label{fig:y equals x}
     \end{subfigure}
     \hfill
     \begin{subfigure}[b]{0.3\textwidth}
         \centering
         \includegraphics[width=\textwidth]{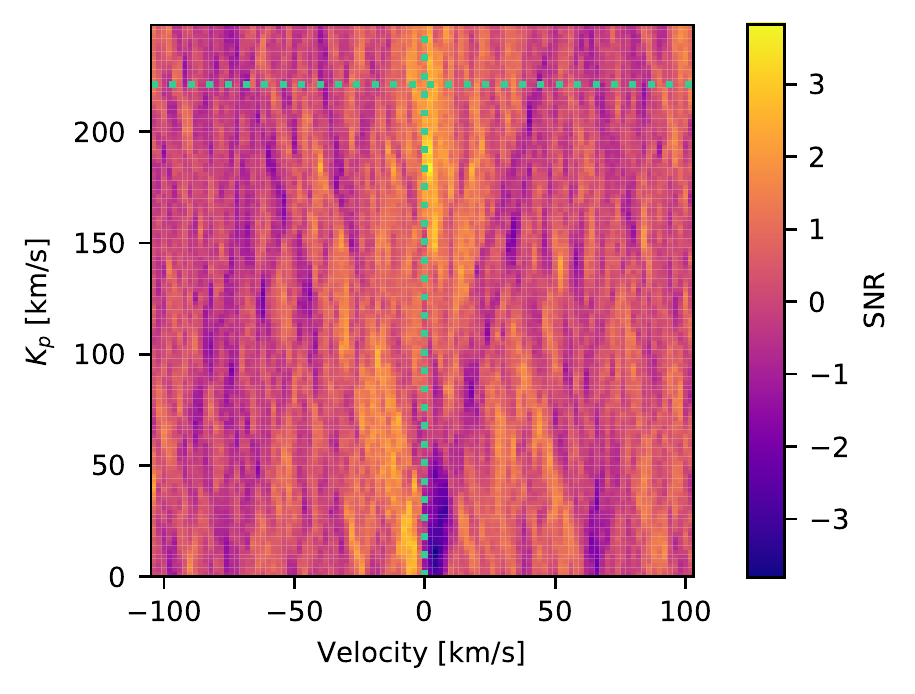}
         \label{fig:three sin x}
     \end{subfigure}
     \hfill
     \begin{subfigure}[b]{0.35\textwidth}
         \centering
         \includegraphics[width=\textwidth]{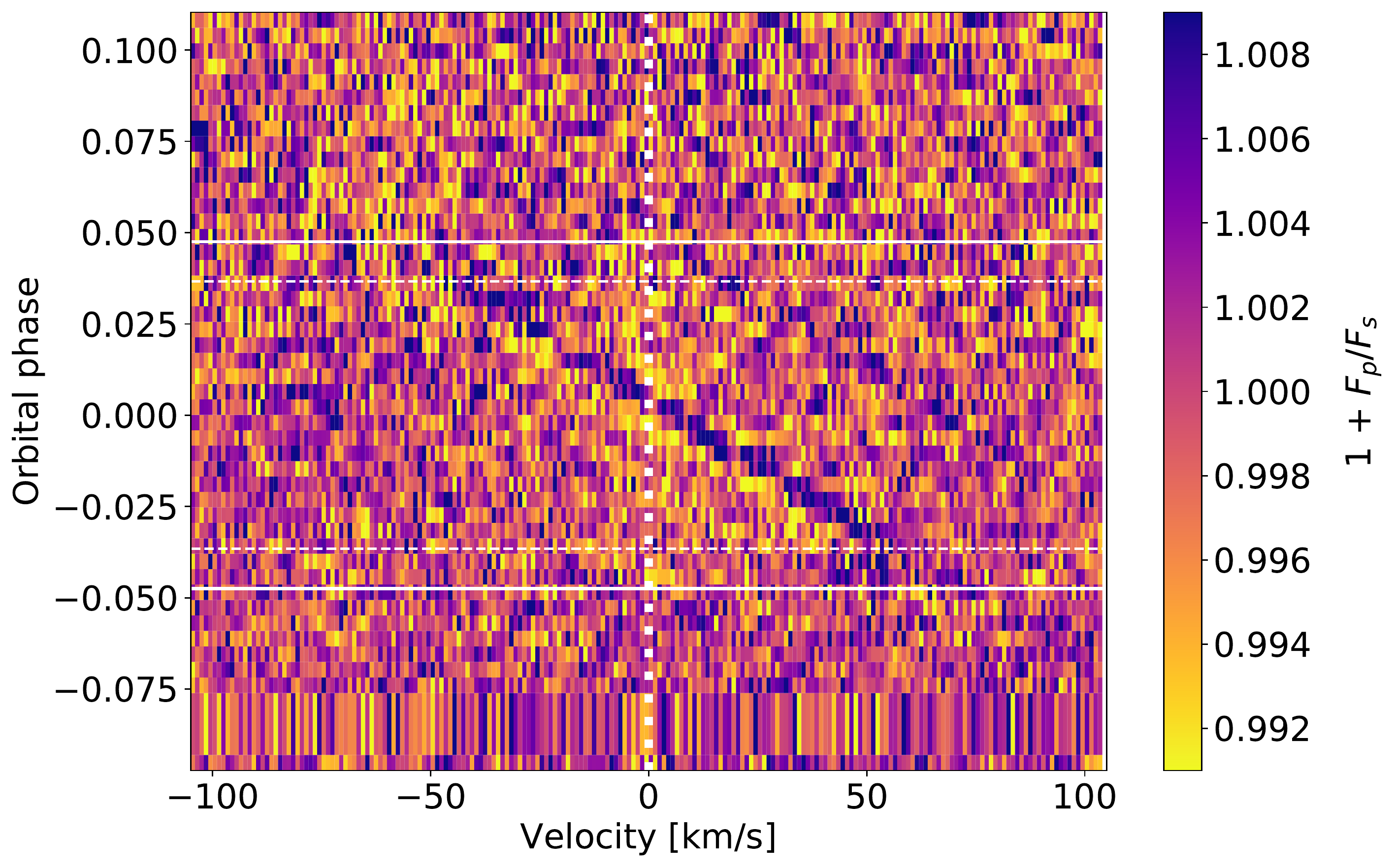}
         \label{fig:five over x}
     \end{subfigure}

     \begin{subfigure}[b]{0.3\textwidth}
         \centering
         \includegraphics[width=\textwidth]{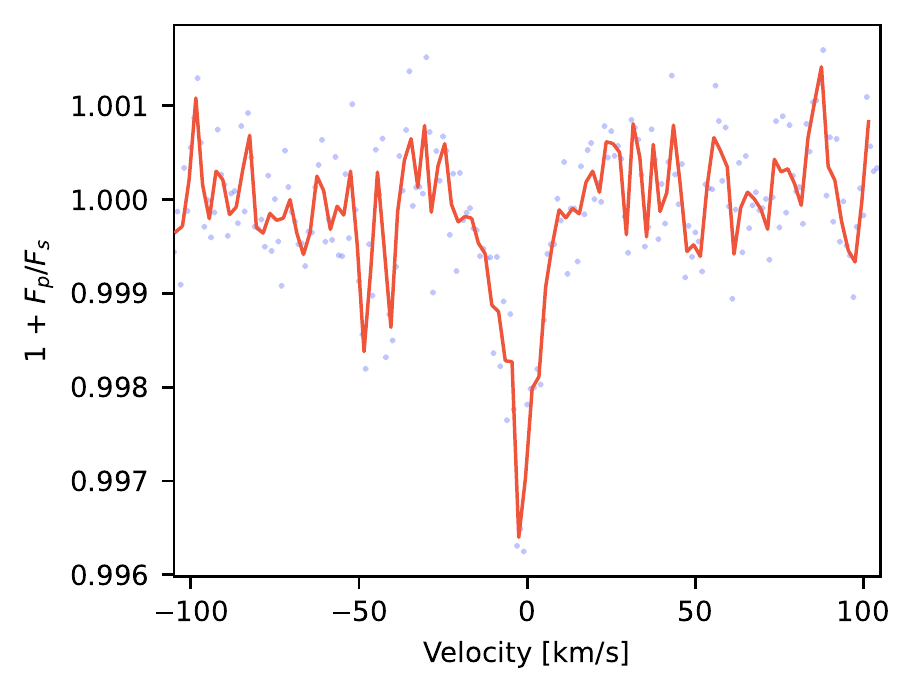}
         \label{fig:y equals x}
     \end{subfigure}
     \hfill
     \begin{subfigure}[b]{0.3\textwidth}
         \centering
         \includegraphics[width=\textwidth]{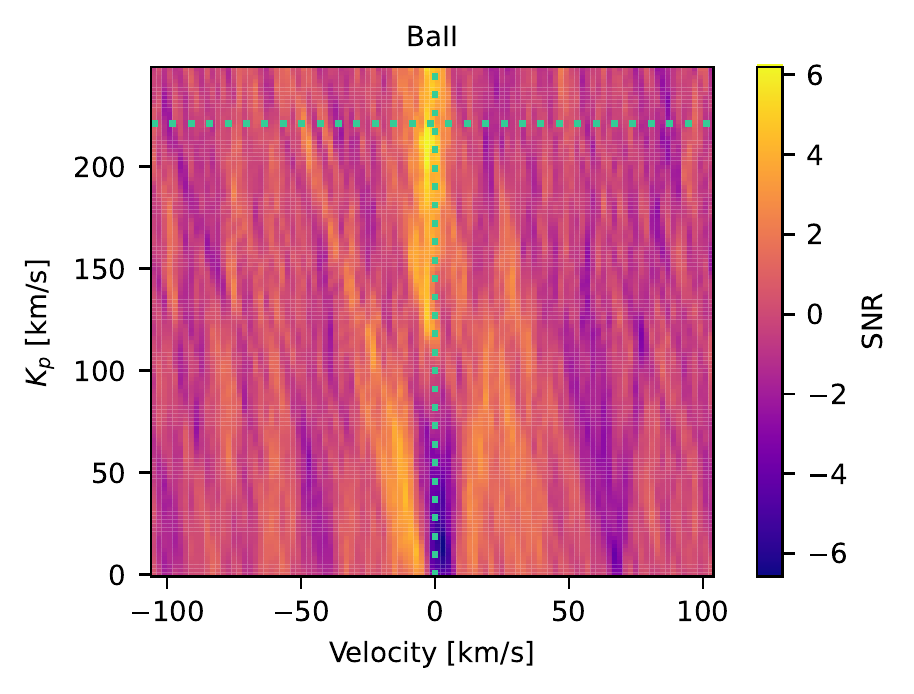}
         \label{fig:three sin x}
     \end{subfigure}
     \hfill
     \begin{subfigure}[b]{0.35\textwidth}
         \centering
         \includegraphics[width=\textwidth]{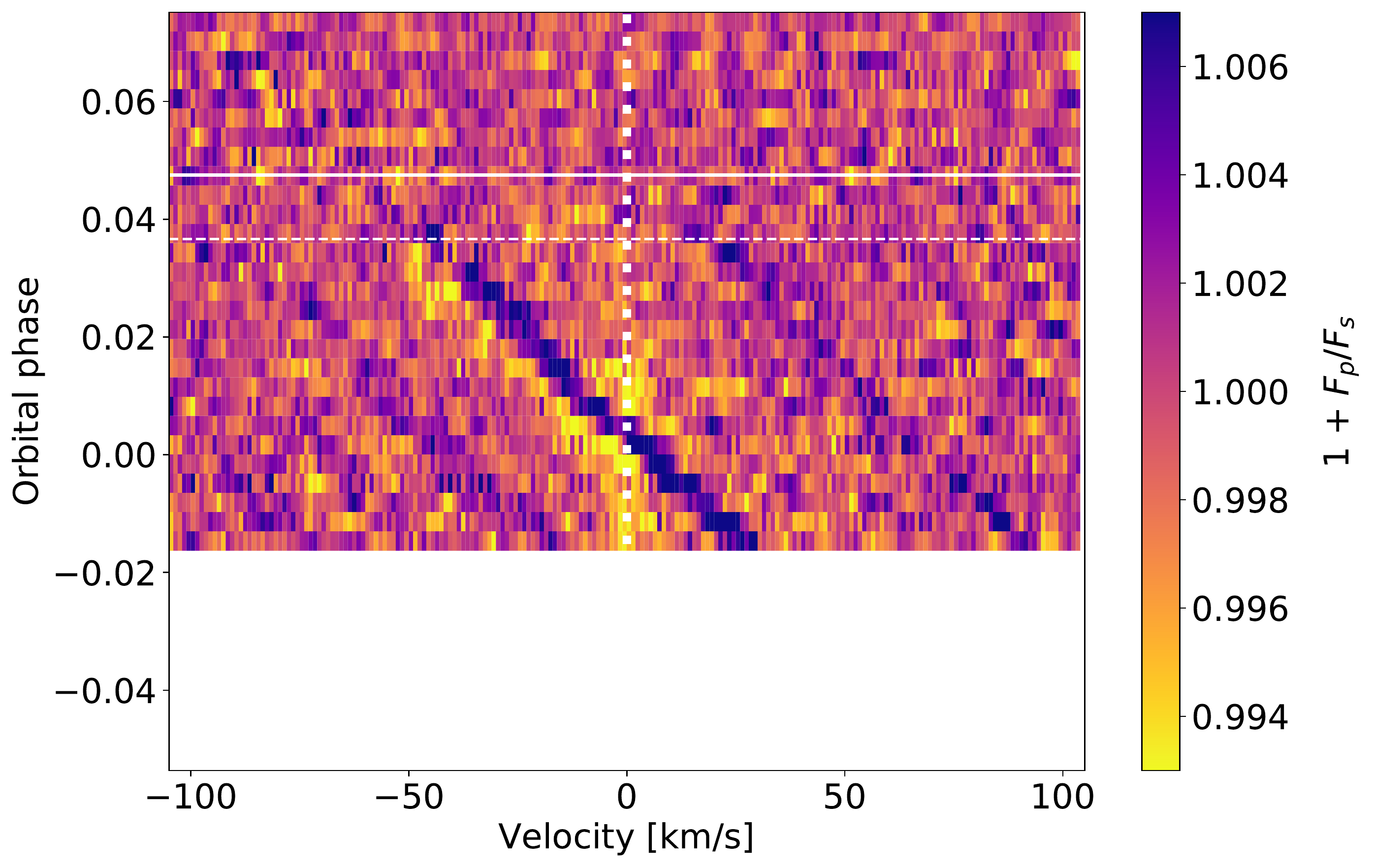}
         \label{fig:five over x}
     \end{subfigure}
        \caption{Same as Fig. \ref{fig:W76 three barium}, but for the WASP-121b observing nights (\textbf{top} in 1UT mode on 2018 November 30, and \textbf{bottom} in 4UT mode  on 2019 January 6). }
        \label{fig:W121 three barium}
\end{figure*}

Another surprising finding is the detection of Ba+ in both planets. This is the heaviest detected element in exoplanetary atmospheres to date. The datasets of WASP-76b and WASP-121b represent some of the highest currently available S/N datasets. That Ba+ is detected in both of the studied planets may indicate that this heavy species can be common in the atmospheres of ultra-hot Jupiters. The CCF plots, tomography plots, and Kp-plots are shown in Figs. \ref{fig:W76 three barium} and \ref{fig:W121 three barium} for the two planets. Close to the expected planet velocity lies a clear absorption signal that follows the planetary Kp in all the exposures. Faint absorption signals directly on the planetary spectrum at the position of the strongest lines of the Ba+ CCF mask are likewise visible (Appendix \ref{Appendix:Individual_Barium}). We also note the ionization trend of these alkaline earth metals (Ca+, Sr+, and Ba+). The presence of these heavy ionized species at high altitudes in the atmospheres of ultra-hot Jupiters may be evidence of unexpected atmospheric dynamics. It is beyond the scope of this paper to describe the mechanisms that would explain the presence of these species in the upper layers of the atmosphere. However, we hope that we encourage further
atmospheric modeling with this discovery.


As studied in \cite{2021A&A...645A..24B}, we also note the much higher S/N and clearer absorption features for the lower-resolution 4UT partial transit (despite the lower resolution: $\sim 70\,000$). This clearly demonstrates the potential of the 16m equivalent mirror or of future high-resolution spectrographs for the Extremely Large Telescope \citep[e.g., ANDES;  ][]{2021Msngr.182...27M}.

\begin{acknowledgements}
This work was supported by Fundação para a Ciência e a Tecnologia (FCT) and Fundo Europeu de Desenvolvimento Regional (FEDER) via COMPETE2020 through the research grants UIDB/04434/2020, UIDP/04434/2020, PTDC/FIS-AST/32113/2017 \& POCI-01-0145-FEDER-032113, PTDC/FIS-AST/28953/2017 \& POCI-01-0145-FEDER-028953.
This work has been carried out in the frame of the National Centre for Competence in Research PlanetS supported by the Swiss National Science Foundation (SNSF). This project has received funding from the European Research Council (ERC) under the European Union's Horizon 2020 research and innovation programme (project {\sc Spice Dune}, grant agreement No 947634). This project has received funding from the European Research Council (ERC) under the European Union’s Horizon 2020 research and innovation programme (project {\sc Four Aces}; grant agreement No 724427).
T.A.S acknowledges support from the Fundação para a Ciência e a
Tecnologia (FCT) through the Fellowship PD/BD/150416/2019 and
POCH/FSE (EC). 
O.D.S.D. is supported in the form of work contract (DL 57/2016/CP1364/CT0004) funded by FCT.
R. A. is a Trottier Postdoctoral Fellow and acknowledges support from the Trottier Family Foundation. This work was supported in part through a grant from FRQNT. This work has been carried out within the framework of the National Centre of Competence in Research PlanetS supported by the Swiss National Science Foundation. The authors acknowledge the financial support of the SNSF.
F.B. acknowledges support from PRIN INAF 2019
CJM acknowledges FCT and POCH/FSE (EC) support through Investigador FCT Contract 2021.01214.CEECIND/CP1658/CT0001.
H.M.T. acknowledges financial support from the Agencia Estatal de Investigaci\'on of the Ministerio de Ciencia, Innovaci\'on y Universidades through project PID2019-109522GB-C51/AEI/10.13039/50110001103
E. E-B. acknowledges financial support from the European Union and the State Agency of Investigation of the Spanish Ministry of Science and Innovation (MICINN) under the grant PRE2020-093107 of the Pre-Doc Program for the Training of Doctors (FPI-SO) through FSE funds.
F.P.E. and C.L.O. would like to acknowledge the Swiss National Science Foundation (SNSF) for supporting research with ESPRESSO through the SNSF grants nr. 140649, 152721, 166227 and 184618. The ESPRESSO Instrument Project was partially funded through SNSF’s FLARE Programme for large infrastructures.
Y.A. acknowledges the support of the Swiss National Fund under grant 200020\_172746.
D.E. acknowledges financial support from the Swiss National Science Foundation for project 200021\_200726.
N.J.N. was financed by projects POCI-01-0145-FEDER-028987, PTDC/FIS-AST/28987/2017, PTDC/FIS-AST/0054/2021 and EXPL/FIS-AST/1368/2021, as well as UIDB/04434/2020 \& UIDP/04434/2020, CERN/FIS-PAR/0037/2019, PTDC/FIS-OUT/29048/2017.
A.S.M. acknowledges financial support from the Spanish Ministry of Science and Innovation (MICINN) under 2018 Juan de la Cierva program IJC2018-035229-I. A. S. M. acknowledges financial support from the MICINN project PID2020-117493GB-I00 and from the Government of the Canary Islands project ProID2020010129.

\end{acknowledgements}

\bibliographystyle{aa} 
\bibliography{aanda.bib} 

\begin{appendix}

\section{Additional table}
\label{Appendix:species_table}

{\renewcommand{\arraystretch}{1.3}
\begin{table*}[h]
\caption{Summary of the Gaussian fit parameters for the detected species in the transits of WASP-76b and WASP-121b.}
\label{table:species}
\centering
\begin{tabular}{lccccc|cccc}
\hline
&
\multicolumn{5}{c}{\textbf{WASP-76}}  & 
\multicolumn{3}{c}{\textbf{WASP-121}} \\ 
\cmidrule(lr){3-6} \cmidrule(lr){7-10}
\multirow{2}{0.5cm}{Species}  & \multirow{2}{0.5cm}{Night} & Amplitude & Center RV & FWHM & \hspace{-0.5cm} \multirow{2}{0.5cm}{S/N\tablefoottext{c}} & Amplitude & Center RV & FWHM & \multirow{2}{0.5cm}{S/N\tablefoottext{c}} \\ \vspace{-0.1cm}
 &  & [ppm] & [\kms] & [\kms] & & [ppm] & [\kms] & [\kms]  \vspace{0.2cm} \\ \hline
\hspace{3mm} \multirow{2}{0.5cm}{Ba+}  & N1 & 1803 $\pm$ 205 & -3.74 $\pm$ 0.74 & 13.3 $\pm$ 1.8 & 7.0 & 2270 $\pm$ 296 & 0.5 $\pm$ 1.6 & 30.5 $\pm$ 5.8 & 6.2 \\
\hspace{3mm}  &  N2 & 1873 $\pm$ 198 & -4.0 $\pm$ 0.58 & 10.7 $\pm$ 1.1 & 7.4 & 2926 $\pm$ 258 & -1.9 $\pm$ 0.5 & 12.0 $\pm$ 1.3 & 10.6  \vspace{0.2cm} \\

\hspace{3mm} \multirow{2}{0.5cm}{Ca}  & N1 & - & - & - & - & 531 $\pm$ 131 & 0.3 $\pm$ 1.4 & 11.1 $\pm$ 2.9 & 2.6  \\
\hspace{3mm}  &  N2 & - & - & - & - & 748 $\pm$ 134 & -2.13 $\pm$ 0.72 & 8.2 $\pm$ 1.7 & 6.5  \vspace{0.2cm} \\

\hspace{3mm} \multirow{2}{0.5cm}{Ca+\tablefoottext{a}}  & N1 & 26517 $\pm$ 1846 & 2.2 $\pm$ 1.4 & 43.1 $\pm$ 4.0 & 26.6 & 29182 $\pm$ 2620 & -18.3 $\pm$ 3.0 & 61.1 $\pm$ 5.5 & 28.6  \\
\hspace{3mm}  &  N2 & 30870 $\pm$ 2527 & 3.0 $\pm$ 0.9 & 23.7 $\pm$ 2.4 &  & 54867 $\pm$ 900 & -10.12 $\pm$ 0.72 & 92.7 $\pm$ 1.8 & 17.2 \vspace{0.2cm} \\

\hspace{3mm} \multirow{2}{0.5cm}{Co}  & N1 & - & - & -  & - & 1182 $\pm$ 146 & 1.8 $\pm$ 0.6 & 9.5 $\pm$ 1.2 & 6.3  \\
\hspace{3mm}  &  N2 & - & - & -  & - & 711 $\pm$ 88 & -1.11 $\pm$ 0.94 & 16.7 $\pm$ 2.7 & 5.7  \vspace{0.2cm} \\

\hspace{3mm} \multirow{2}{0.5cm}{Cr}  & N1 & 383 $\pm$ 51 & -5.44 $\pm$ 0.67 & 10.5 $\pm$ 1.7 & 10.0 & 9648 $\pm$ 72 & -1.1 $\pm$ 1.0 & 18.3 $\pm$ 2.3 & 6.4  \\
\hspace{3mm}  &  N2 & 314 $\pm$ 36 & -7.1 $\pm$ 1.6 & 30.4 $\pm$ 4.9 & 9.1 & 449 $\pm$ 60 & -2.66 $\pm$ 0.88 & 12.2 $\pm$ 1.7 & 10.9 \vspace{0.2cm}  \\

\hspace{3mm} \multirow{2}{0.5cm}{Fe}  & N1 & 667 $\pm$ 47 & -3.42 $\pm$ 0.34 & 9.01 $\pm$ 0.68 & 13.5 & 1300 $\pm$ 72 & -0.27 $\pm$ 0.32 & 11.1 $\pm$ 0.64 & 13.7  \\
\hspace{3mm}  &  N2 & 724 $\pm$ 36 & -3.92 $\pm$ 0.34 & 13.88 $\pm$ 0.77 & 22.2 & 1141 $\pm$ 65 & -2.6 $\pm$ 0.3 & 10.7 $\pm$ 0.68 & 16.3  \vspace{0.2cm} \\

\hspace{3mm} \multirow{2}{0.5cm}{Fe+}  & N1 & - & -  & - & - & 2395 $\pm$ 366 & 1.1 $\pm$ 0.8 & 9.8 $\pm$ 1.5 & 4.8  \\
\hspace{3mm}  &  N2 & - & - & - &  - & 2021 $\pm$ 299 & -2.46 $\pm$ 0.63 & 8.5 $\pm$ 1.3 & 8.3  \vspace{0.2cm} \\

\hspace{3mm} \multirow{2}{0.5cm}{H\tablefoottext{b}}  & N1 & 1861 $\pm$ 319 & 3.1 $\pm$ 4.8 & 43 $\pm$ 10 & 5.5 & 11085 $\pm$ 698 & -2.6 $\pm$ 1.0 & 31.3 $\pm$ 2.0 & 14.8  \\
\hspace{3mm}  &  N2 & 1488 $\pm$ 373 & -3.0 $\pm$ 2.7 & 19.2 $\pm$ 4.0 & 2.4 & 15692 $\pm$ 458 & -1.57 $\pm$ 0.46 & 33.5 $\pm$ 1.2 & 46.4  \vspace{0.2cm} \\

\hspace{3mm} \multirow{2}{0.5cm}{Li}  & N1 & 1606 $\pm$ 153 & -1.4 $\pm$ 1.2 & 27.6 $\pm$ 3.0 & 6.8 & 1063 $\pm$ 437 & -2.9 $\pm$ 2.6 & 11.3 $\pm$ 4.3 & 2.1  \\
\hspace{3mm}  &  N2 & 1558 $\pm$ 157 & -2.41 $\pm$ 0.92 & 17.2 $\pm$ 1.8 & 6.4 & 2163 $\pm$ 305 & -8.5 $\pm$ 1.4 & 20.6 $\pm$ 3.2 & 5.2 \vspace{0.2cm} \\

\hspace{3mm} \multirow{2}{0.5cm}{Mg}  & N1 & 873 $\pm$ 193 & -1.3 $\pm$ 2.1 & 17.7 $\pm$ 3.7 & 4.4 & 2821 $\pm$ 579 & -2.4 $\pm$ 1.4 & 10.3 $\pm$ 2.8 & 6.2  \\
\hspace{3mm}  &  N2 & 2578 $\pm$ 237 & -3.25 $\pm$ 0.57 & 12.2 $\pm$ 1.1 & 9.5 & 2293 $\pm$ 316 & -2.37 $\pm$ 0.82 & 11.4 $\pm$ 1.5 & 6.9 \vspace{0.2cm}  \\

\hspace{3mm} \multirow{2}{0.5cm}{Mn}  & N1 & 1307 $\pm$ 119 & -3.82 $\pm$ 0.57 & 12.4 $\pm$ 1.2 & 5.9 & 1005 $\pm$ 242 & -1.4 $\pm$ 1.0 & 8.0 $\pm$ 1.9 & 2.5 \\
\hspace{3mm}  &  N2 & 1406 $\pm$ 163 & -3.94 $\pm$ 0.39 & 7.1 $\pm$ 1.0 & 7.1 & 1169 $\pm$ 135 & -2.0 $\pm$ 0.8 & 13.3 $\pm$ 1.5 & 6.0 \vspace{0.2cm}  \\

\hspace{3mm} \multirow{2}{0.5cm}{Na}  & N1 & 1015 $\pm$ 88 & 1.59 $\pm$ 0.95 & 20.7 $\pm$ 1.9 & 10.4 & 1406 $\pm$ 184 & -0.98 $\pm$ 0.97 & 14.8 $\pm$ 2.2 & 7.8  \\
\hspace{3mm}  &  N2 & 884 $\pm$ 79 & -0.0 $\pm$ 1.0 & 22.5 $\pm$ 2.2 & 8.9 & 1245 $\pm$ 111 & -1.44 $\pm$ 0.68 & 15.1 $\pm$ 1.4 & 10.1 \vspace{0.2cm}  \\

\hspace{3mm} \multirow{2}{0.5cm}{Ni}  & N1 & - & - & -  & - & 1663 $\pm$ 522 & -1.52 $\pm$ 0.24 & 1.94 $\pm$ 0.98 & 4.2  \\
\hspace{3mm}  &  N2 & - & - & - & - & 611 $\pm$ 132 & -2.9 $\pm$ 1.2 & 11.4 $\pm$ 2.5 & 5.7 \vspace{0.2cm}  \\

\hspace{3mm} \multirow{2}{0.5cm}{Sr+}  & N1 & - & - & -  & - & 4548 $\pm$ 1092 & 1.1 $\pm$ 3.5 & 24.0 $\pm$ 4.4 & 5.7 \\
\hspace{3mm}  &  N2 & - & - & - & - & 7659 $\pm$ 781 & -0.24 $\pm$ 0.85 & 15.6 $\pm$ 1.6 & 8.2 \vspace{0.2cm}  \\

\hspace{3mm} \multirow{2}{0.5cm}{V}  & N1 & 346 $\pm$ 32 & -4.88 $\pm$ 0.48 & 10.3 $\pm$ 1.0 & 6.2 & 618 $\pm$ 63 & 1.94 $\pm$ 0.65 & 12.8 $\pm$ 1.7 & 7.7  \\
\hspace{3mm}  &  N2 & 360 $\pm$ 26 & -4.42 $\pm$ 0.53 & 14.4 $\pm$ 1.1 & 10.6 & 587 $\pm$ 41 & -2.13 $\pm$ 0.44 & 13.0 $\pm$ 1.1 & 9.9 \vspace{0.2cm}  \\

\hline

\end{tabular}

\tablefoot{\tablefoottext{a}{The Ca+ signal is highly asymmetric (see Fig. \ref{fig:Calcium}), therefore the Gaussian fit may not be accurate.} \tablefoottext{b}{The H signal corresponds to the direct fit of the observed H$\alpha$ line in the planetary transmitted spectrum and not to a CCF, as was done with remaining species.} \newline \tablefoottext{c}{The S/N\,columns correspond to the significance of the detection. This significance was computed as the ratio of the fit amplitude to the standard deviation of the binned adjacent continuum (with the binning width equal to the fit FWHM).}}

\end{table*}}

\FloatBarrier

\section{Parameters}
\label{Appendix:Parameters}

{\renewcommand{\arraystretch}{1.2}
\begin{table*}[h!]
\caption{Stellar and planetary parameters.}
\label{table:system-params}
\centering
\begin{tabular}{lcc|cc}
\hline
&
\multicolumn{2}{c}{\textbf{WASP-76}}  & 
\multicolumn{2}{c}{\textbf{WASP-121}} \\ 
\cmidrule(lr){2-3} \cmidrule(lr){4-5}
Parameter & Value & Source & Value & Source  \\ \hline
Stellar Information\\
\hspace{3mm} $T_\textrm{eff}$ (K) &  6329 $\pm$ 65  & Ehr2020 & 6586 $\pm$ 59 & Bou2021 \\
\hspace{3mm} log $g$ (cgs) &  4.196 $\pm$ 0.106 & Ehr2020 & 4.47 $\pm$ 0.08 & Bou2021 \\
\hspace{3mm} [Fe/H] (dex) &  0.366 $\pm$ 0.053  & Ehr2020 & 0.13 $\pm$ 0.04 & Bou2021 \\
\hspace{3mm} $M_*$ ($M_\odot$) &  1.458 $\pm$ 0.021  & Ehr2020 & 1.38 $\pm$ 0.02 & Bou2021 \\
\hspace{3mm} $R_*$ ($R_\odot$) &  1.756 $\pm$ 0.071  & Ehr2020 & 1.44 $\pm$ 0.03 & Bou2021 \\
\hspace{3mm} Age  (Gyr) & 1.816 $\pm$ 0.274  & Ehr2020 & 1.03 $\pm$ 0.43 & Bou2021 \\
\hspace{3mm} V (mag) &  9.52 $\pm$ 0.03  & Tycho-2 & 10.51 $\pm$ 0.04 & Tycho-2 \\
\hspace{3mm} $K_s$ (\ms) &  116.02 $\pm$ 1.35  & Ehr2020 & 0.177 $\pm$ 0.008 &  Bor2020\\

\hline

Planet Information\\

\hspace{3mm} Period [days] & 1.80988198 $\pm$ \, 0.00000064 & Ehr2020 & 1.27492504 $\pm$ 0.00000015 & \\
\hspace{3mm} $T_{0}$ [\text{BJD{\text -}2400000}] & 58080.626165 $\pm$ \, 0.0004188 & Ehr2020 & 58119.72074 $\pm$ 0.00017 & Bor2020 \\
\hspace{3mm} $R_{p}/R_*$ &  0.10852 $\pm$ \; 0.00096  & Ehr2020 & 0.12534 $\pm$ 0.00005 & Bor2020 \\
\hspace{3mm} $a/R_*$ &  4.08 $\pm$ \; 0.06 & Ehr2020 & 3.8131 $\pm$ 0.0075 & Bor2020 \\
\hspace{3mm} $i$ [º] & 89.623 $\pm$ \; 0.034 & Ehr2020 & 88.49 $\pm$ 0.16 & Bor2020 \\
\hspace{3mm} $e$ & 0 (fixed) & Ehr2020 & 0 (fixed) & Bor2020 \\
\hspace{3mm} $M_p$ ($M_{Jup}$) &  0.894 $\pm$ 0.014  & Ehr2020 & 1.157 $\pm$ 0.070 & Bor2020 \\
\hspace{3mm} $R_p$ ($R_{Jup}$) &  1.854 $\pm$ 0.077  & Ehr2020 & 1.753 $\pm$ 0.036 & Bor2020 \\
\hspace{3mm} $K_p$ (\kms) &  196.52 $\pm$ 0.94 & Ehr2020 & 221.15 $\pm$ 17.00 & Calculated\\

\hline
\\\end{tabular}

\tablefoot{Ehr2020 - \cite{2020Natur.580..597E} , Bou2021 - {\cite{2021A&A...645A..24B}}, Tycho-2 - \cite{2000A&A...355L..27H}, Bor2020 - \cite{2020A&A...635A.205B}
}

\end{table*}}

\FloatBarrier

\section{Telluric correction.  Input parameters and spectrum comparison}
\label{Appendix:Telluric}

\begin{table}[h!]
\caption{Some of the key input parameters used in \texttt{Molecfit}.}
\label{table:measur_rvs}
\centering
\begin{tabular}{m{0.2\textwidth}>{\centering}m{0.3\textwidth}>{\centering\arraybackslash}m{0.4\textwidth}}
\hline
Parameter key & Value & Description \\ \hline
\texttt{LIST\_MOLEC} & H2O, O2 & List of molecules to be included in the model. \\[0.1cm]
\texttt{WAVE\_INCLUDE} & 0.6890,0.6900, 0.7160,0.7340, 0.7590,0.7770 & Wavelength ranges to be included. \\[0.1cm]
\texttt{WLG\_TO\_MICRON} & 0.0001 & Multiplicative factor applied to the wavelength to express it in micrometres. \\[0.1cm]
\texttt{WAVELENGTH\_FRAME} & VAC\_RV & Wavelength reference frame. \\[0.1cm]
\texttt{FTOL} & 1e-8 & Relative chi-square convergence criterion. \\[0.1cm]
\texttt{XTOL} & 1e-8 & Relative parameter convergence criterion. \\
\texttt{FIT\_CONTINUUM} & 1 & Flag to enable/disable the polynomial fit of the continuum. \\[0.1cm]
\texttt{CONTINUUM\_N} & 4 & Degree of the polynomial continuum fit to use per range. \\[0.1cm]
\texttt{FIT\_RES\_GAUSS} & True & Fit resolution by Gaussian. \\[0.1cm]
\texttt{RES\_GAUSS} & 3.5 & Initial value for FWHM of the Gaussian in pixels, at the centre of the spectrum. \\[0.1cm]
\texttt{LNFL\_LINE\_DB} & updated\_aer\_v\_3.6$^{a}$ & File name of the line list  \\[0.1cm]
\hline
\\\end{tabular}
\tablefoot{Descriptions taken from the \texttt{Molecfit} Pipeline User
Manual at \url{https://ftp.eso.org/pub/dfs/pipelines/instruments/molecfit/molecfit-pipeline-manual-4.1.pdf}. $^{a}$From private communication with Romain Allart due to duplicated lines in the aer3.6 line list \citep{2017A&A...606A.144A, 2020A&A...644A.155A}. 
}
\end{table}

\FloatBarrier

\vspace{3cm}

\section{Wiggles}
\label{Appendix:Wiggles}

\begin{figure*}[tbh]
\captionsetup[subfigure]{labelformat=empty}
\centering
\begin{subfigure}[t]{\hsize}
    \includegraphics[width=0.9\linewidth]{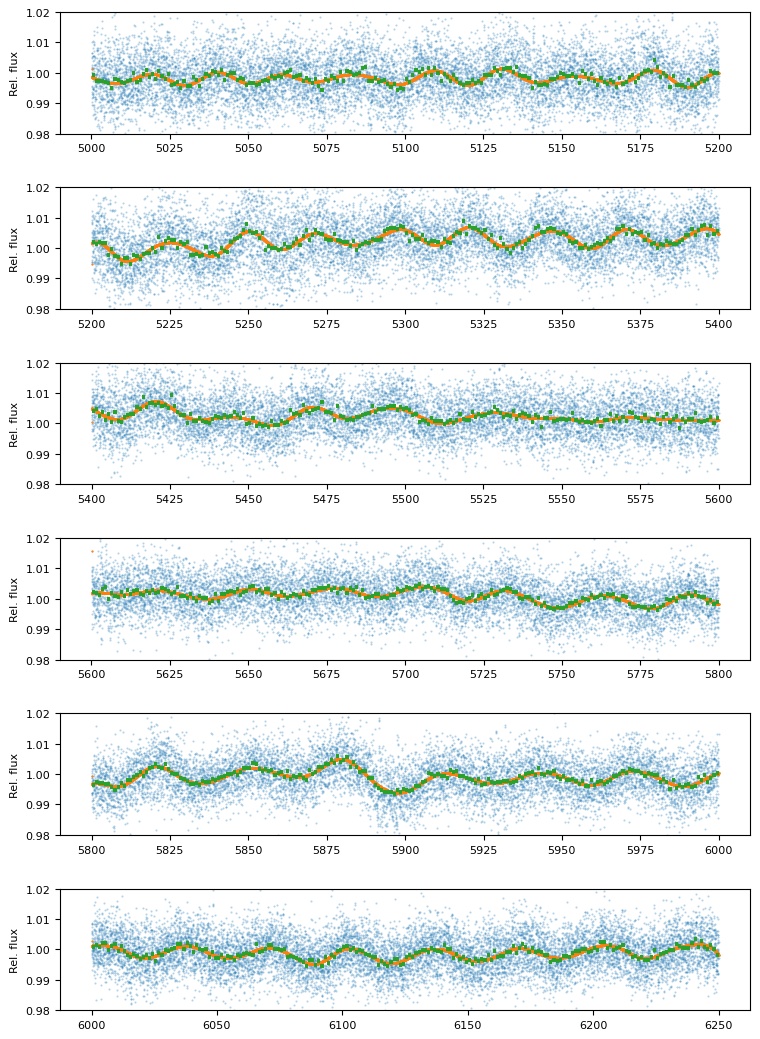}
    \caption{}
\end{subfigure}   
    \caption{(Figure continues on the next page.)}
    \end{figure*}
\clearpage   
\begin{figure*}[tb]\ContinuedFloat
\captionsetup[subfigure]{labelformat=empty}
\centering
\begin{subfigure}[t]{\hsize}
    \includegraphics[width=0.9\linewidth]{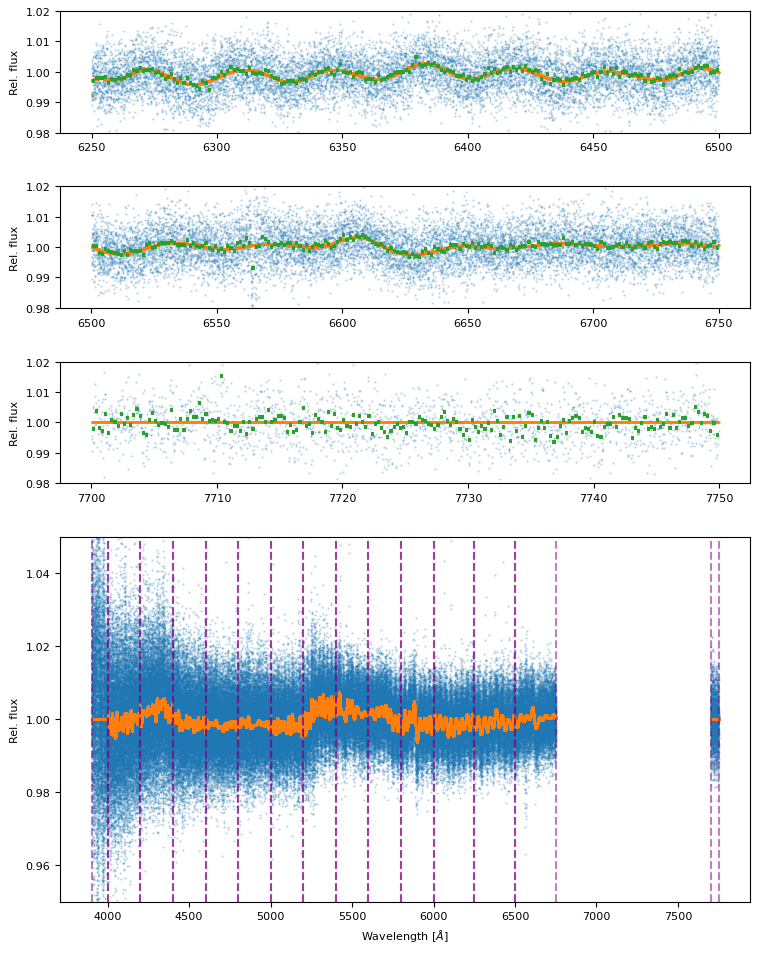}
    \caption{}
\end{subfigure}
    \caption{Fitting procedure during the wiggle-correction step for some of the wavelength regions in the first exposure of the 4UT observing night of WASP-121b. The fit of the entire spectral range is shown in the last panel. We divide the spectrum into regions (vertical purple lines) in which we define the priors and bounds of the fitting functions. For some regions near the lowest and highest wavelengths, we did not apply the spline fitting because it would not be a correct match to the observed data.   }
    \label{fig:wiggles}
\end{figure*}

\FloatBarrier

\section{Lines per species used from the \texttt{Mantis} masks.}
\label{Appendix:lines_mask}

{\renewcommand{\arraystretch}{1.}
\begin{table}[h]
\caption{Number of lines per species we used for our CCFs using the \texttt{Mantis} masks at a temperature of 2500 K.}
\label{table:lines_mask}
\centering
\begin{tabular}{cc}
\hline
Chemical species & Number of lines  \\ \hline
Li  & 9  \\
Na  & 43  \\
Mg & 52  \\
K & 58  \\
Ca & 486  \\
Ca+ & 2  \\
Ti+ & 9  \\
V & 2543  \\
Cr & 2096  \\
Mn & 362  \\
Fe & 3827  \\
Fe+ & 74\tablefoottext{a}  \\
Co & 984  \\
Ni & 580  \\
Sr+ & 5  \\
Ba+ & 12  \\

\hline
\\\end{tabular}
\tablefoot{$^{a}$ Lines for the 3000 K template because no lines are available at 2500 K.
}
\end{table}}

\section{Stellar signal masking}
\label{Appendix: 15Masking}

\begin{figure*}[th]
     \centering
     \begin{subfigure}[b]{0.4\textwidth}
         \centering
         \includegraphics[width=\textwidth]{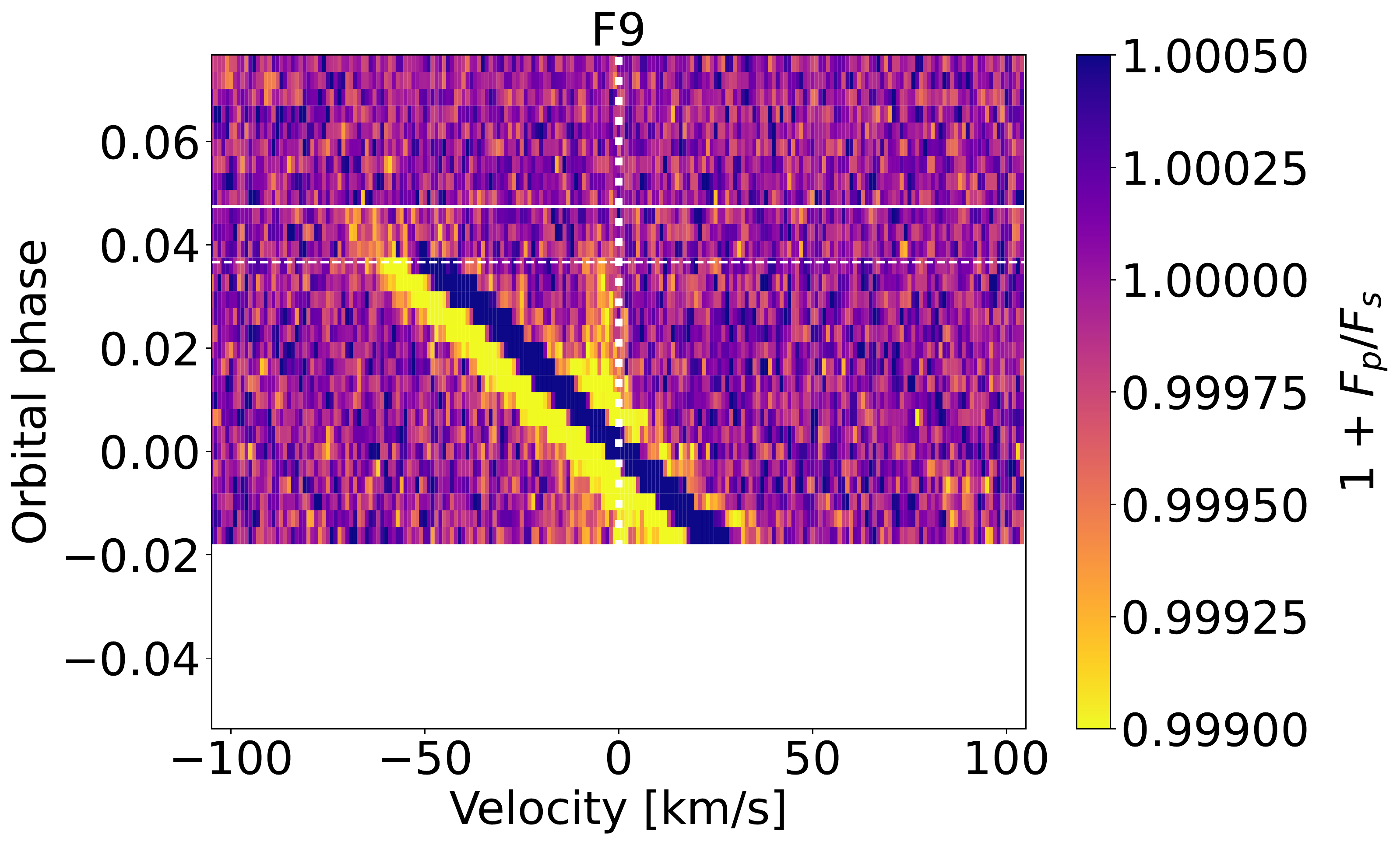}
         \label{fig:y equals x}
     \end{subfigure}
     \hfill
     \begin{subfigure}[b]{0.4\textwidth}
         \centering
         \includegraphics[width=\textwidth]{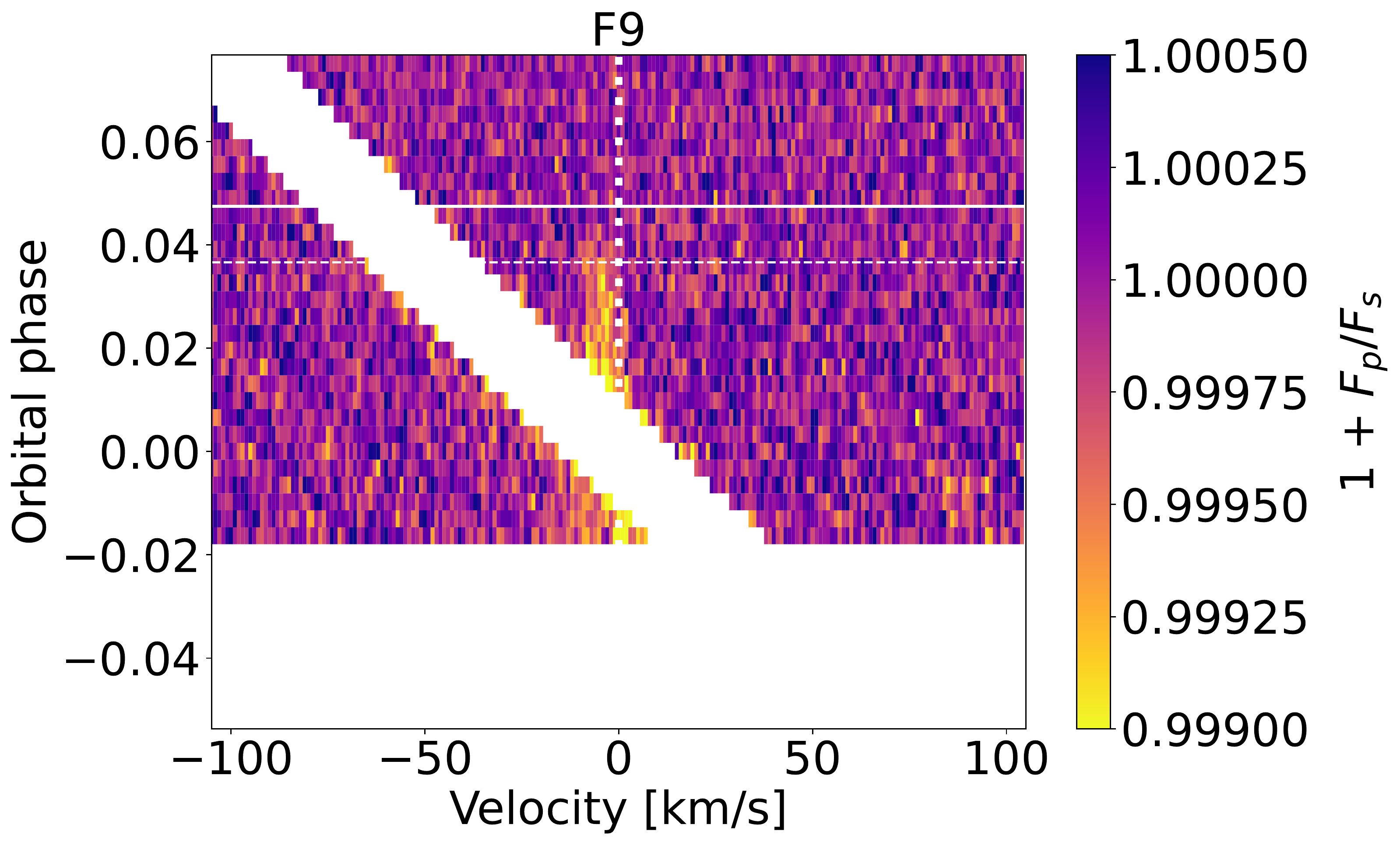}
         \label{fig:three sin x}
     \end{subfigure}
     \hfill
     \begin{subfigure}[b]{0.1\textwidth}
         \centering
         \includegraphics[width=\textwidth]{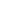}
         \label{fig:five over x}
     \end{subfigure}
        
        \caption{Tomography plots in the planetary rest frame built from the WASP-121b CCFs (night 2, 4UTs) of the planetary spectrum with the F9 ESPRESSO mask. \textbf{Left}: Without masking the stellar signal. \textbf{Right}: By masking the stellar velocities ranging from -15 km/s to +15 km/s in the stellar rest frame. The white horizontal band for the early orbital phases originates from the lack of data at the start of transit.}
        \label{fig:15tomo}
\end{figure*}

\FloatBarrier

\section{Calcium II and titanium II CCFs}
\label{Appendix:Calcium}

\FloatBarrier

\FloatBarrier

\section{CCF and Kp-plots for WASP-76b and WASP-121b}
\label{Appendix:WASP-all}

\begin{figure*}[b]
     \centering
     \begin{subfigure}[b]{0.33\textwidth}
         \centering
         \includegraphics[width=\textwidth]{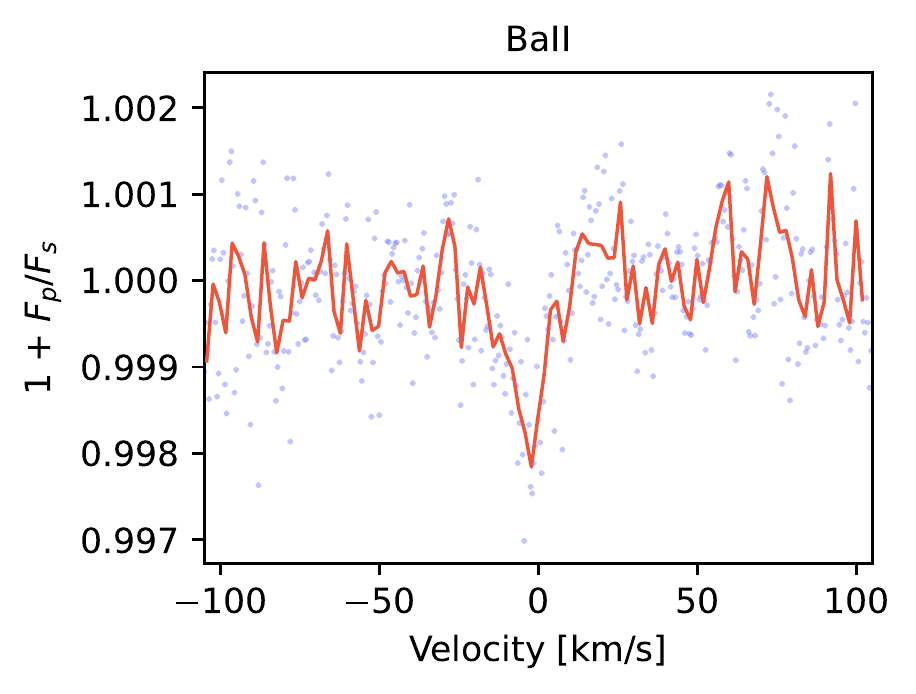}
         \label{fig:y equals x}
     \end{subfigure}
     \hfill
     \begin{subfigure}[b]{0.33\textwidth}
         \centering
         \includegraphics[width=\textwidth]{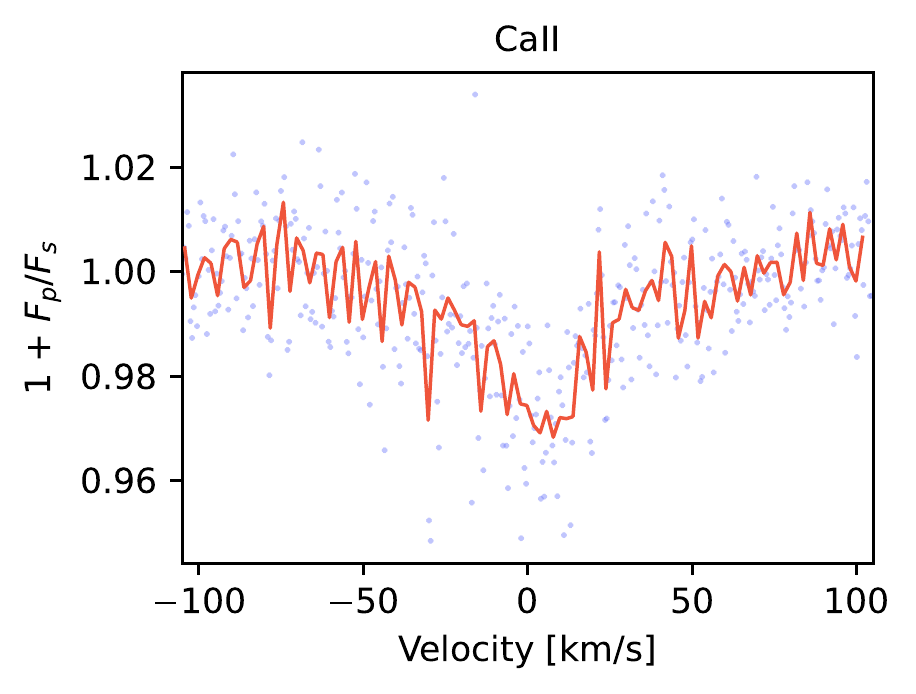}
         \label{fig:three sin x}
     \end{subfigure}
     \hfill
     \begin{subfigure}[b]{0.33\textwidth}
         \centering
         \includegraphics[width=\textwidth]{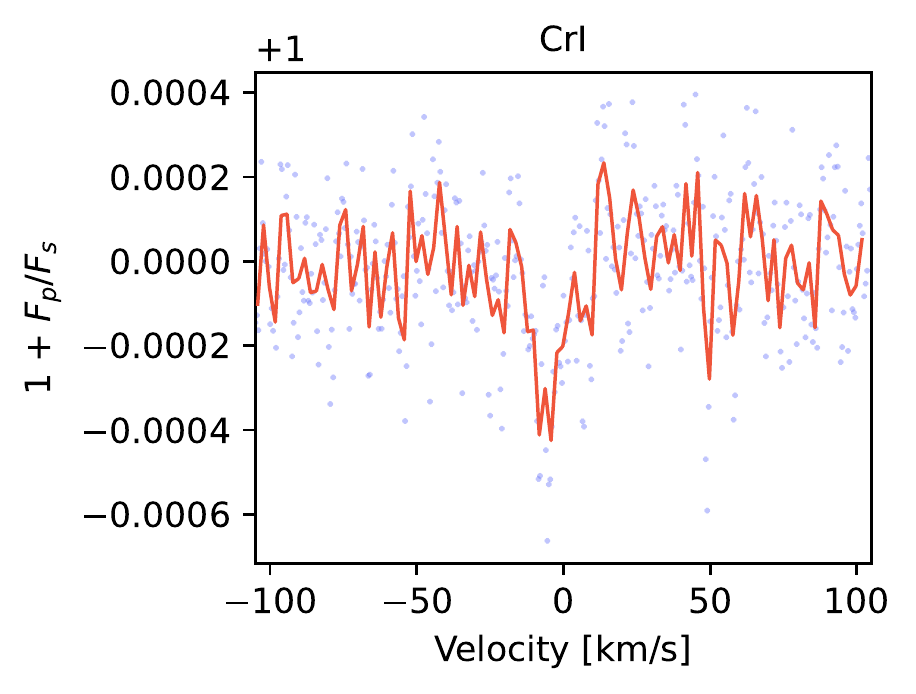}
         \label{fig:five over x}
     \end{subfigure}
     
     \begin{subfigure}[b]{0.33\textwidth}
         \centering
         \includegraphics[width=\textwidth]{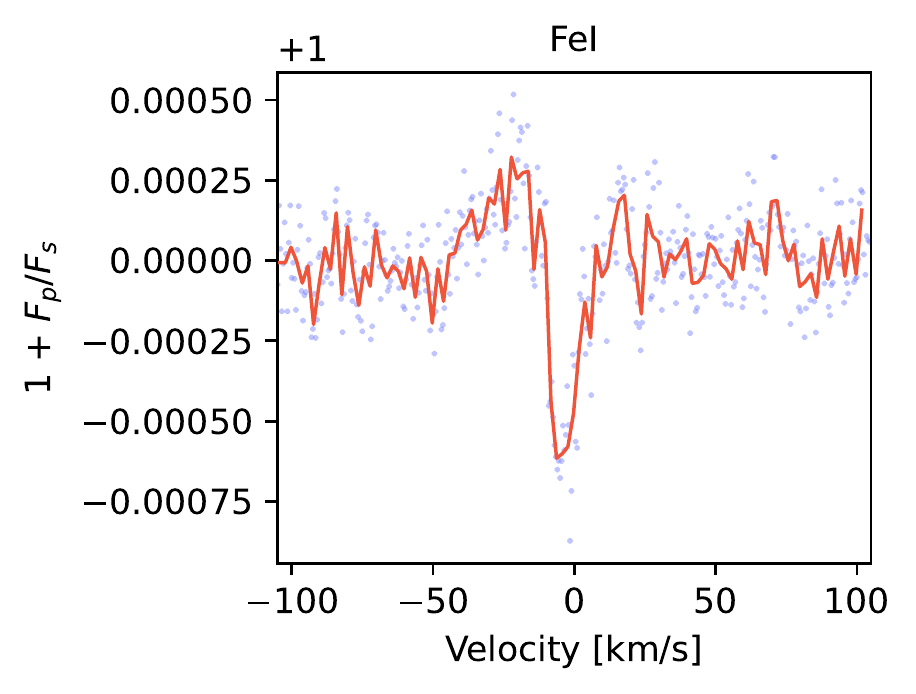}
         \label{fig:y equals x}
     \end{subfigure}
     \hfill
     \begin{subfigure}[b]{0.33\textwidth}
         \centering
         \includegraphics[width=\textwidth]{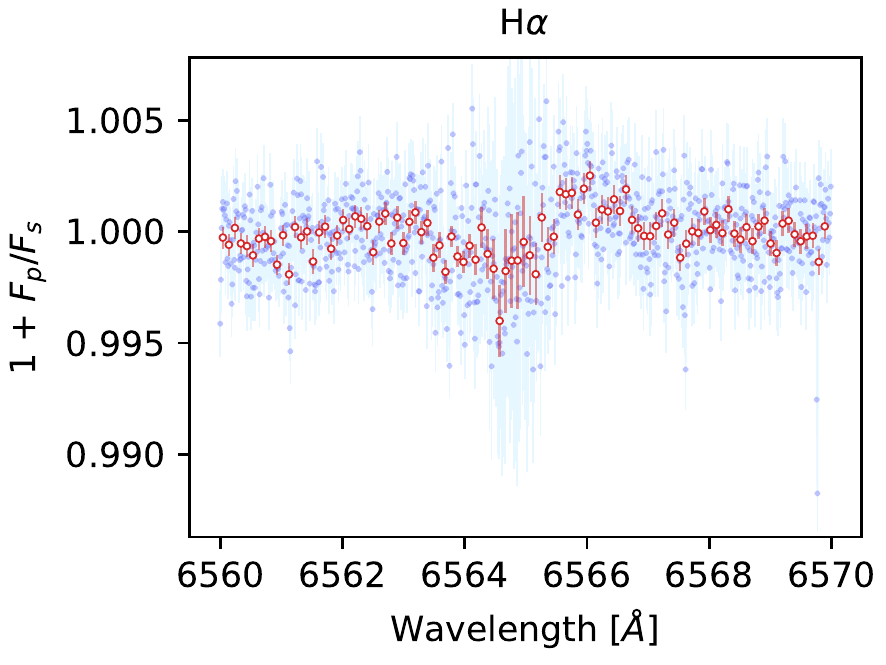}
         \label{fig:three sin x}
     \end{subfigure}
     \hfill
     \begin{subfigure}[b]{0.33\textwidth}
         \centering
         \includegraphics[width=\textwidth]{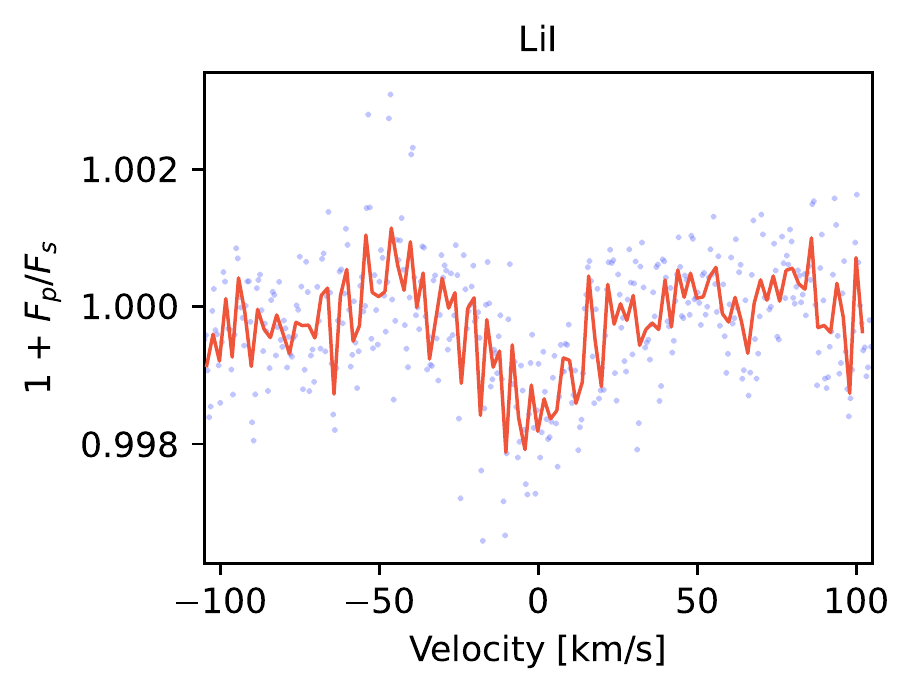}
         \label{fig:five over x}
     \end{subfigure}
        
     \begin{subfigure}[b]{0.33\textwidth}
         \centering
         \includegraphics[width=\textwidth]{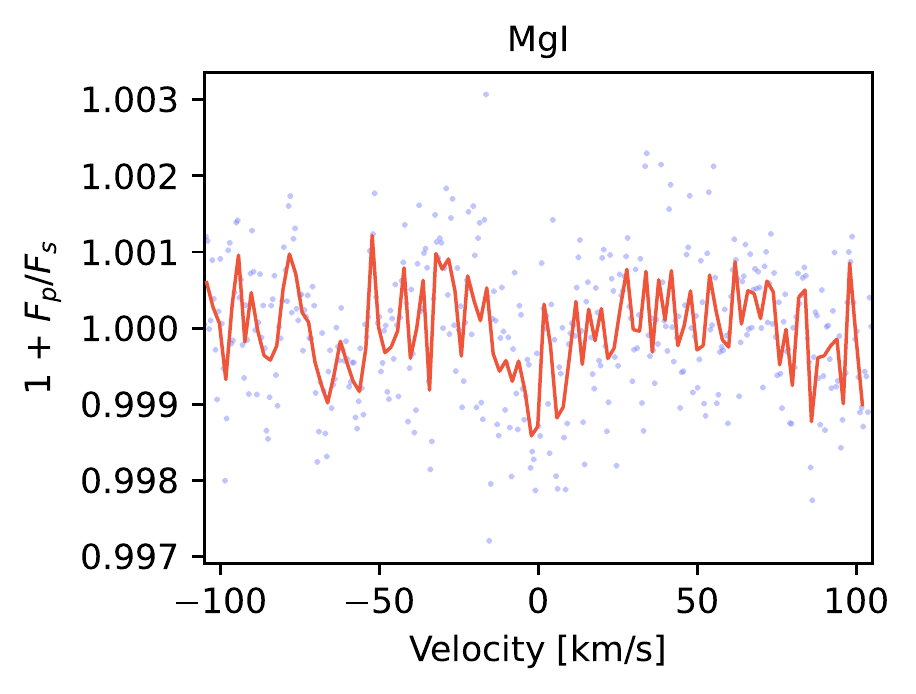}
         \label{fig:y equals x}
     \end{subfigure}
     \hfill
     \begin{subfigure}[b]{0.33\textwidth}
         \centering
         \includegraphics[width=\textwidth]{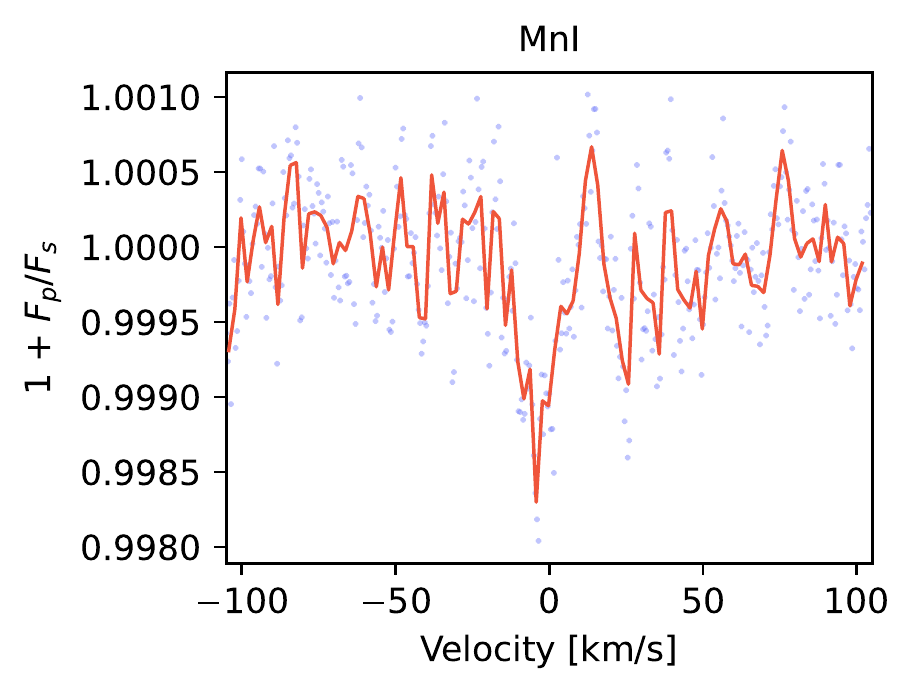}
         \label{fig:three sin x}
     \end{subfigure}
     \hfill
     \begin{subfigure}[b]{0.33\textwidth}
         \centering
         \includegraphics[width=\textwidth]{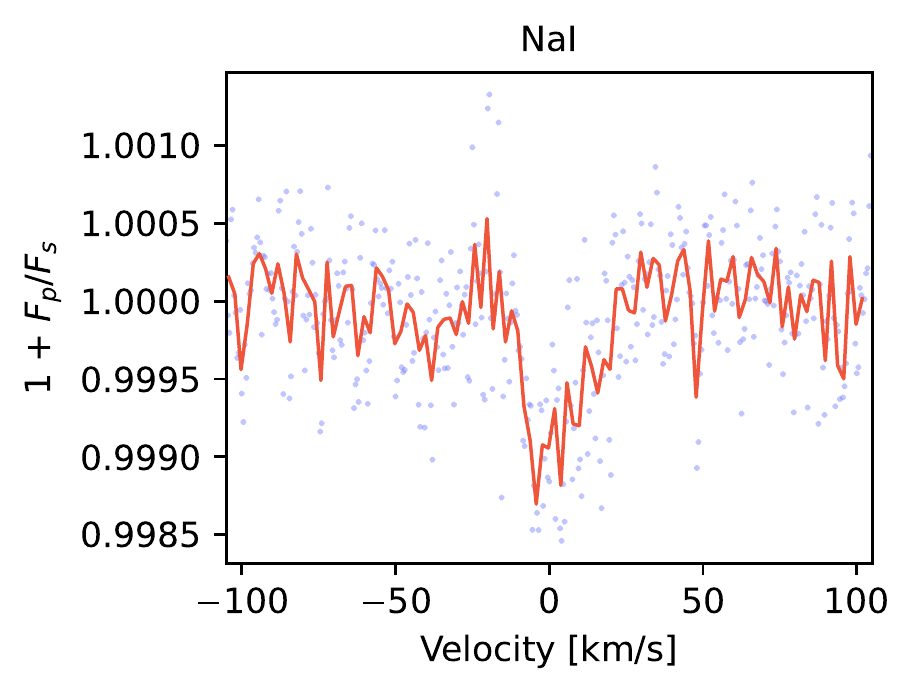}
         \label{fig:five over x}
     \end{subfigure}
        
     \begin{subfigure}[b]{0.33\textwidth}
         \centering
         \includegraphics[width=\textwidth]{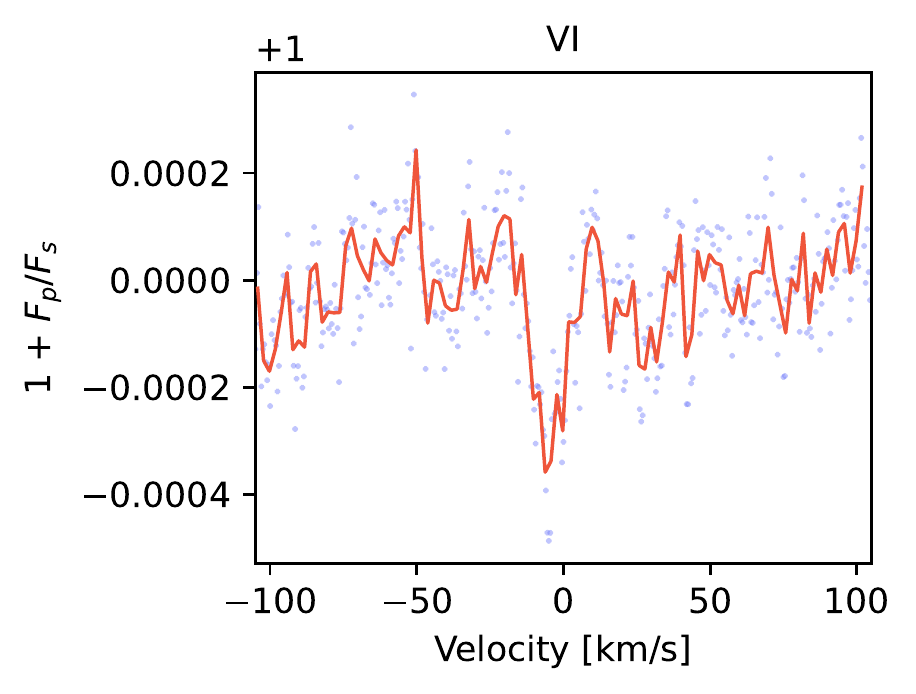}
         \label{fig:y equals x}
     \end{subfigure}
     \hfill
     \begin{subfigure}[b]{0.33\textwidth}
         \centering
         \includegraphics[width=\textwidth]{nothing.png}
         \label{fig:three sin x}
     \end{subfigure}
     \hfill
     \begin{subfigure}[b]{0.33\textwidth}
         \centering
         \includegraphics[width=\textwidth]{nothing.png}
         \label{fig:five over x}
     \end{subfigure}
        
        \caption{WASP-76b, night 1 (2018 September 3). CCFs of the averaged in-transit exposures in the planetary frame of reference for the detected chemical species, except for the H$\alpha$ line, which is present directly in the retrieved planetary spectrum. }
        \label{fig:W76n1ccf}
\end{figure*}

\begin{figure*}
     \centering
     \begin{subfigure}[b]{0.33\textwidth}
         \centering
         \includegraphics[width=\textwidth]{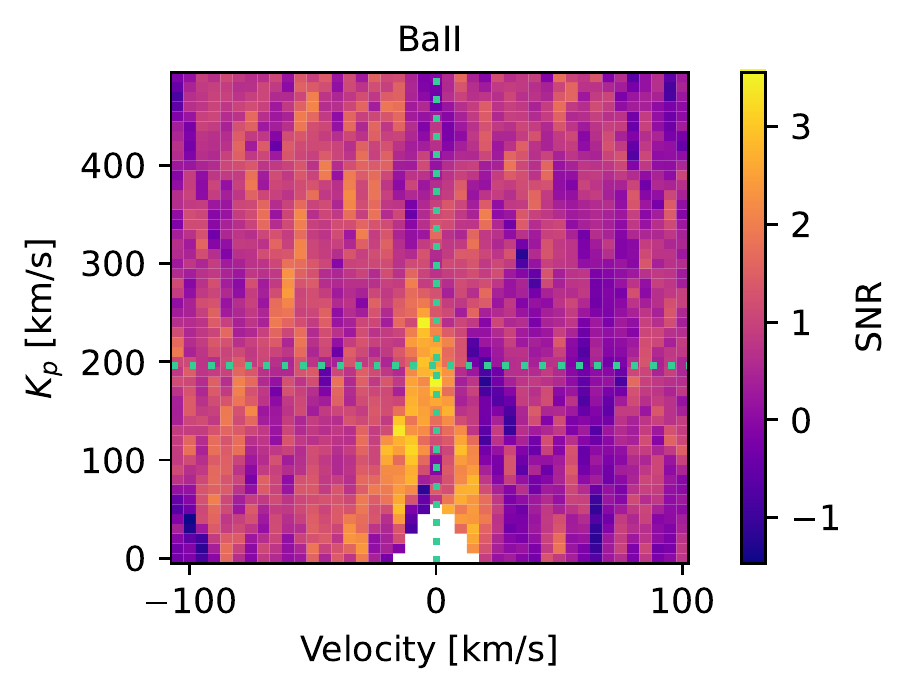}
         \label{fig:y equals x}
     \end{subfigure}
     \hfill
     \begin{subfigure}[b]{0.33\textwidth}
         \centering
         \includegraphics[width=\textwidth]{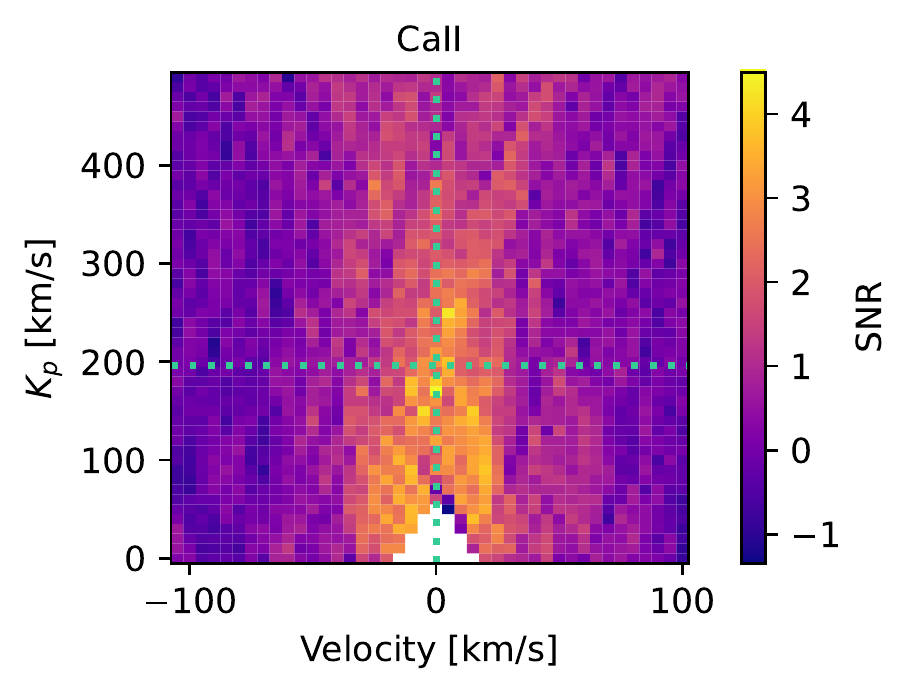}
         \label{fig:three sin x}
     \end{subfigure}
     \hfill
     \begin{subfigure}[b]{0.33\textwidth}
         \centering
         \includegraphics[width=\textwidth]{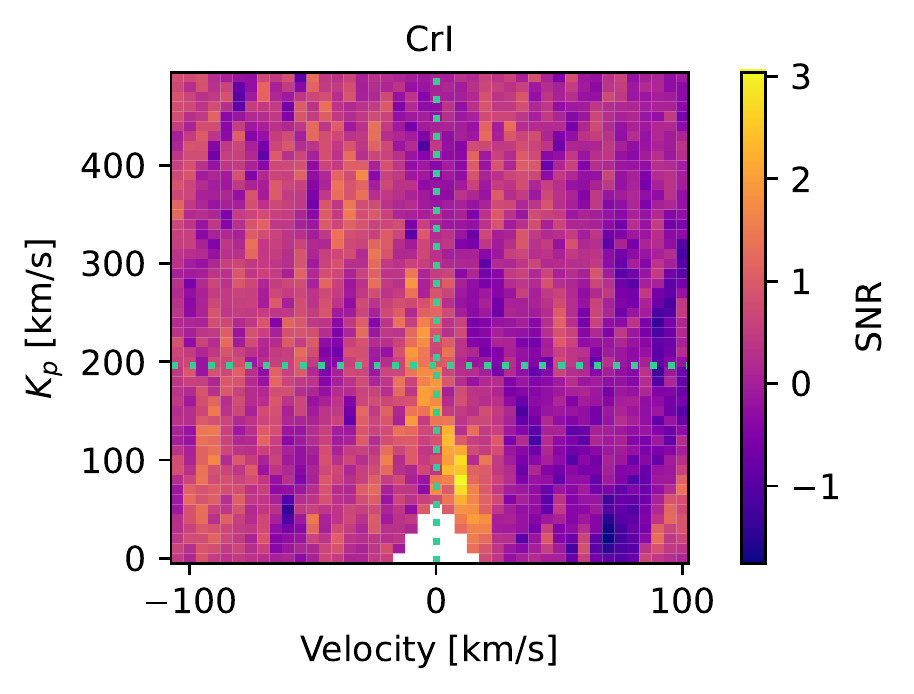}
         \label{fig:five over x}
     \end{subfigure}
     
     \begin{subfigure}[b]{0.33\textwidth}
         \centering
         \includegraphics[width=\textwidth]{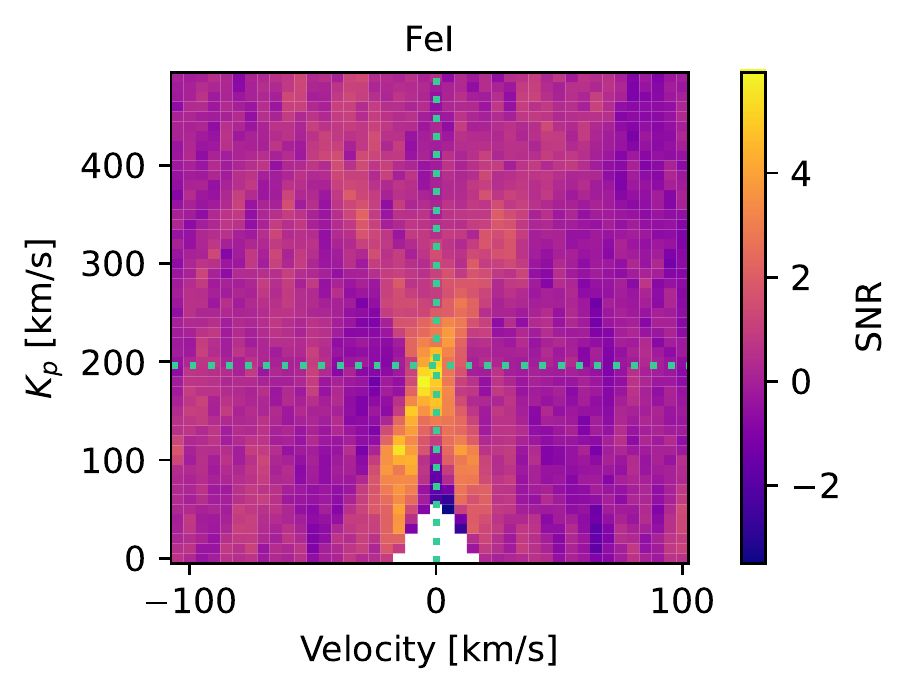}
         \label{fig:y equals x}
     \end{subfigure}
     \hfill
     \begin{subfigure}[b]{0.33\textwidth}
         \centering
         \includegraphics[width=\textwidth]{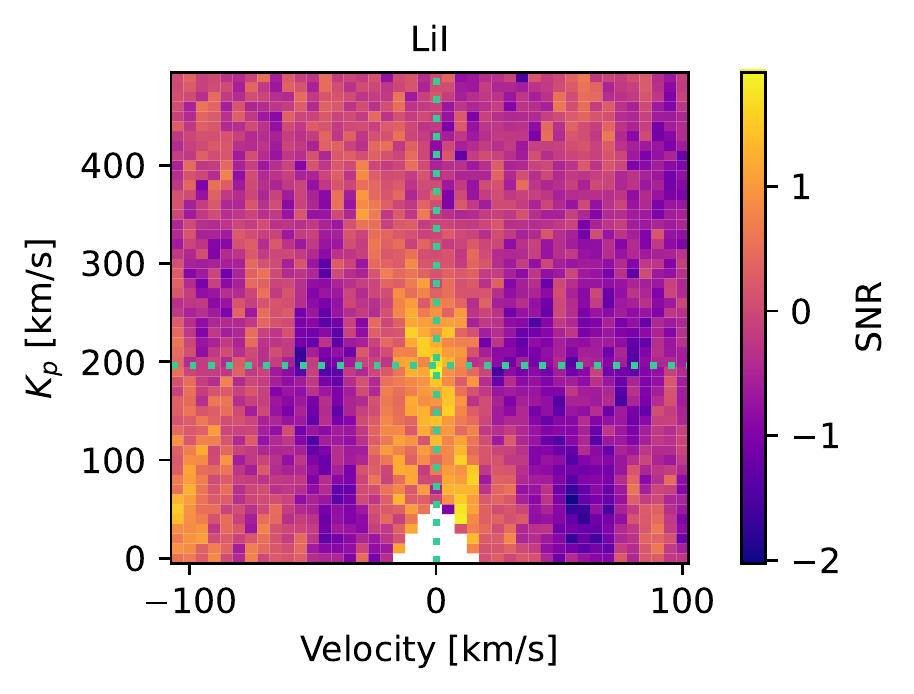}
         \label{fig:three sin x}
     \end{subfigure}
     \hfill
     \begin{subfigure}[b]{0.33\textwidth}
         \centering
         \includegraphics[width=\textwidth]{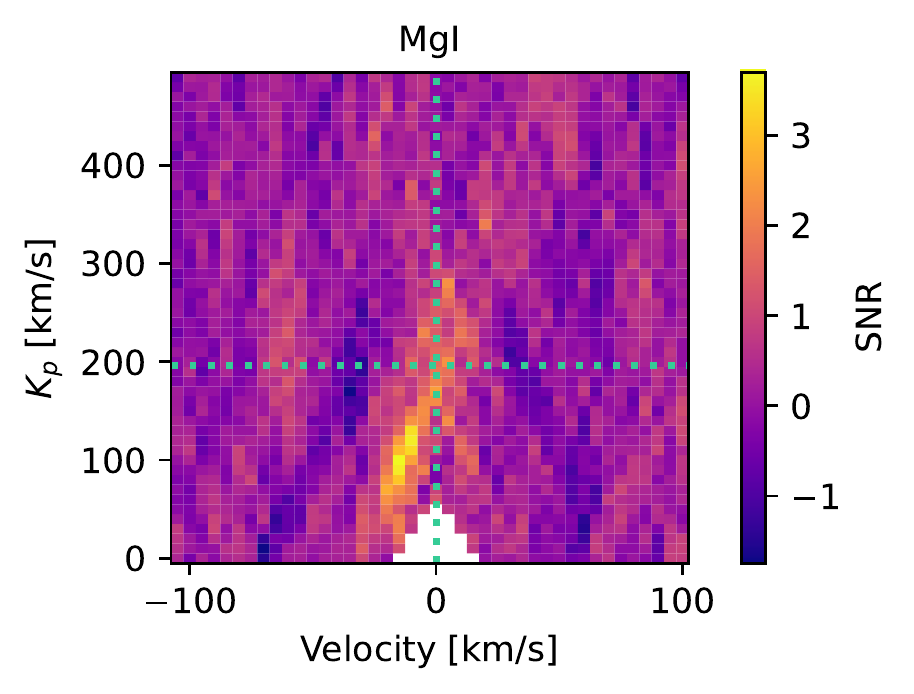}
         \label{fig:five over x}
     \end{subfigure}
     
     \begin{subfigure}[b]{0.33\textwidth}
         \centering
         \includegraphics[width=\textwidth]{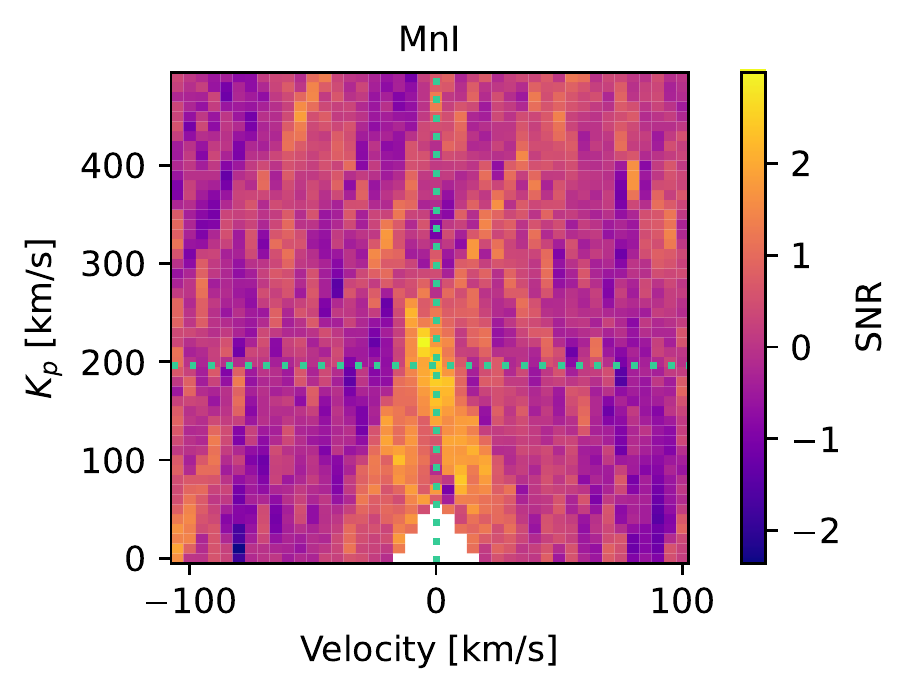}
         \label{fig:y equals x}
     \end{subfigure}   
     \hfill
     \begin{subfigure}[b]{0.33\textwidth}
         \centering
         \includegraphics[width=\textwidth]{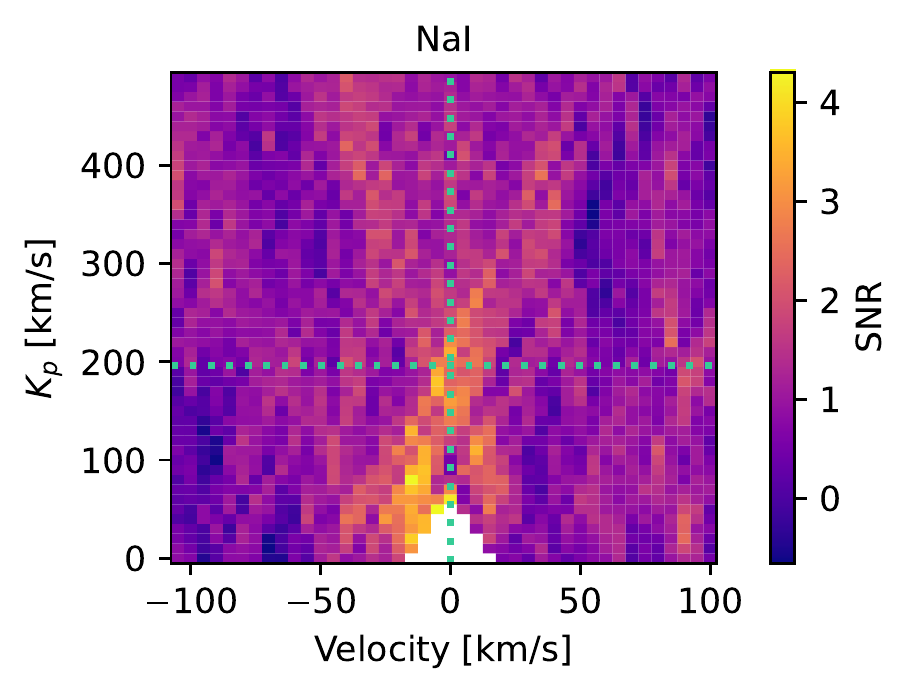}
         \label{fig:three sin x}
     \end{subfigure}
     \hfill
     \begin{subfigure}[b]{0.33\textwidth}
         \centering
         \includegraphics[width=\textwidth]{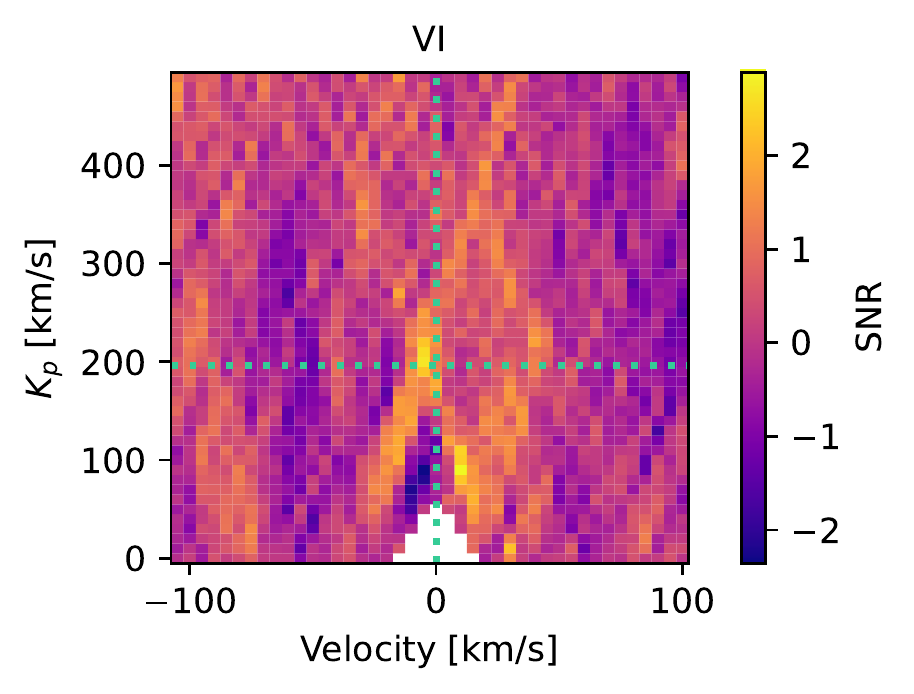}
         \label{fig:five over x}
     \end{subfigure}
        
        \caption{WASP-76b, night 1 (2018 September 3). \textit{\textup{Kp-plots.}}  Map of the sum of all the individual exposures in the planetary rest frame across different values in the \textit{\textup{Kp - planetary velocity}} plane. We mask the signal that matches the stellar velocities ranging from -15 km/s to +15 km/s in the stellar rest frame. The dashed green lines represent the expected position of the planetary signal in this map. }
        \label{fig:W76n1kp}
\end{figure*}

\begin{figure*}
     \centering
     \begin{subfigure}[b]{0.33\textwidth}
         \centering
         \includegraphics[width=\textwidth]{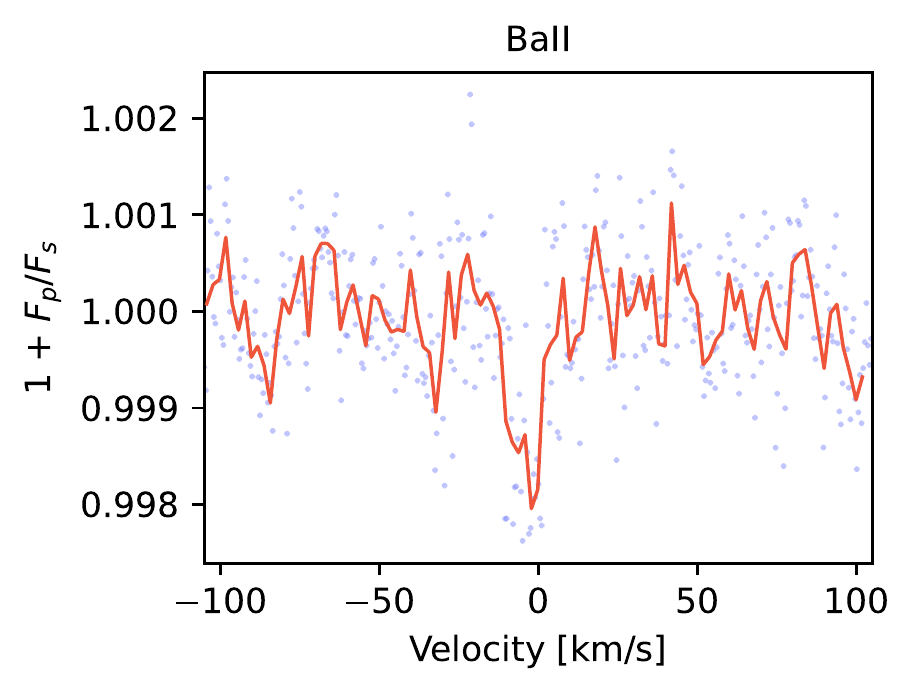}
         \label{fig:y equals x}
     \end{subfigure}
     \hfill
     \begin{subfigure}[b]{0.33\textwidth}
         \centering
         \includegraphics[width=\textwidth]{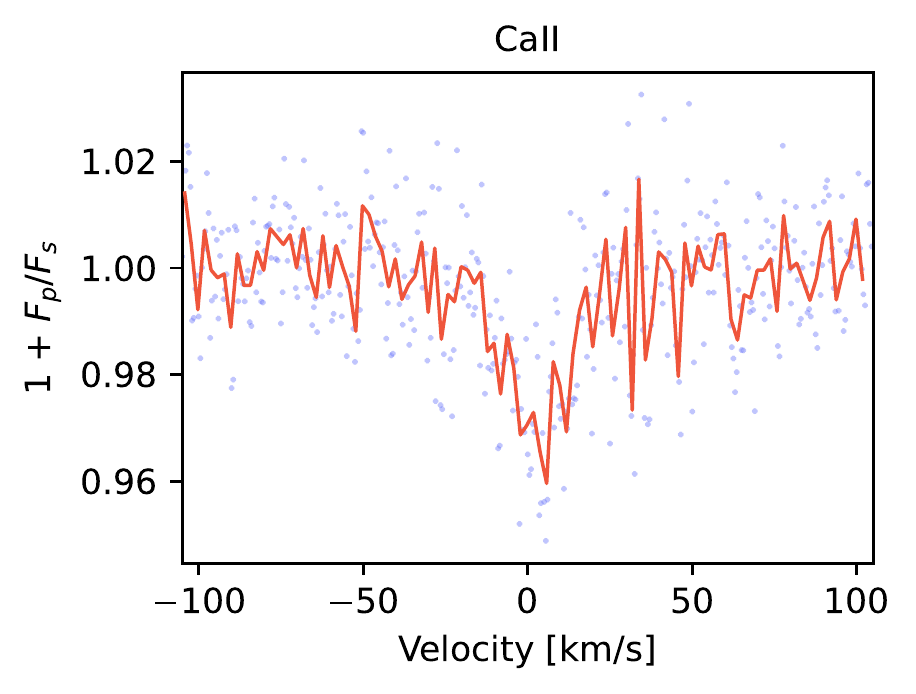}
         \label{fig:three sin x}
     \end{subfigure}
     \hfill
     \begin{subfigure}[b]{0.33\textwidth}
         \centering
         \includegraphics[width=\textwidth]{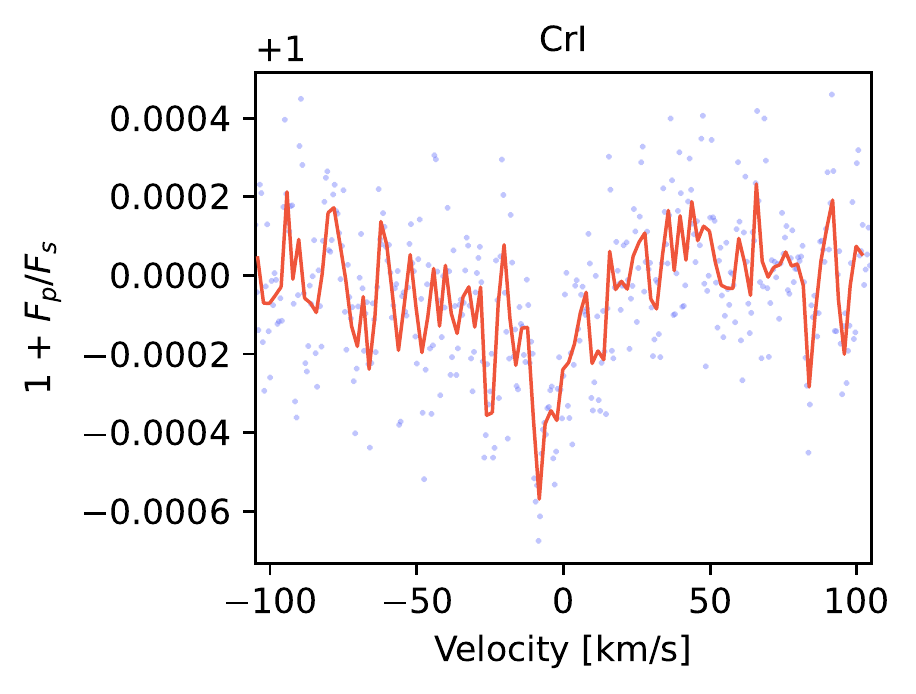}
         \label{fig:five over x}
     \end{subfigure}
     
     \begin{subfigure}[b]{0.33\textwidth}
         \centering
         \includegraphics[width=\textwidth]{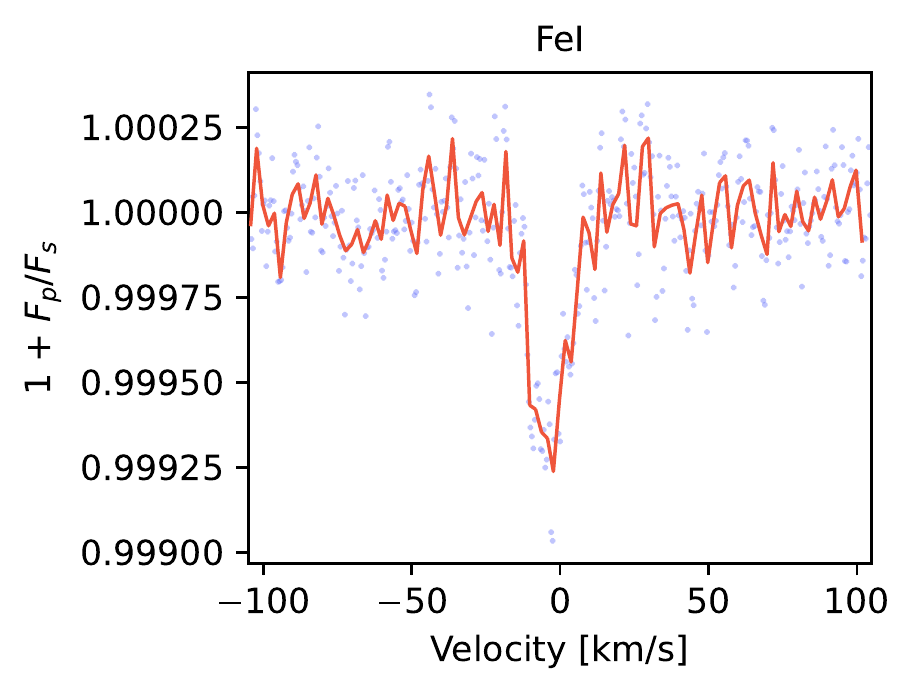}
         \label{fig:y equals x}
     \end{subfigure}
     \hfill
     \begin{subfigure}[b]{0.33\textwidth}
         \centering
         \includegraphics[width=\textwidth]{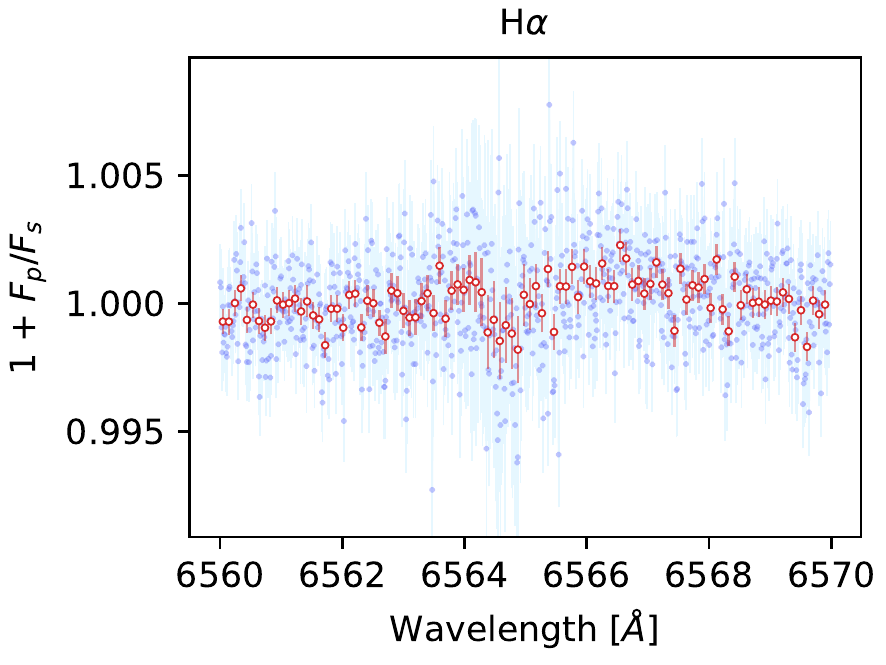}
         \label{fig:three sin x}
     \end{subfigure}
     \hfill
     \begin{subfigure}[b]{0.33\textwidth}
         \centering
         \includegraphics[width=\textwidth]{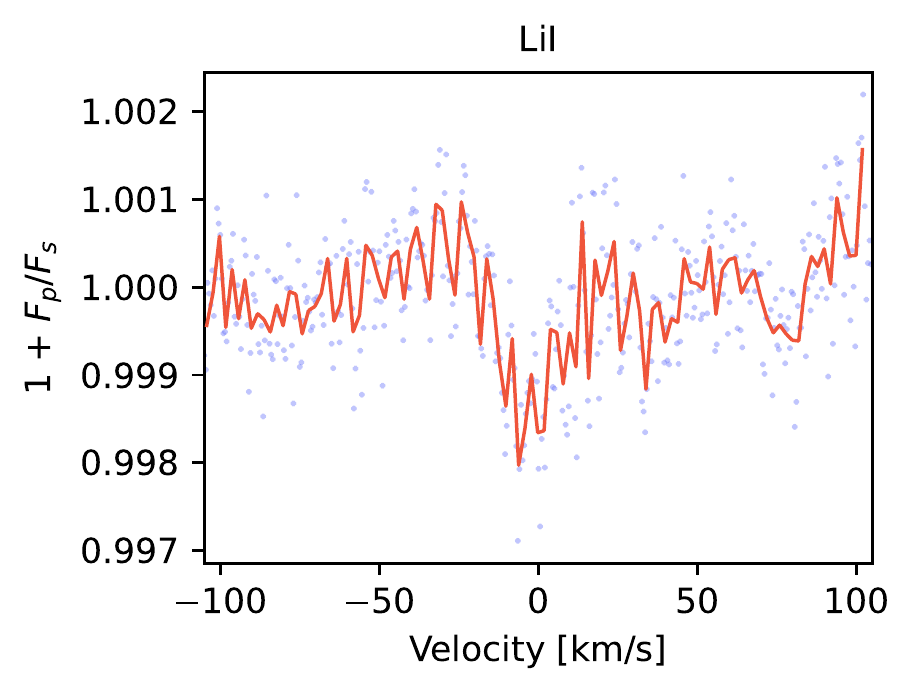}
         \label{fig:five over x}
     \end{subfigure}
        
     \begin{subfigure}[b]{0.33\textwidth}
         \centering
         \includegraphics[width=\textwidth]{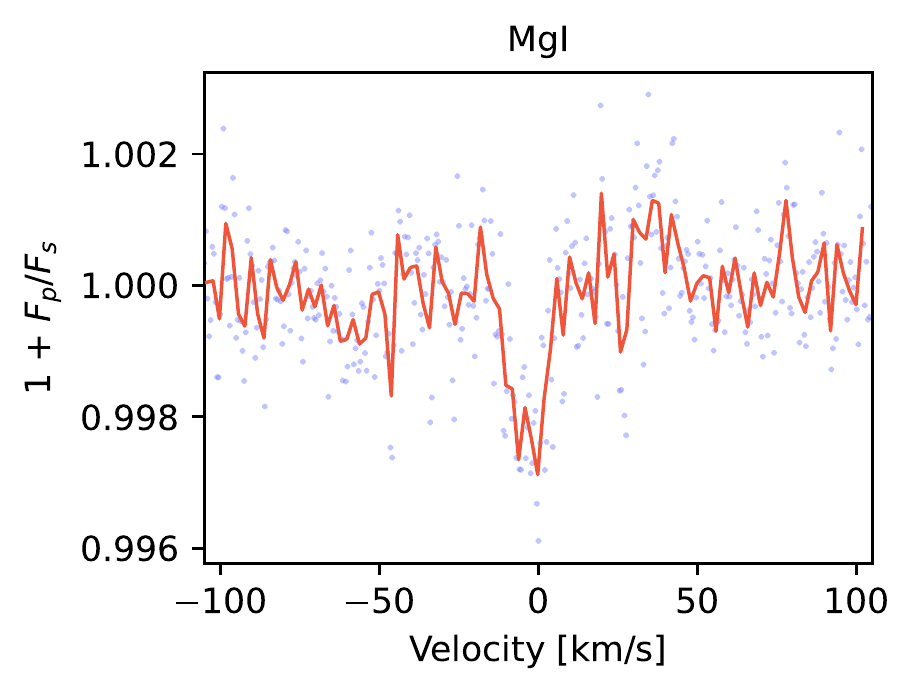}
         \label{fig:y equals x}
     \end{subfigure}
     \hfill
     \begin{subfigure}[b]{0.33\textwidth}
         \centering
         \includegraphics[width=\textwidth]{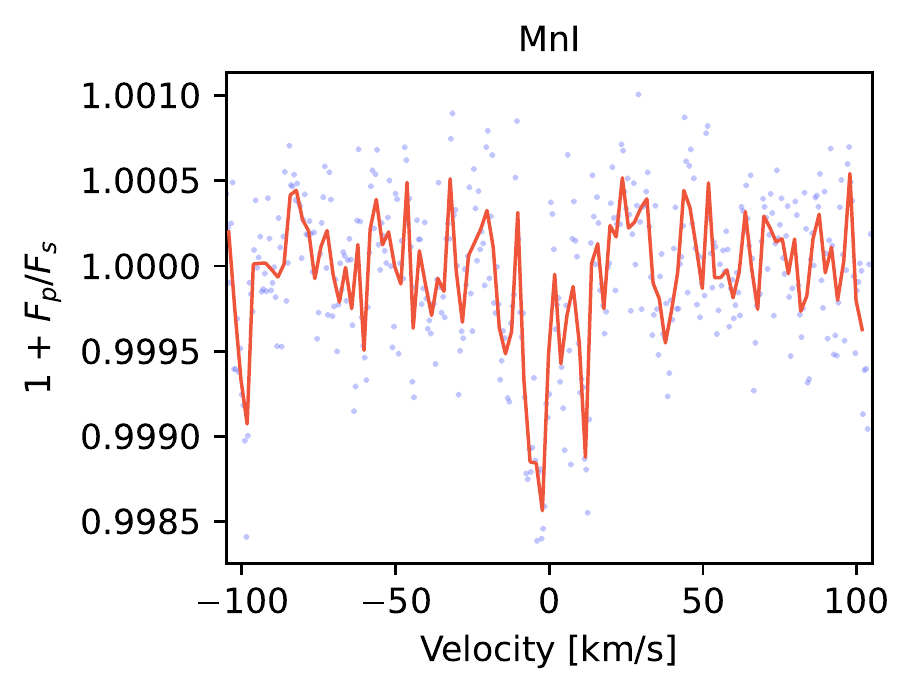}
         \label{fig:three sin x}
     \end{subfigure}
     \hfill
     \begin{subfigure}[b]{0.33\textwidth}
         \centering
         \includegraphics[width=\textwidth]{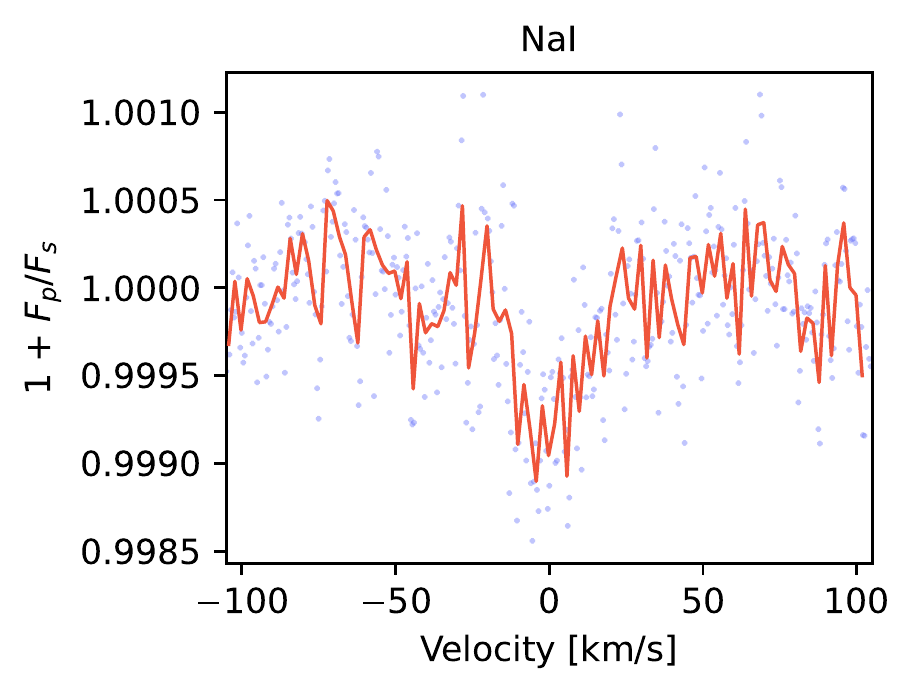}
         \label{fig:five over x}
     \end{subfigure}
        
     \begin{subfigure}[b]{0.33\textwidth}
         \centering
         \includegraphics[width=\textwidth]{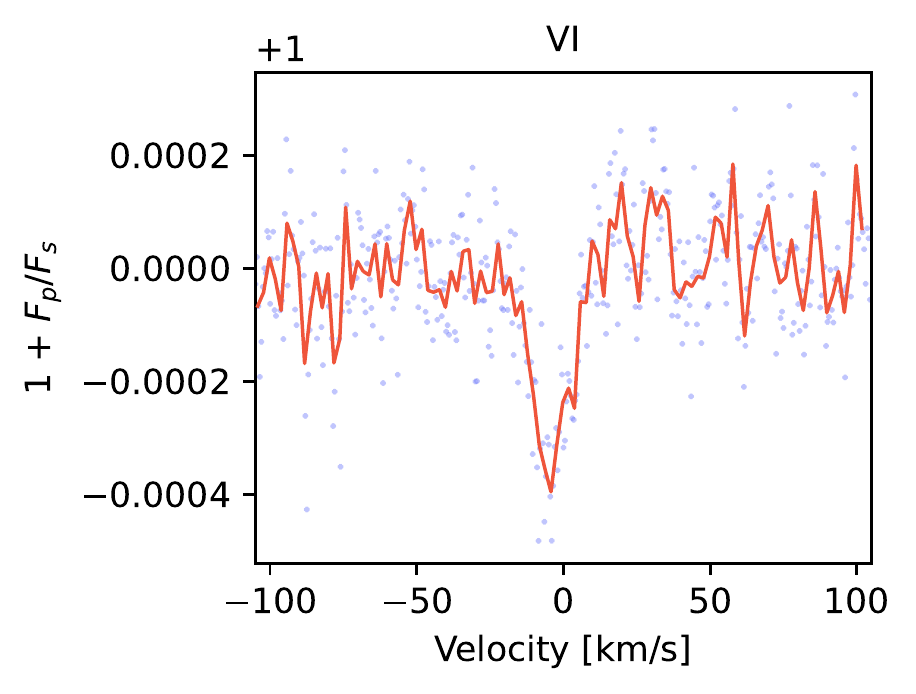}
         \label{fig:y equals x}
     \end{subfigure}
     \hfill
     \begin{subfigure}[b]{0.33\textwidth}
         \centering
         \includegraphics[width=\textwidth]{nothing.png}
         \label{fig:three sin x}
     \end{subfigure}
     \hfill
     \begin{subfigure}[b]{0.33\textwidth}
         \centering
         \includegraphics[width=\textwidth]{nothing.png}
         \label{fig:five over x}
     \end{subfigure}
        
        \caption{Same as Fig. \ref{fig:W76n1ccf} for WASP-76b, night 2 (2018 October 31). }
        \label{fig:W76n2ccf}
\end{figure*}

\begin{figure*}
     \centering
     \begin{subfigure}[b]{0.33\textwidth}
         \centering
         \includegraphics[width=\textwidth]{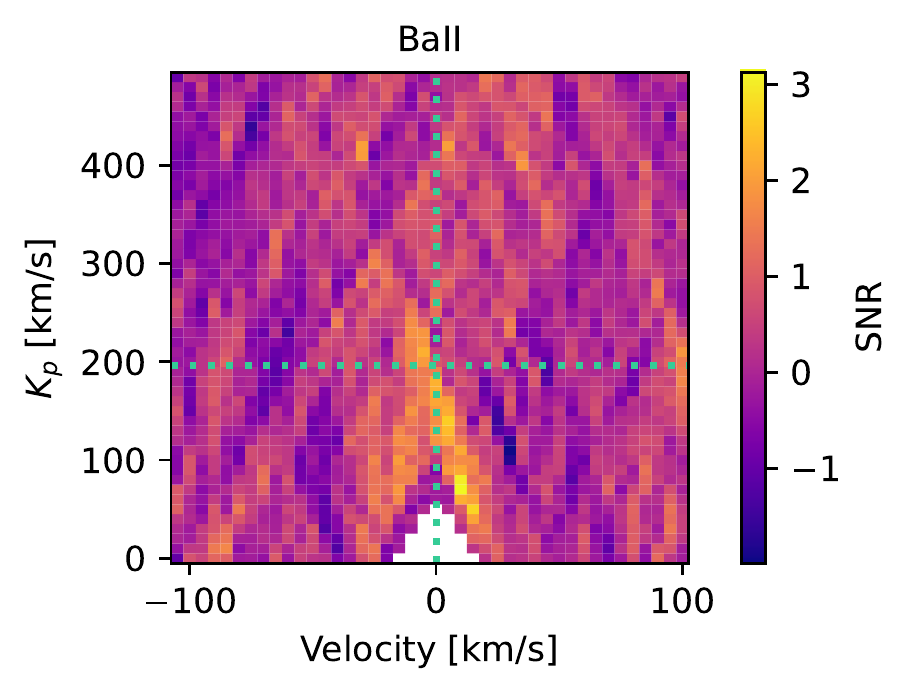}
         \label{fig:y equals x}
     \end{subfigure}
     \hfill
     \begin{subfigure}[b]{0.33\textwidth}
         \centering
         \includegraphics[width=\textwidth]{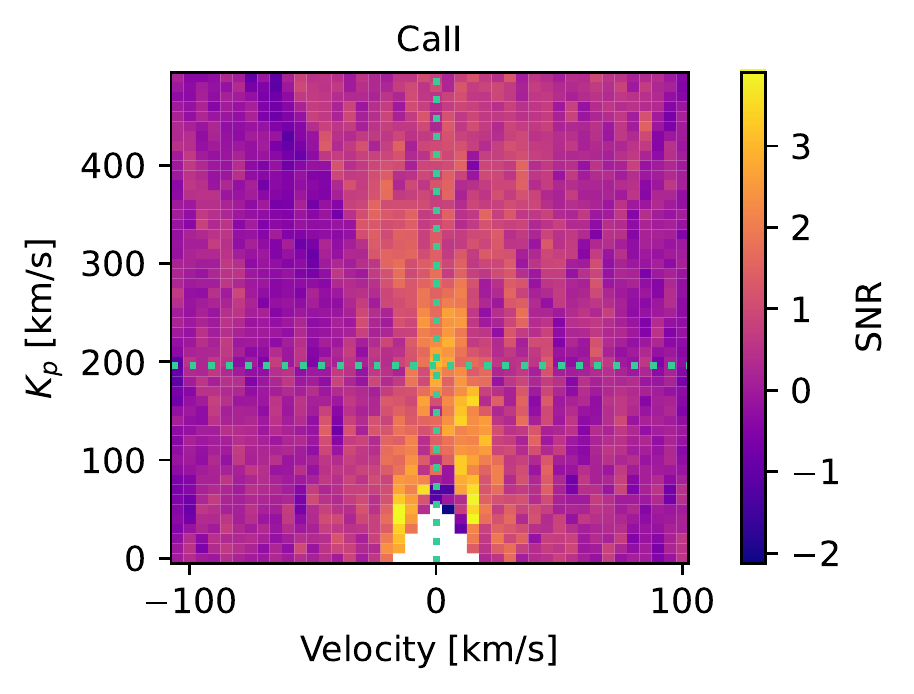}
         \label{fig:three sin x}
     \end{subfigure}
     \hfill
     \begin{subfigure}[b]{0.33\textwidth}
         \centering
         \includegraphics[width=\textwidth]{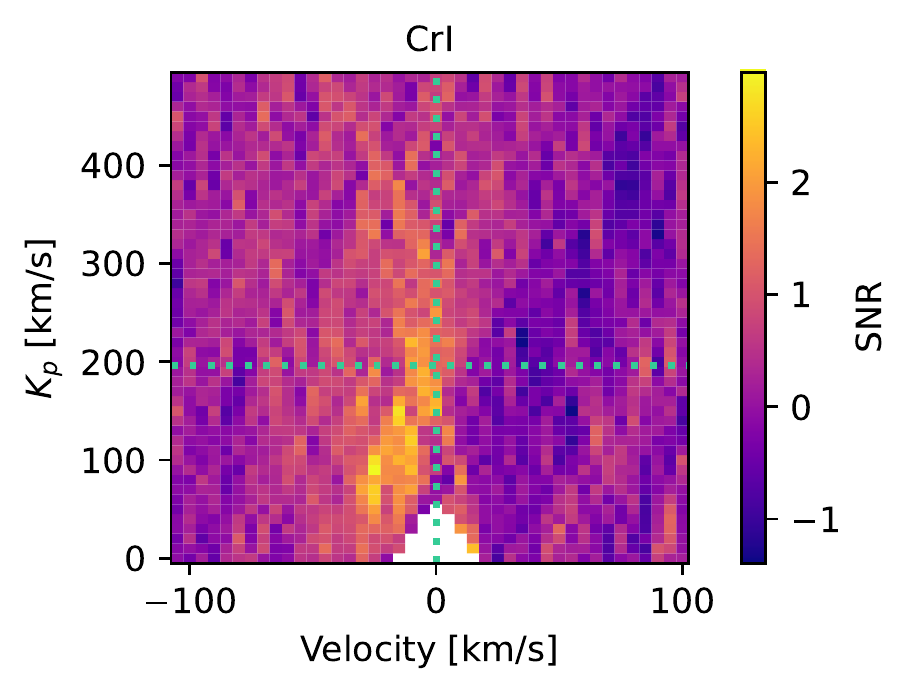}
         \label{fig:five over x}
     \end{subfigure}
     
     \begin{subfigure}[b]{0.33\textwidth}
         \centering
         \includegraphics[width=\textwidth]{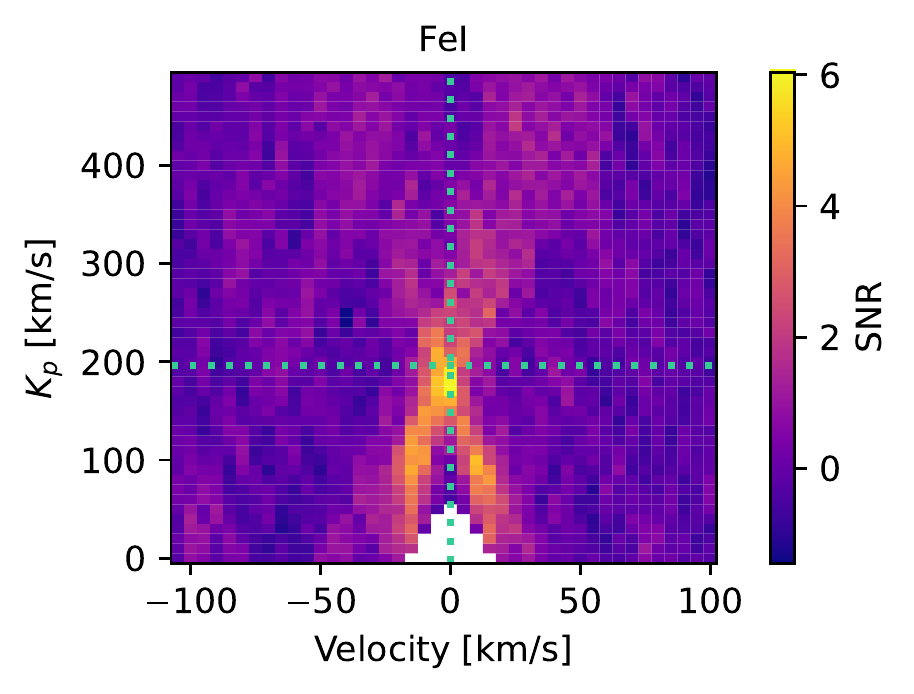}
         \label{fig:y equals x}
     \end{subfigure}
     \hfill
     \begin{subfigure}[b]{0.33\textwidth}
         \centering
         \includegraphics[width=\textwidth]{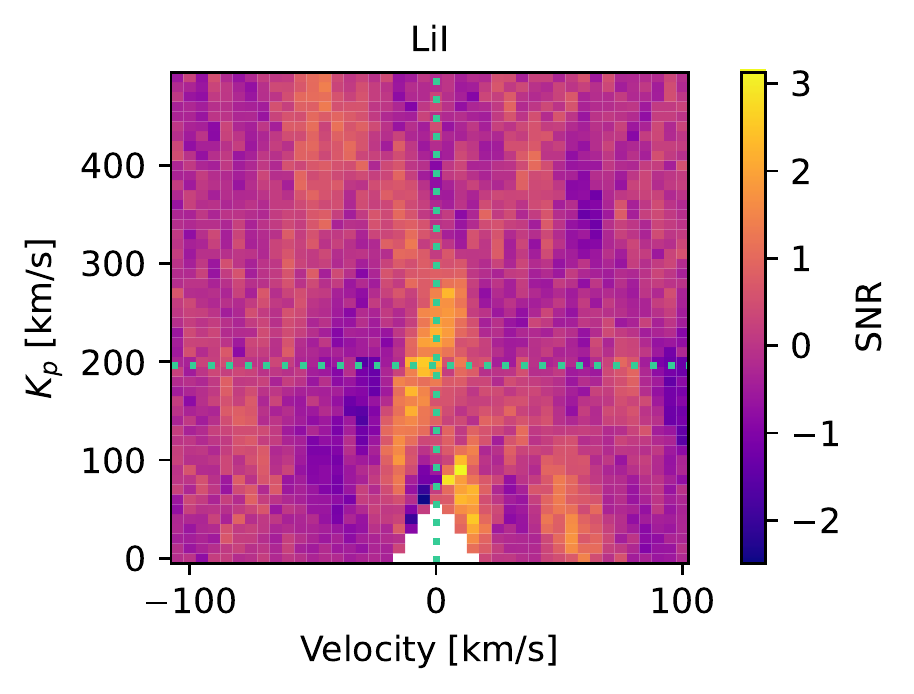}
         \label{fig:three sin x}
     \end{subfigure}
     \hfill
     \begin{subfigure}[b]{0.33\textwidth}
         \centering
         \includegraphics[width=\textwidth]{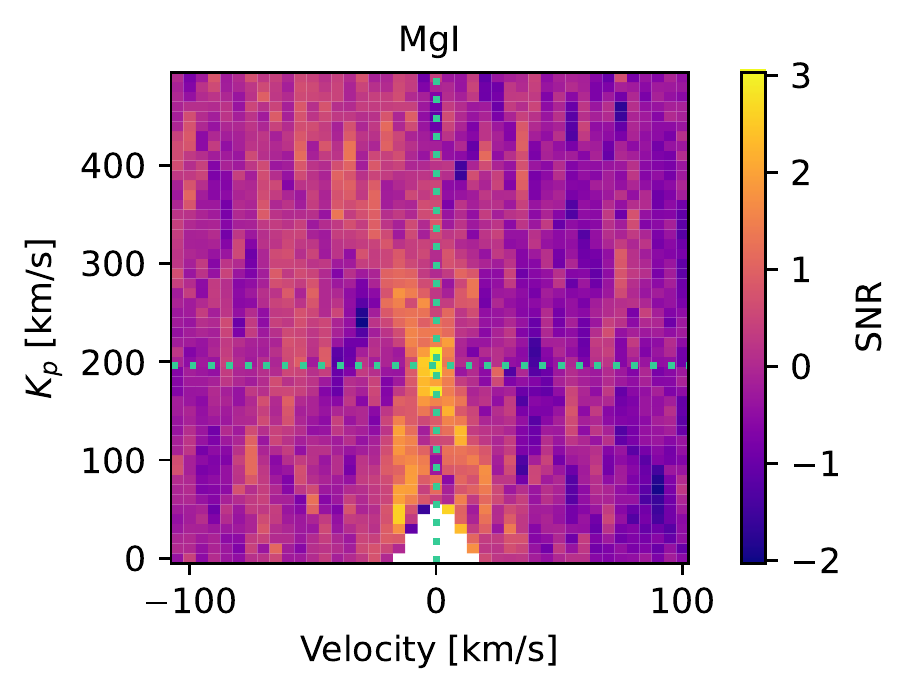}
         \label{fig:five over x}
     \end{subfigure}
     
     \begin{subfigure}[b]{0.33\textwidth}
         \centering
         \includegraphics[width=\textwidth]{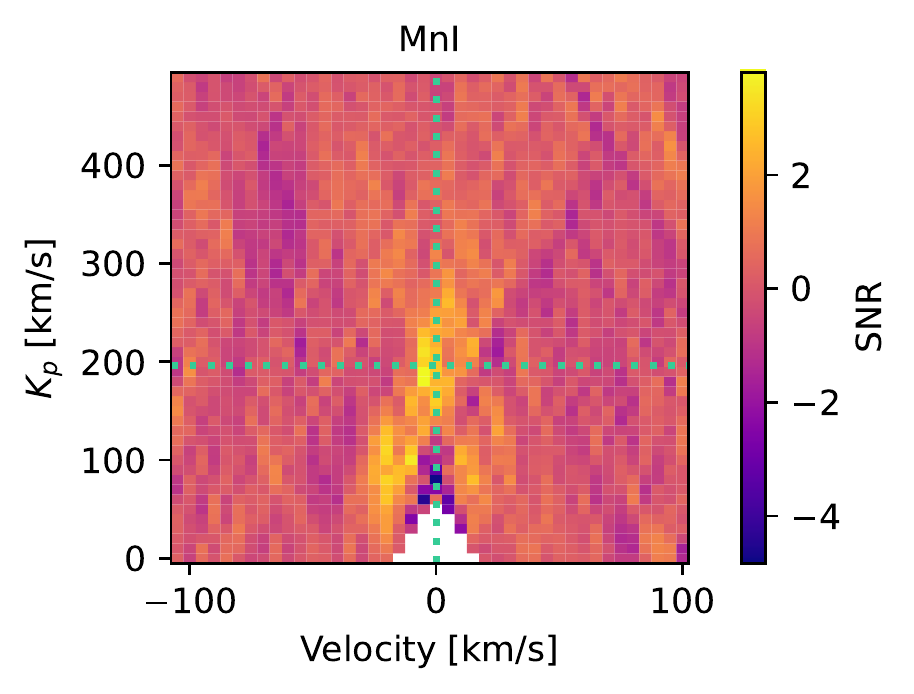}
         \label{fig:y equals x}
     \end{subfigure}   
     \hfill
     \begin{subfigure}[b]{0.33\textwidth}
         \centering
         \includegraphics[width=\textwidth]{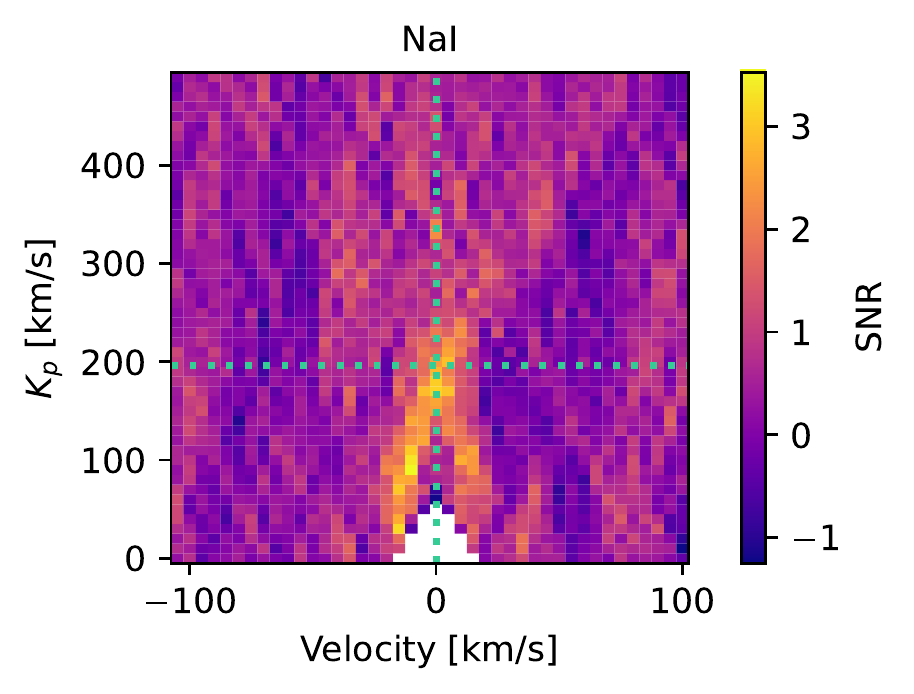}
         \label{fig:three sin x}
     \end{subfigure}
     \hfill
     \begin{subfigure}[b]{0.33\textwidth}
         \centering
         \includegraphics[width=\textwidth]{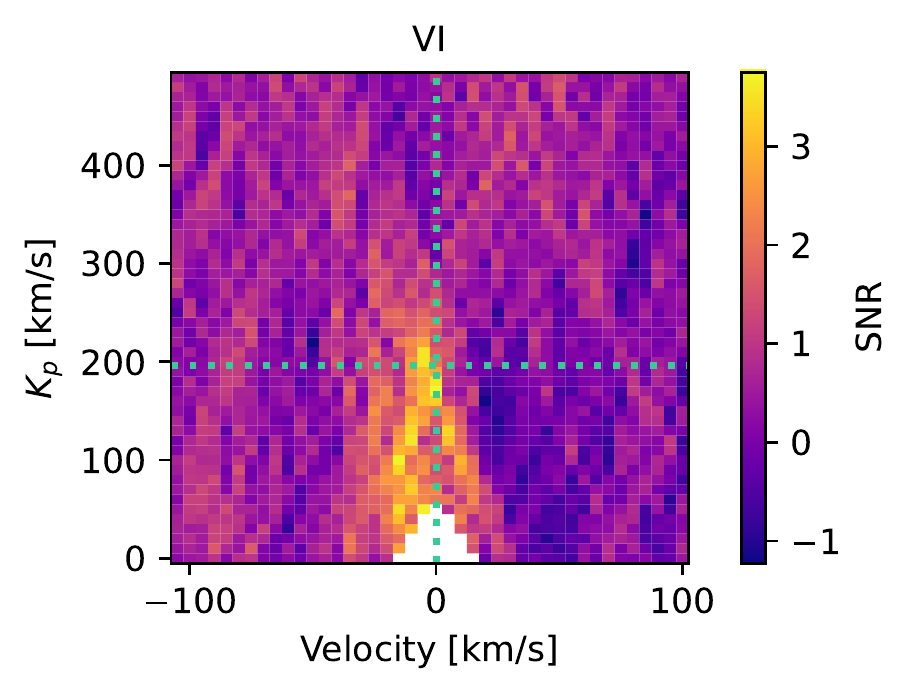}
         \label{fig:five over x}
     \end{subfigure}
        
        \caption{Same as Fig. \ref{fig:W76n1kp} for WASP-76b, night 2 (2018 October 31). }
        \label{fig:W76n2kp}
\end{figure*}


\FloatBarrier
\begin{figure*}
     \centering
     \begin{subfigure}[b]{0.33\textwidth}
         \centering
         \includegraphics[width=\textwidth]{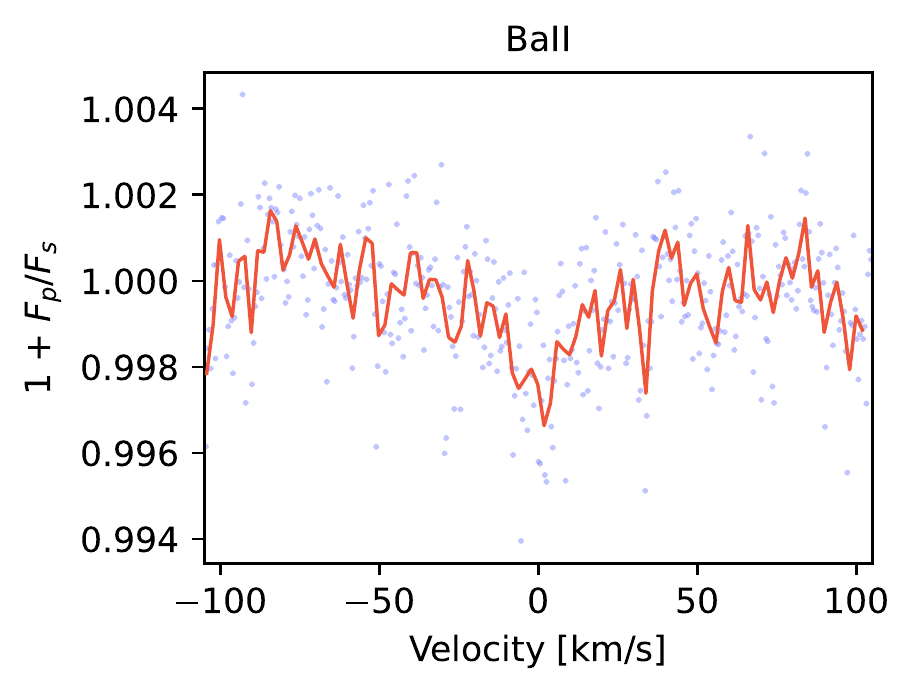}
         \label{fig:y equals x}
     \end{subfigure}
     \hfill
     \begin{subfigure}[b]{0.33\textwidth}
         \centering
         \includegraphics[width=\textwidth]{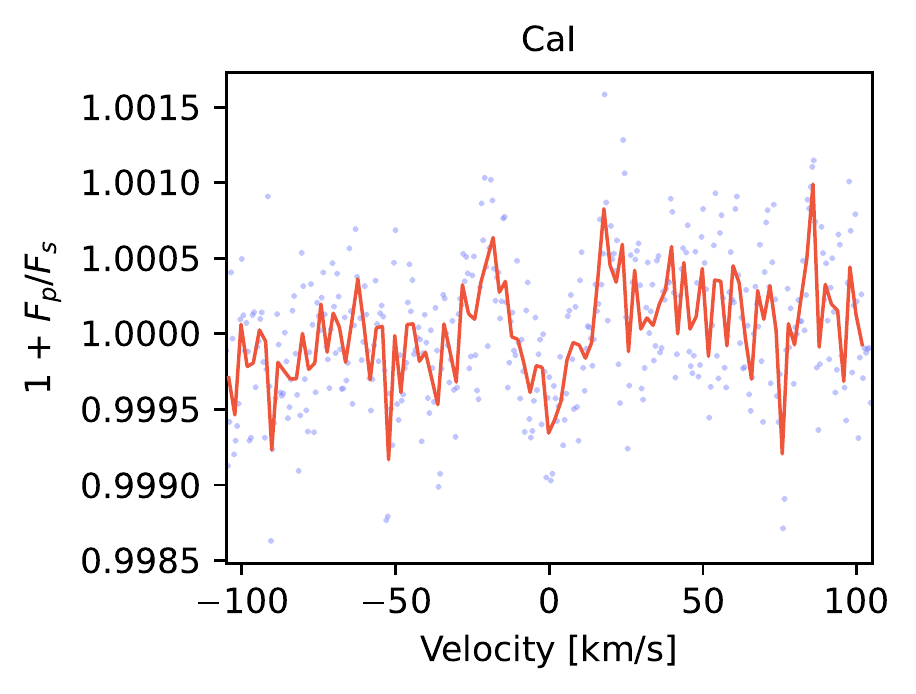}
         \label{fig:three sin x}
     \end{subfigure}
     \hfill
     \begin{subfigure}[b]{0.33\textwidth}
         \centering
         \includegraphics[width=\textwidth]{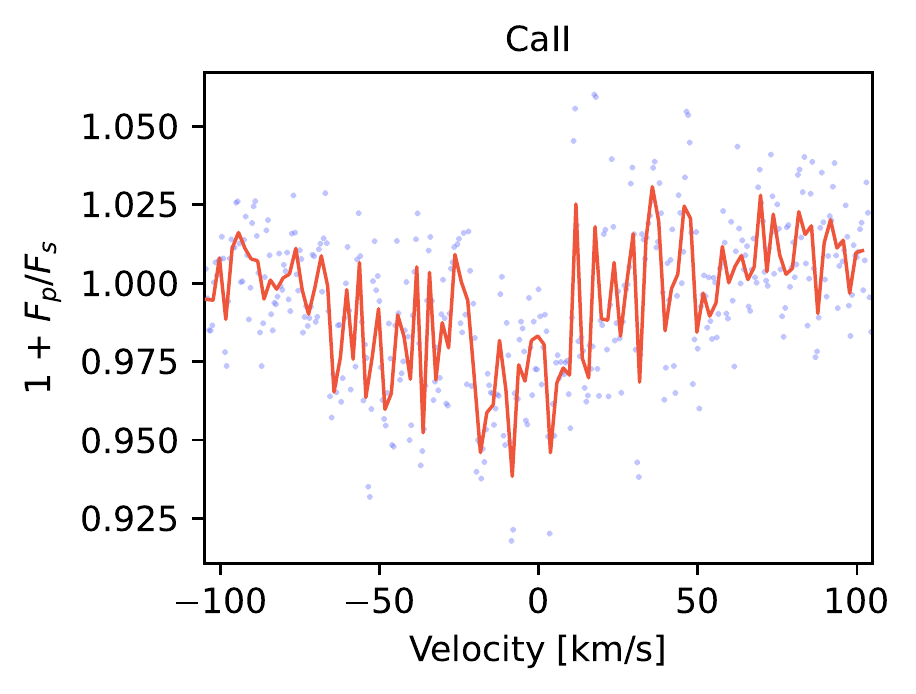}
         \label{fig:five over x}
     \end{subfigure}
     
     \begin{subfigure}[b]{0.33\textwidth}
         \centering
         \includegraphics[width=\textwidth]{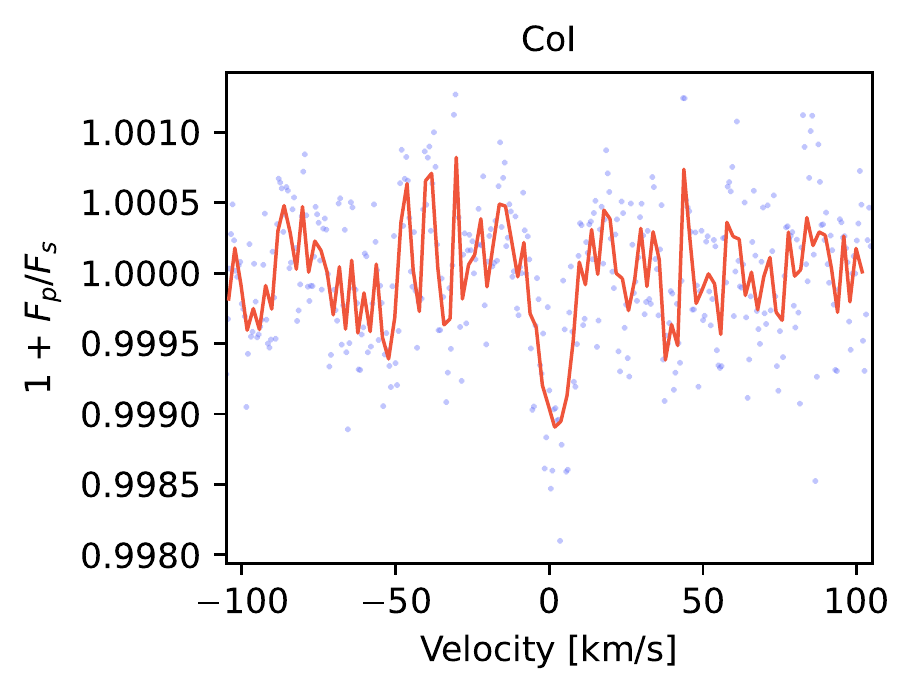}
         \label{fig:y equals x}
     \end{subfigure}
     \hfill
     \begin{subfigure}[b]{0.33\textwidth}
         \centering
         \includegraphics[width=\textwidth]{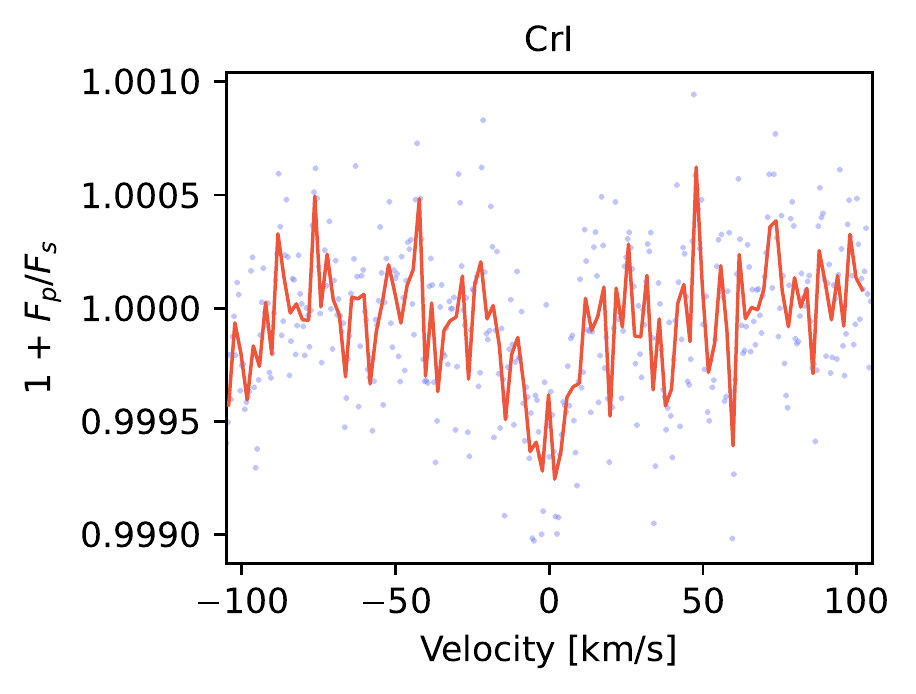}
         \label{fig:three sin x}
     \end{subfigure}
     \hfill
     \begin{subfigure}[b]{0.33\textwidth}
         \centering
         \includegraphics[width=\textwidth]{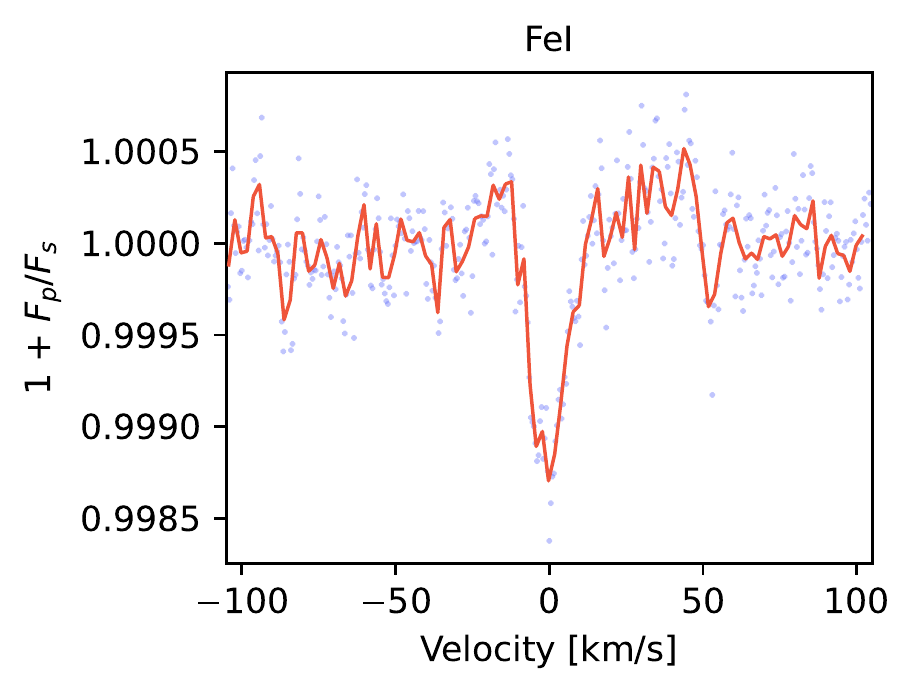}
         \label{fig:five over x}
     \end{subfigure}
        
     \begin{subfigure}[b]{0.33\textwidth}
         \centering
         \includegraphics[width=\textwidth]{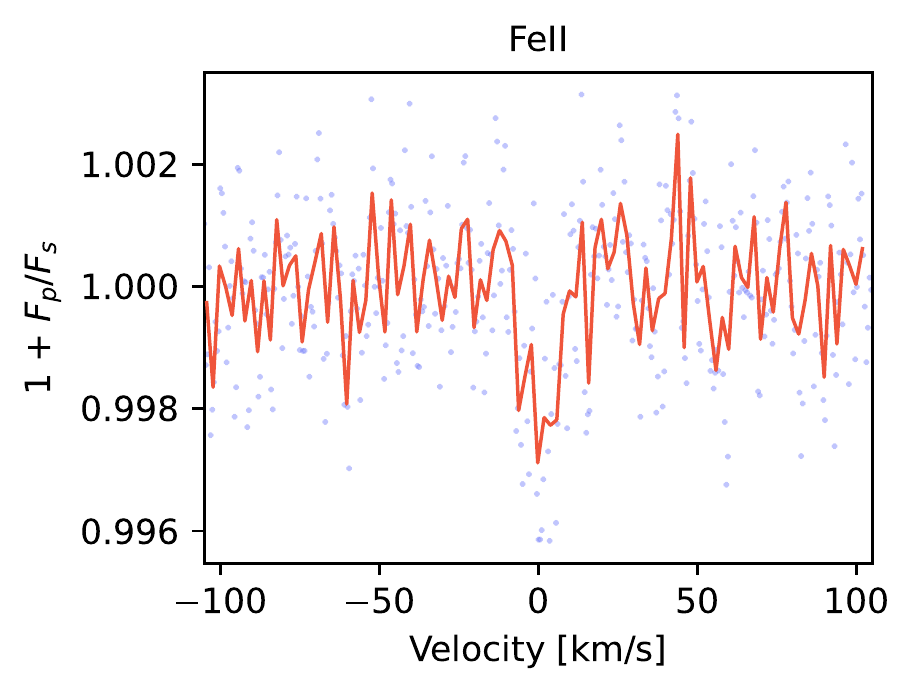}
         \label{fig:y equals x}
     \end{subfigure}
     \hfill
     \begin{subfigure}[b]{0.33\textwidth}
         \centering
         \includegraphics[width=\textwidth]{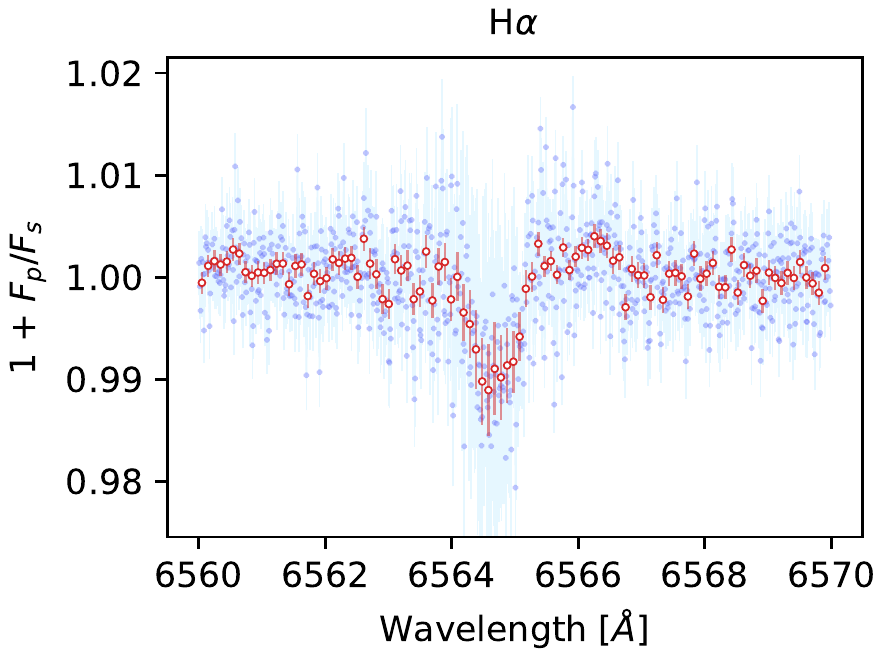}
         \label{fig:three sin x}
     \end{subfigure}
     \hfill
     \begin{subfigure}[b]{0.33\textwidth}
         \centering
         \includegraphics[width=\textwidth]{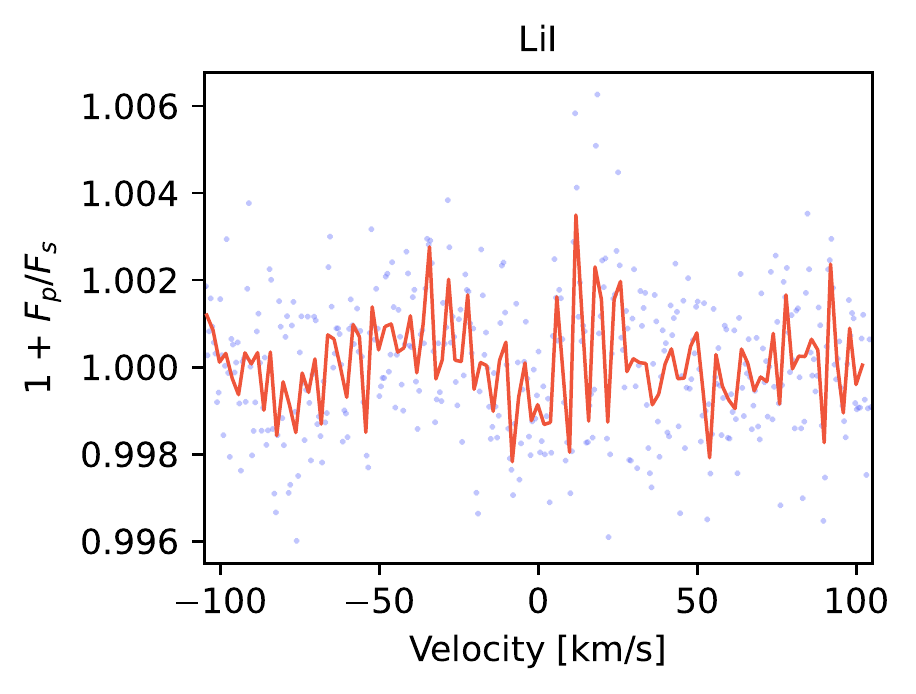}
         \label{fig:five over x}
     \end{subfigure}
        
     \begin{subfigure}[b]{0.33\textwidth}
         \centering
         \includegraphics[width=\textwidth]{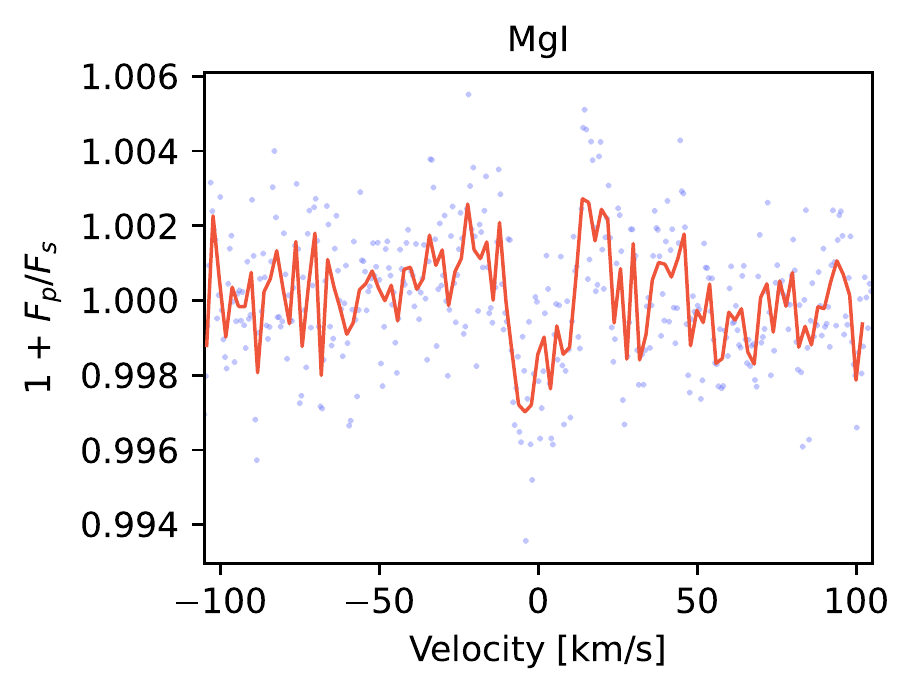}
         \label{fig:y equals x}
     \end{subfigure}
     \hfill
     \begin{subfigure}[b]{0.33\textwidth}
         \centering
         \includegraphics[width=\textwidth]{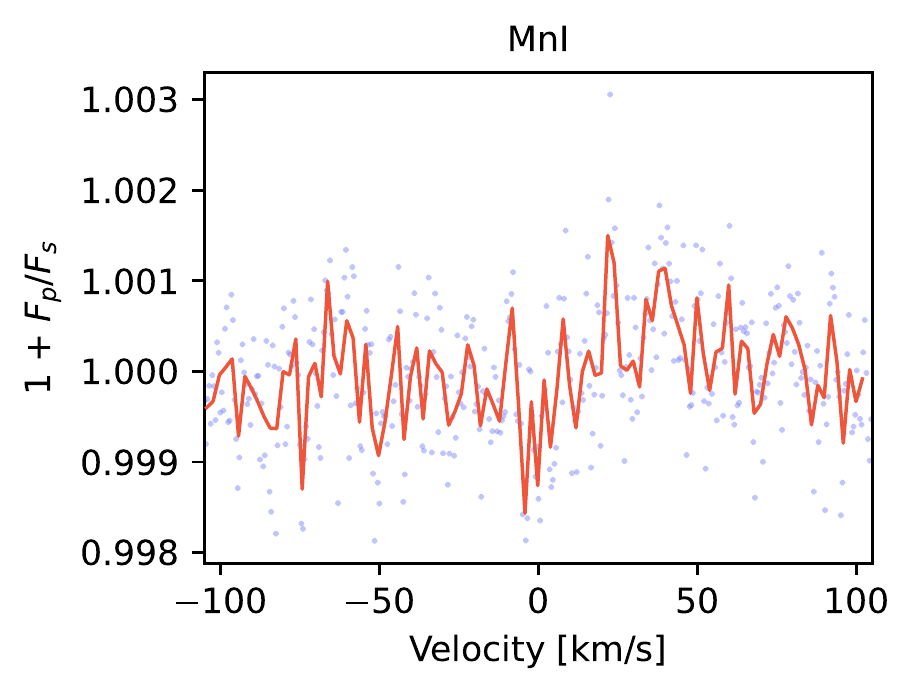}
         \label{fig:three sin x}
     \end{subfigure}
     \hfill
     \begin{subfigure}[b]{0.33\textwidth}
         \centering
         \includegraphics[width=\textwidth]{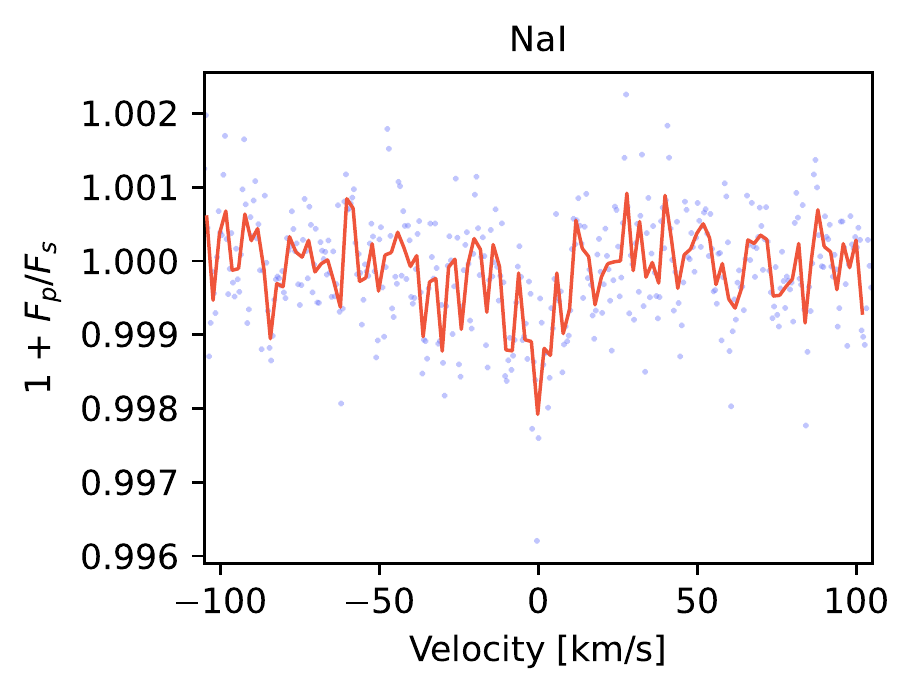}
         \label{fig:five over x}
     \end{subfigure}
        
     \begin{subfigure}[b]{0.33\textwidth}
         \centering
         \includegraphics[width=\textwidth]{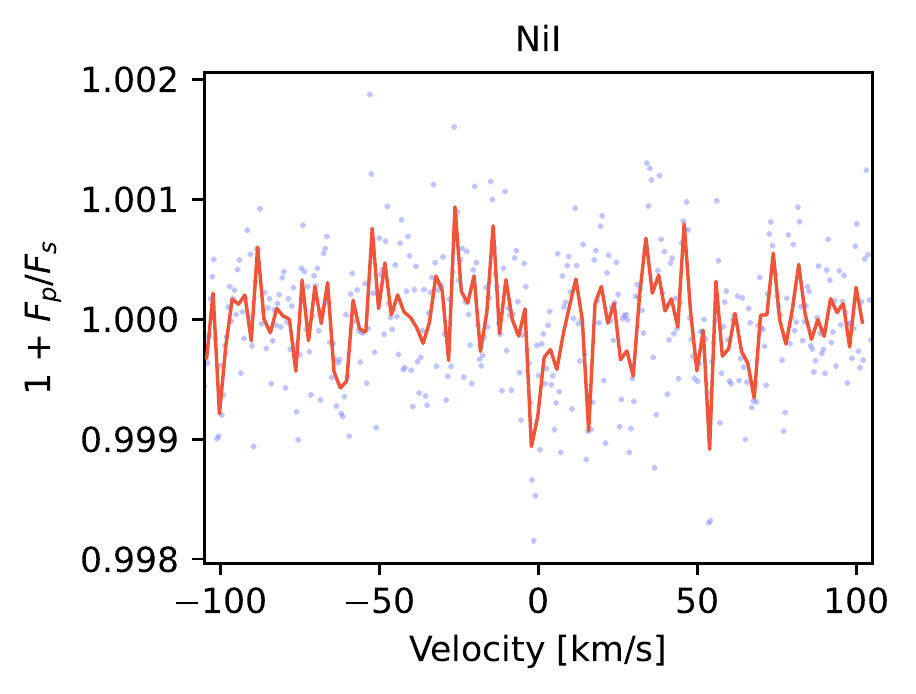}
         \label{fig:y equals x}
     \end{subfigure}
     \hfill
     \begin{subfigure}[b]{0.33\textwidth}
         \centering
         \includegraphics[width=\textwidth]{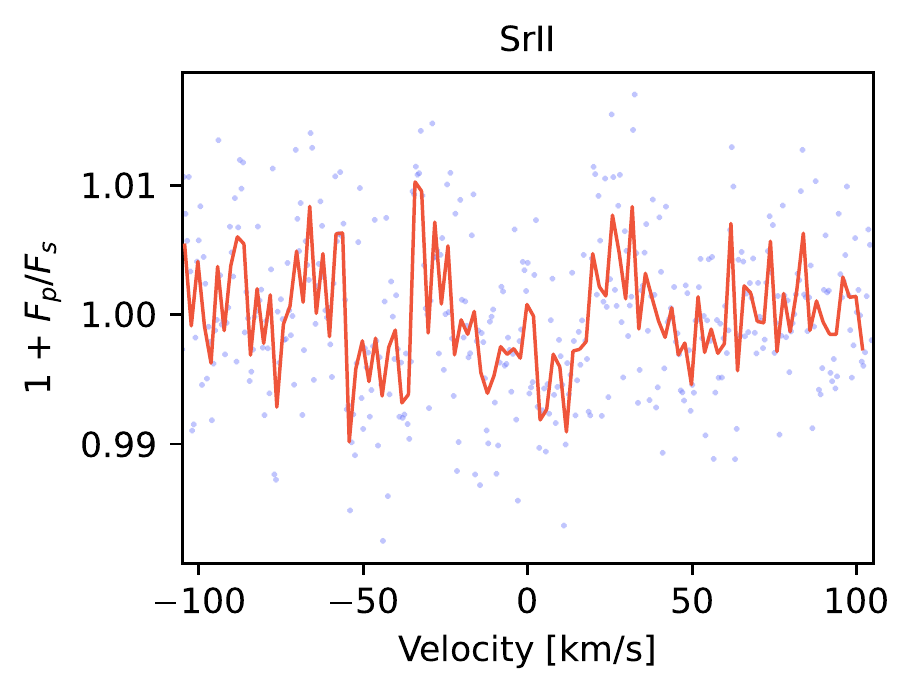}
         \label{fig:three sin x}
     \end{subfigure}
     \hfill
     \begin{subfigure}[b]{0.33\textwidth}
         \centering
         \includegraphics[width=\textwidth]{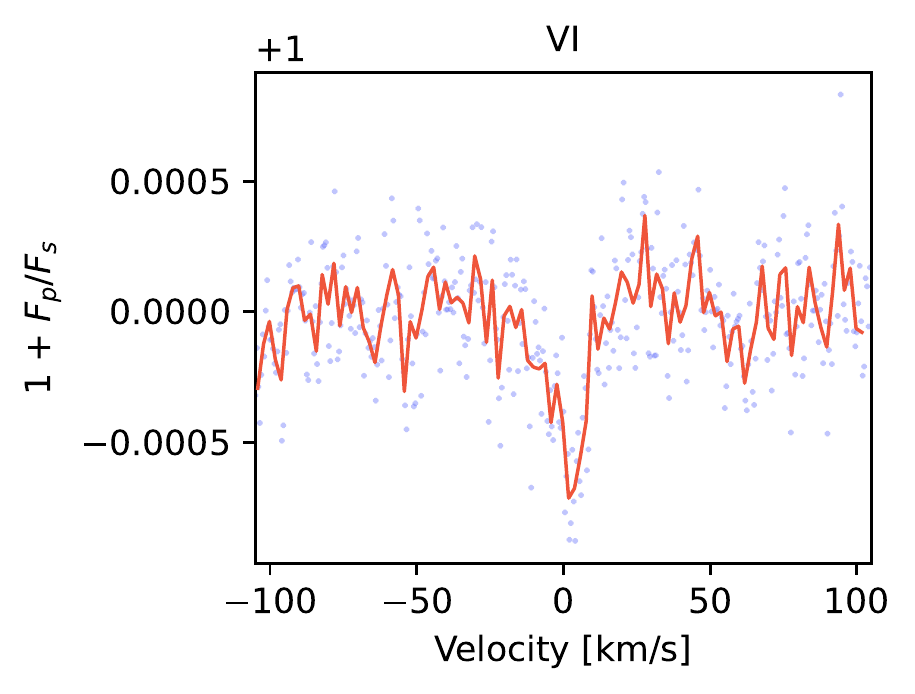}
         \label{fig:five over x}
     \end{subfigure}
        \caption{Same as Fig. \ref{fig:W76n1ccf} for WASP-121b, night 1 (1UT - 2018 November 30). }
        \label{fig:W121n1ccf}
\end{figure*}

\begin{figure*}
     \centering
     \begin{subfigure}[b]{0.33\textwidth}
         \centering
         \includegraphics[width=\textwidth]{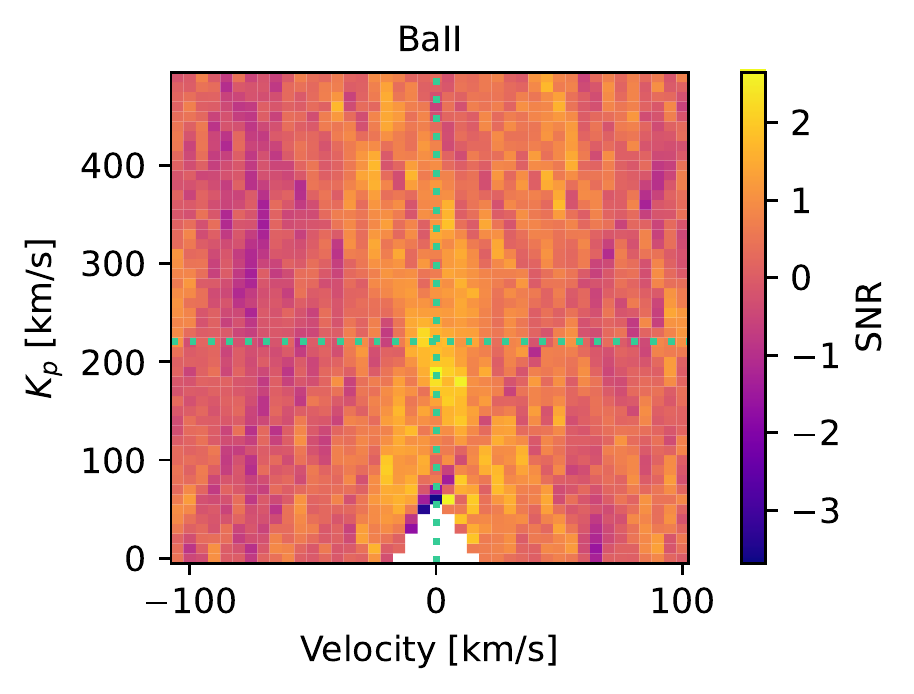}
         \label{fig:y equals x}
     \end{subfigure}
     \hfill
     \begin{subfigure}[b]{0.33\textwidth}
         \centering
         \includegraphics[width=\textwidth]{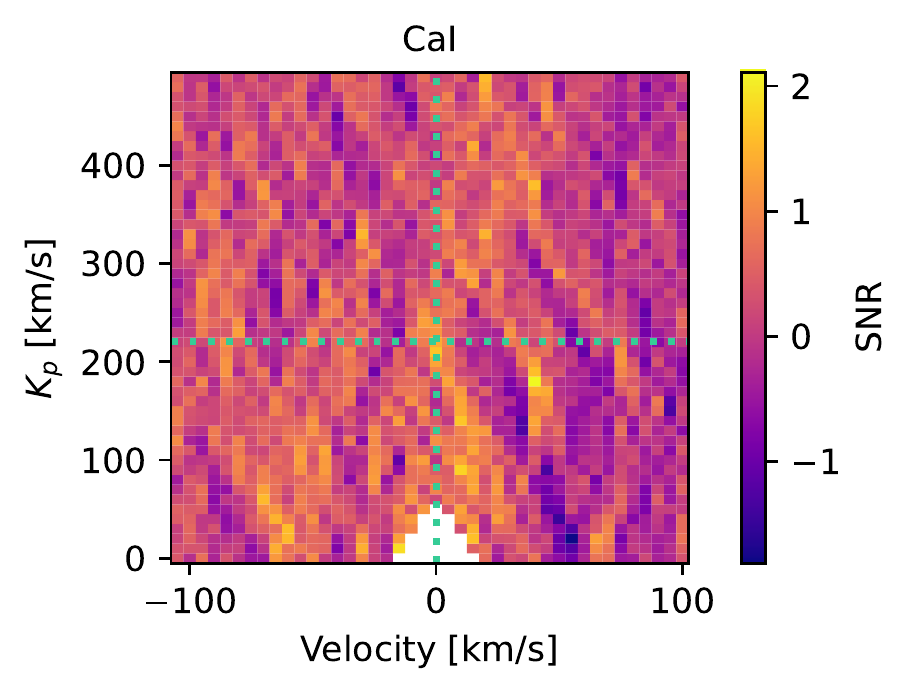}
         \label{fig:three sin x}
     \end{subfigure}
     \hfill
     \begin{subfigure}[b]{0.33\textwidth}
         \centering
         \includegraphics[width=\textwidth]{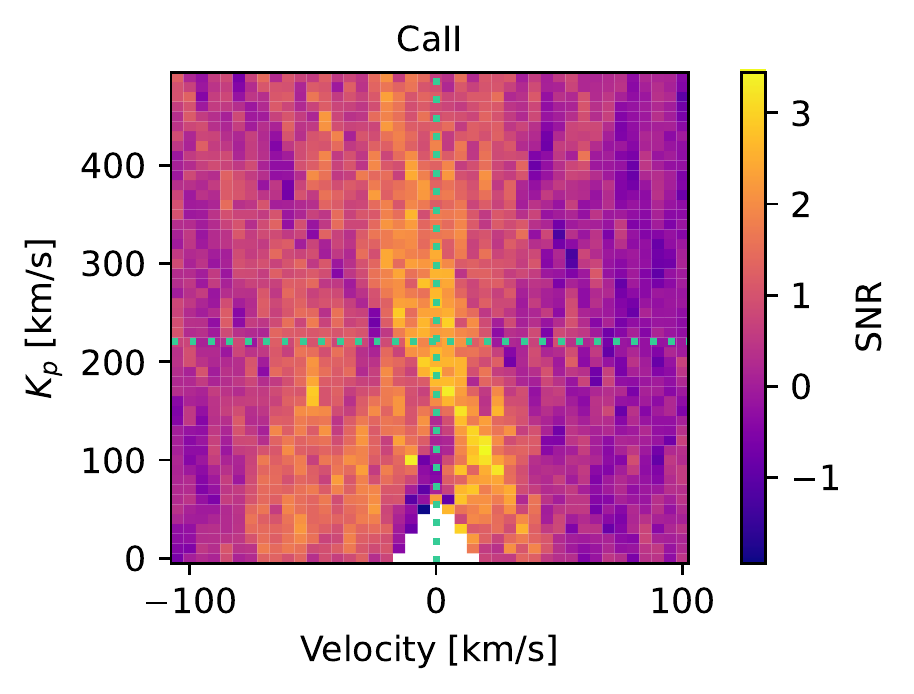}
         \label{fig:five over x}
     \end{subfigure}
     
     \begin{subfigure}[b]{0.33\textwidth}
         \centering
         \includegraphics[width=\textwidth]{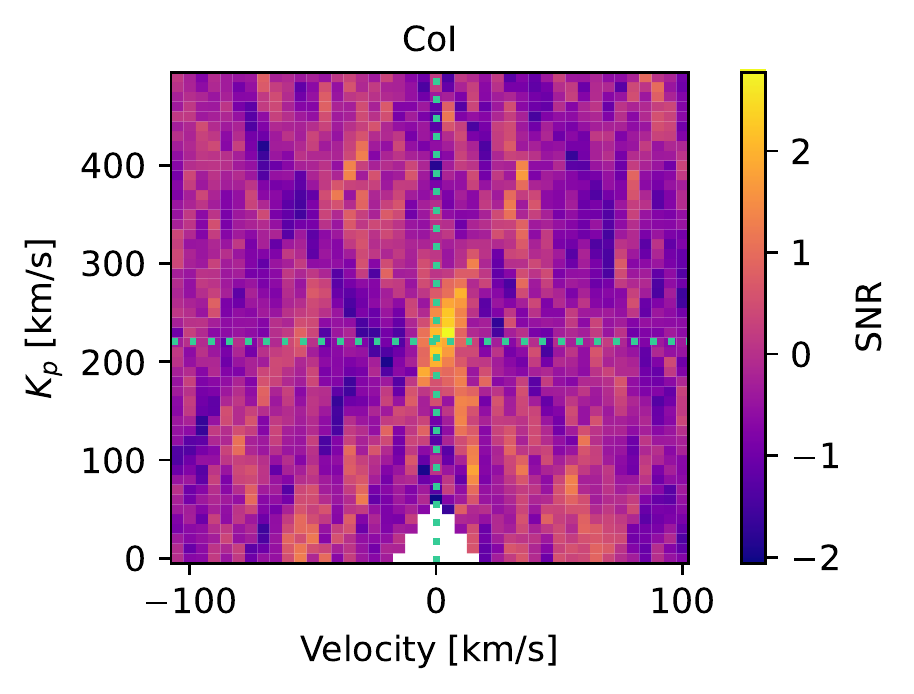}
         \label{fig:y equals x}
     \end{subfigure}
     \hfill
     \begin{subfigure}[b]{0.33\textwidth}
         \centering
         \includegraphics[width=\textwidth]{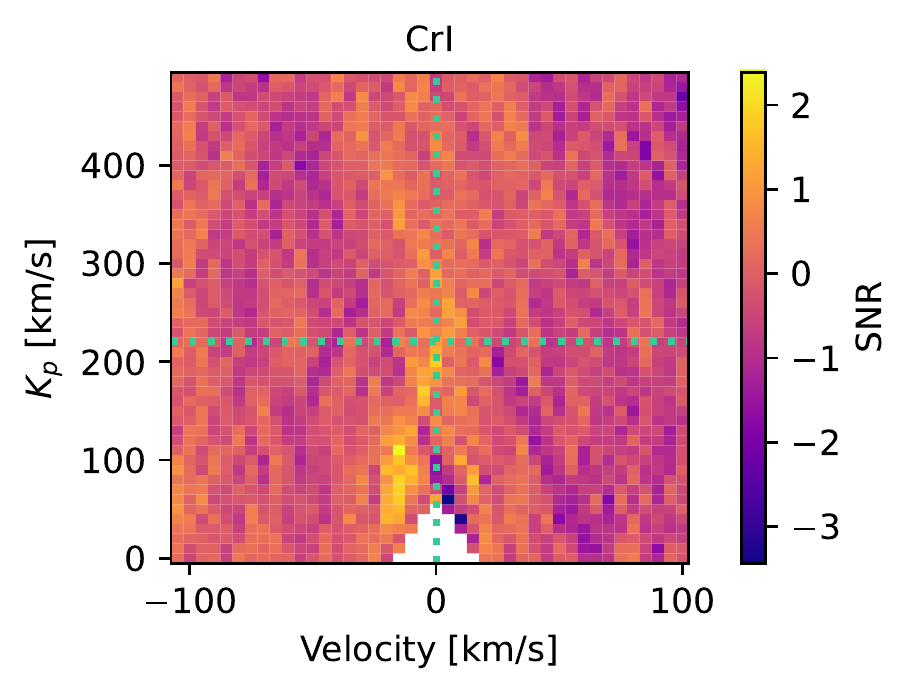}
         \label{fig:three sin x}
     \end{subfigure}
     \hfill
     \begin{subfigure}[b]{0.33\textwidth}
         \centering
         \includegraphics[width=\textwidth]{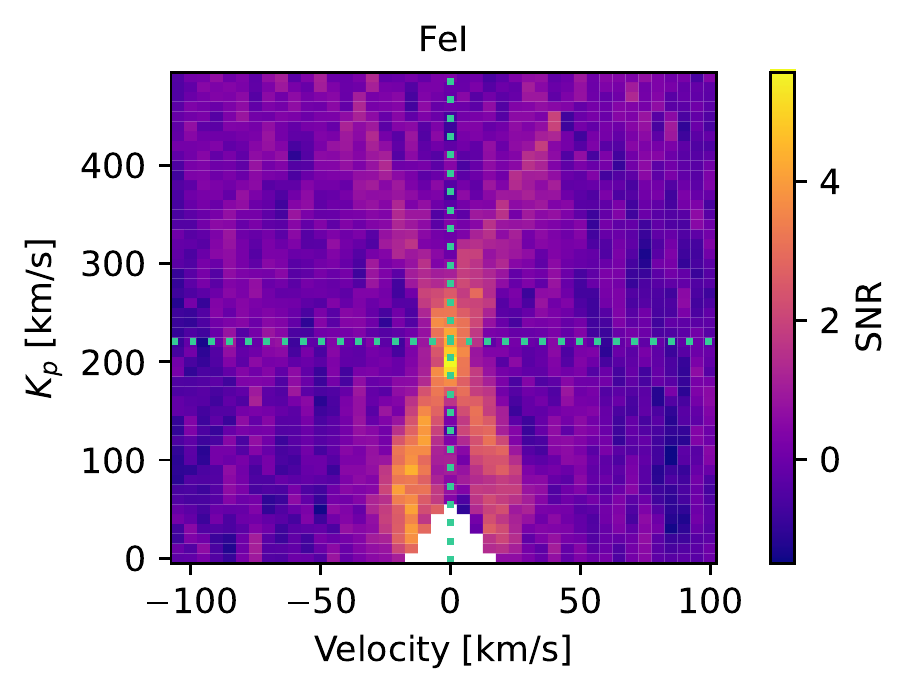}
         \label{fig:five over x}
     \end{subfigure}
     
     \begin{subfigure}[b]{0.33\textwidth}
         \centering
         \includegraphics[width=\textwidth]{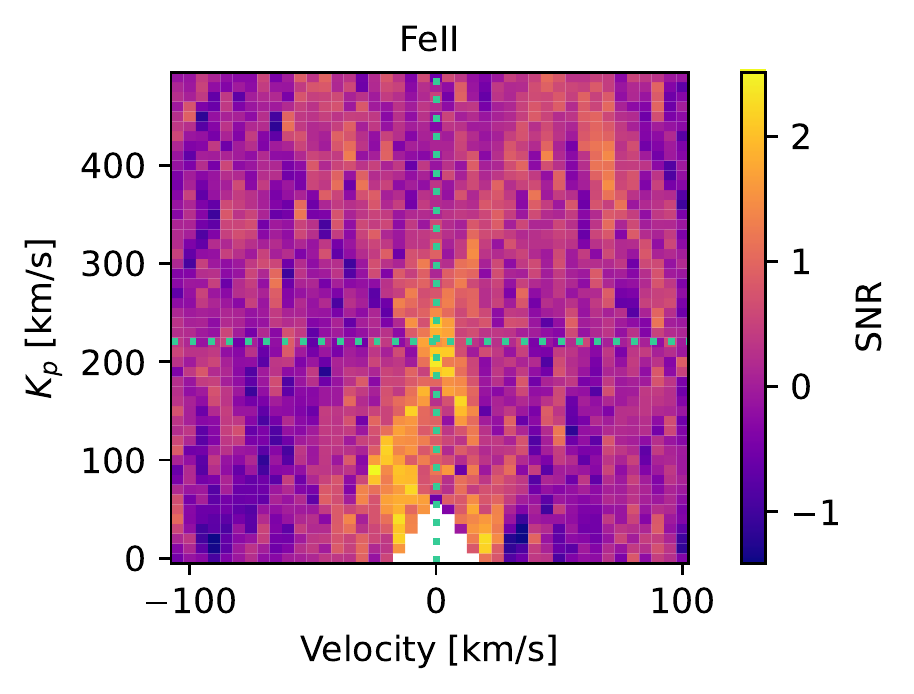}
         \label{fig:y equals x}
     \end{subfigure}   
     \hfill
     \begin{subfigure}[b]{0.33\textwidth}
         \centering
         \includegraphics[width=\textwidth]{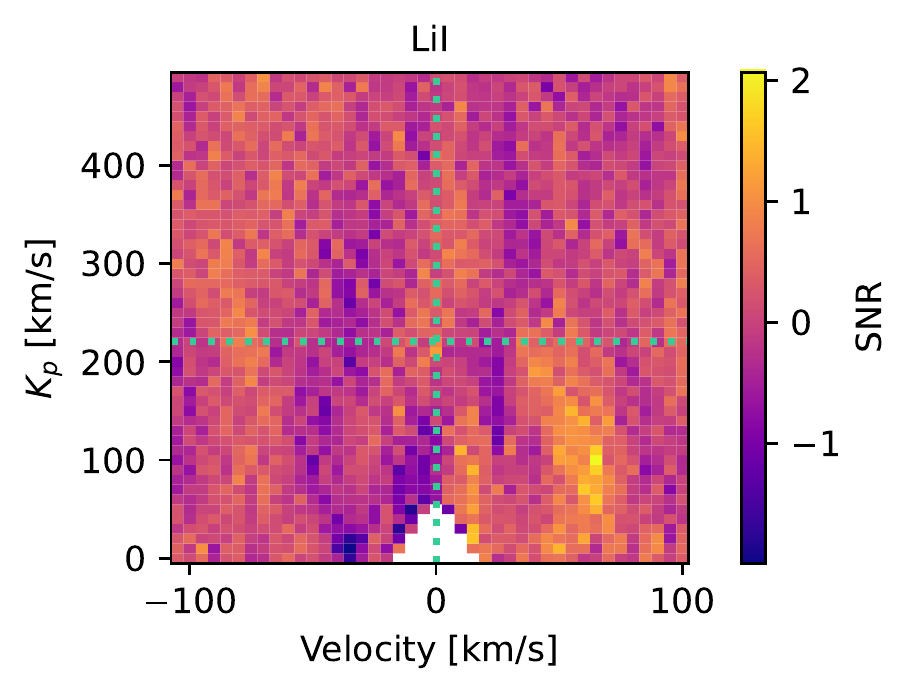}
         \label{fig:three sin x}
     \end{subfigure}
     \hfill
     \begin{subfigure}[b]{0.33\textwidth}
         \centering
         \includegraphics[width=\textwidth]{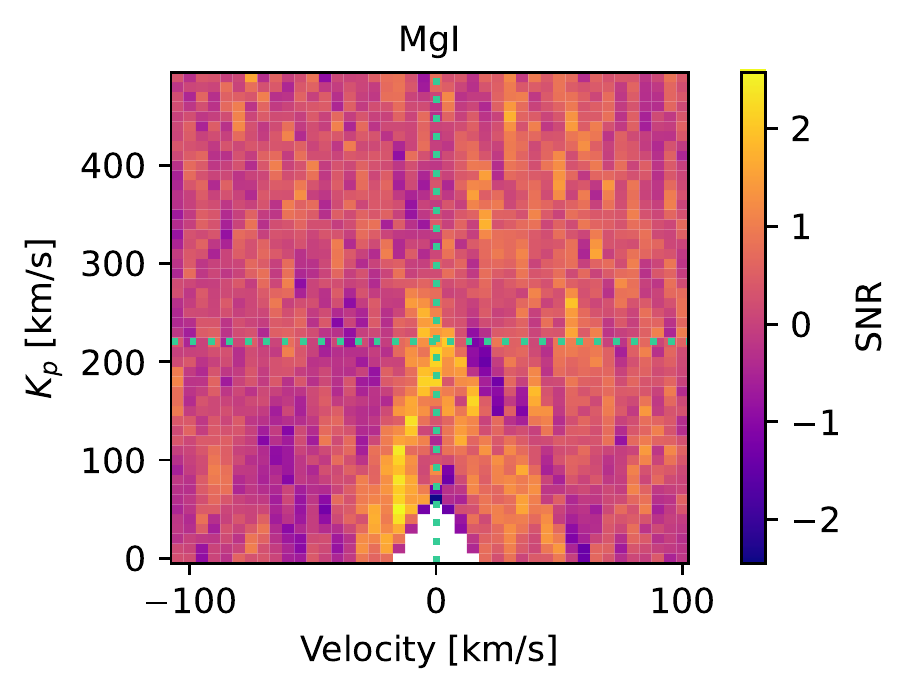}
         \label{fig:five over x}
     \end{subfigure}
        
     \begin{subfigure}[b]{0.33\textwidth}
         \centering
         \includegraphics[width=\textwidth]{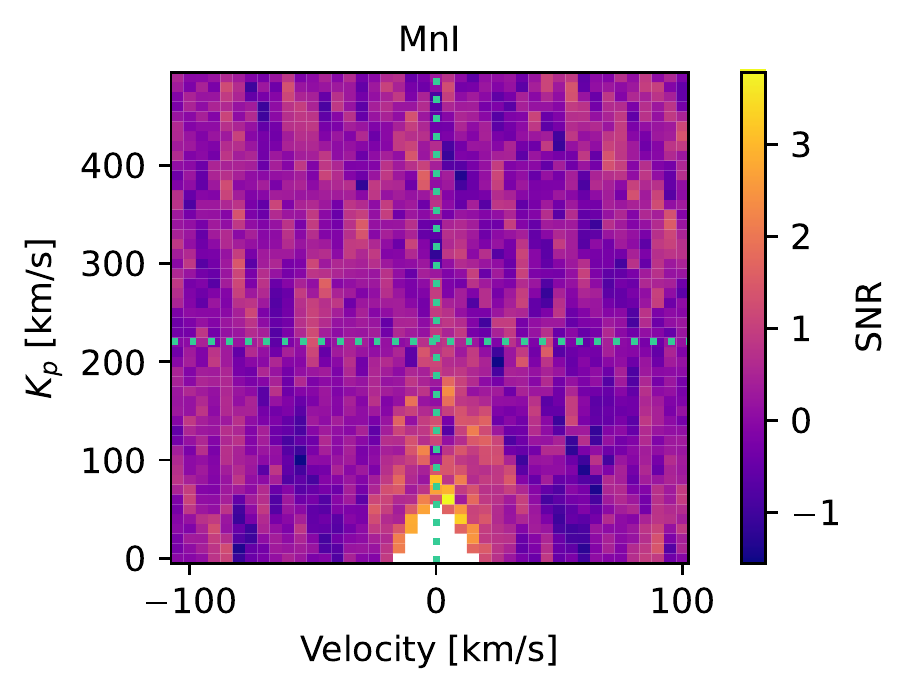}
         \label{fig:y equals x}
     \end{subfigure}
     \hfill
     \begin{subfigure}[b]{0.33\textwidth}
         \centering
         \includegraphics[width=\textwidth]{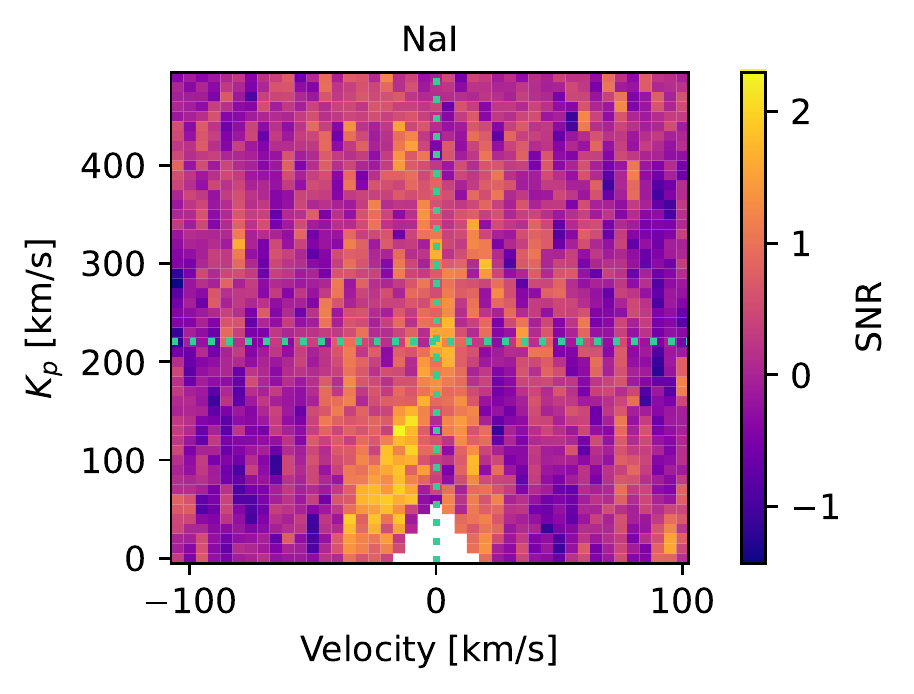}
         \label{fig:three sin x}
     \end{subfigure}
     \hfill
     \begin{subfigure}[b]{0.33\textwidth}
         \centering
         \includegraphics[width=\textwidth]{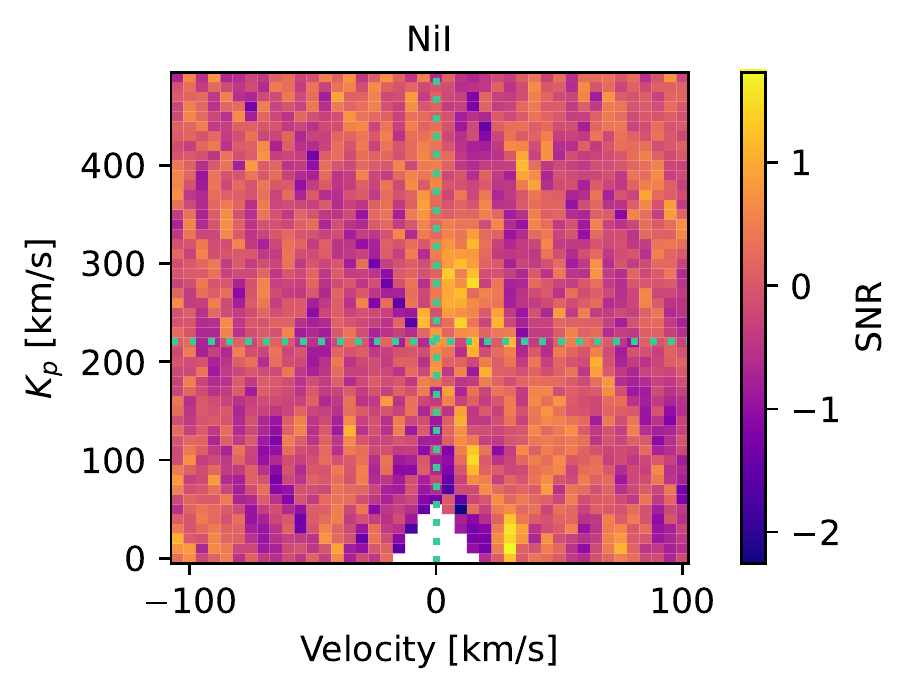}
         \label{fig:five over x}
     \end{subfigure}
     
          \begin{subfigure}[b]{0.33\textwidth}
         \centering
         \includegraphics[width=\textwidth]{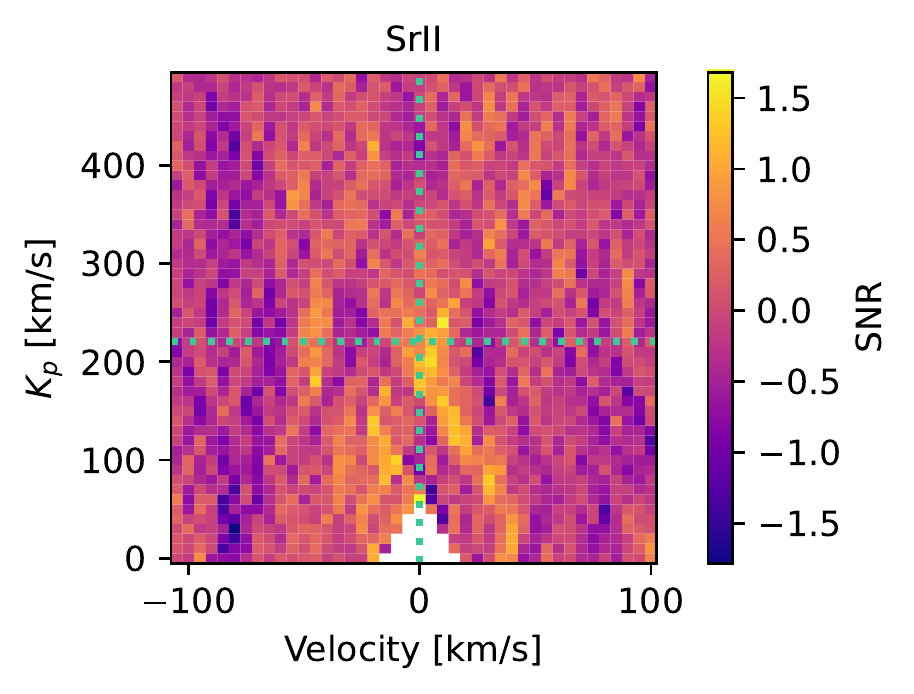}
         \label{fig:y equals x}
     \end{subfigure}
     \hfill
     \begin{subfigure}[b]{0.33\textwidth}
         \centering
         \includegraphics[width=\textwidth]{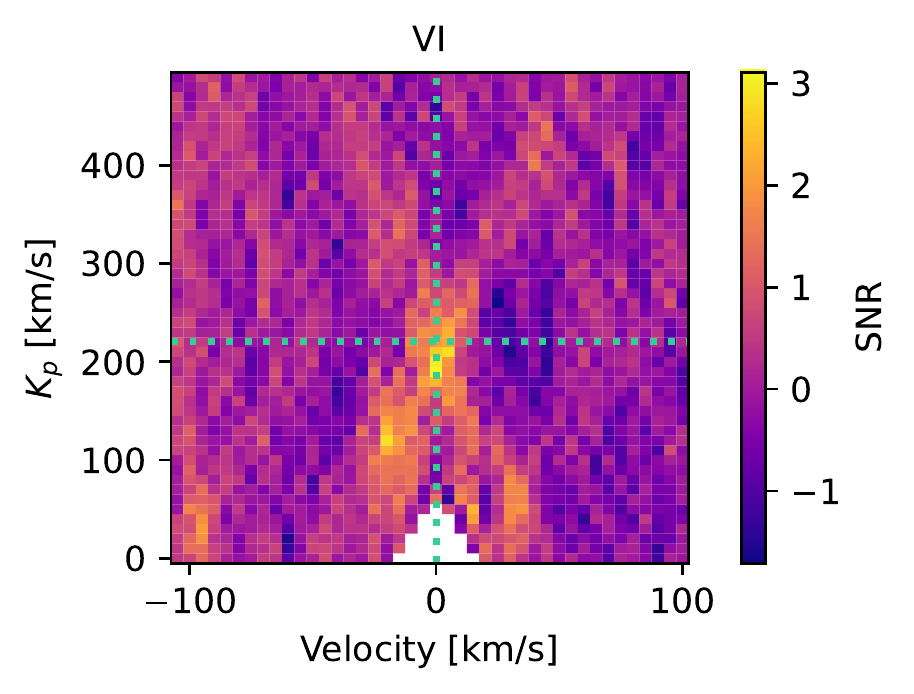}
         \label{fig:three sin x}
     \end{subfigure}
     \hfill
     \begin{subfigure}[b]{0.33\textwidth}
         \centering
         \includegraphics[width=\textwidth]{nothing.png}
         \label{fig:five over x}
     \end{subfigure}
        \caption{Same as Fig. \ref{fig:W76n1kp} for WASP-121b, night 1 (1UT - 2018 November 30). }
        \label{fig:W121n1kp}
\end{figure*}

\begin{figure*}
     \centering
     \begin{subfigure}[b]{0.33\textwidth}
         \centering
         \includegraphics[width=\textwidth]{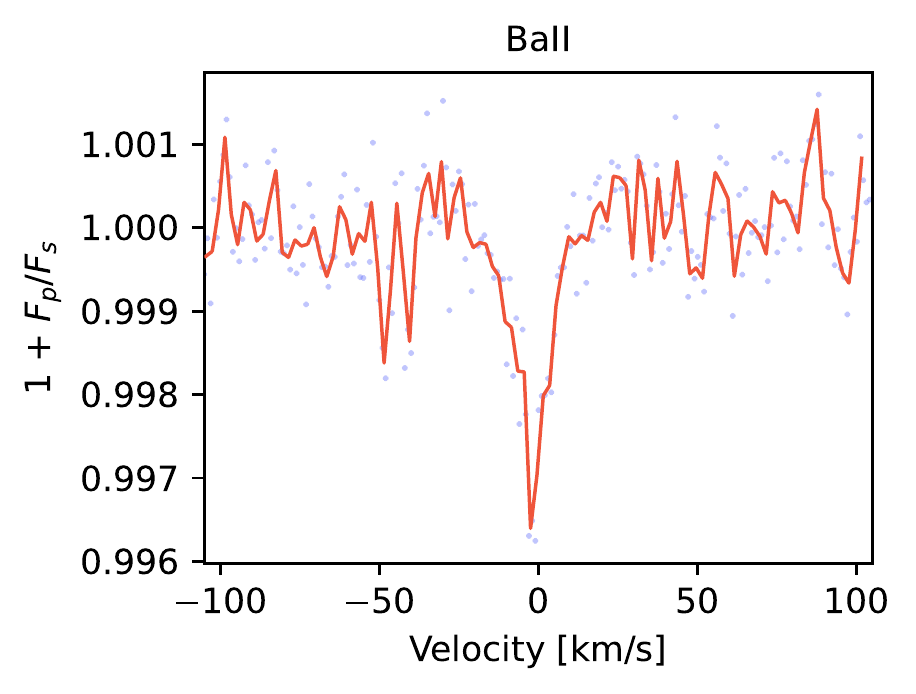}
         \label{fig:y equals x}
     \end{subfigure}
     \hfill
     \begin{subfigure}[b]{0.33\textwidth}
         \centering
         \includegraphics[width=\textwidth]{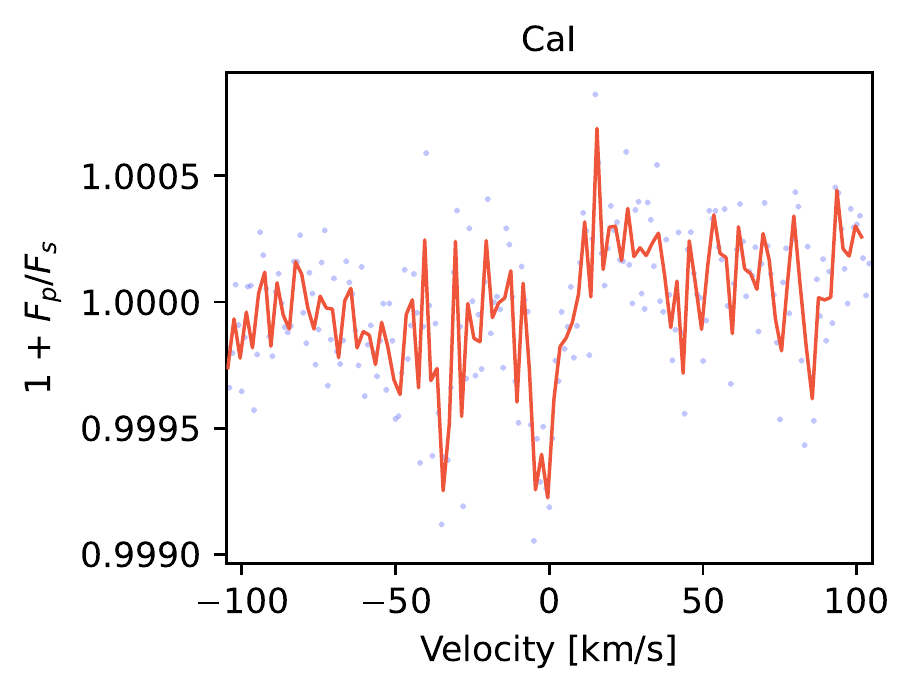}
         \label{fig:three sin x}
     \end{subfigure}
     \hfill
     \begin{subfigure}[b]{0.33\textwidth}
         \centering
         \includegraphics[width=\textwidth]{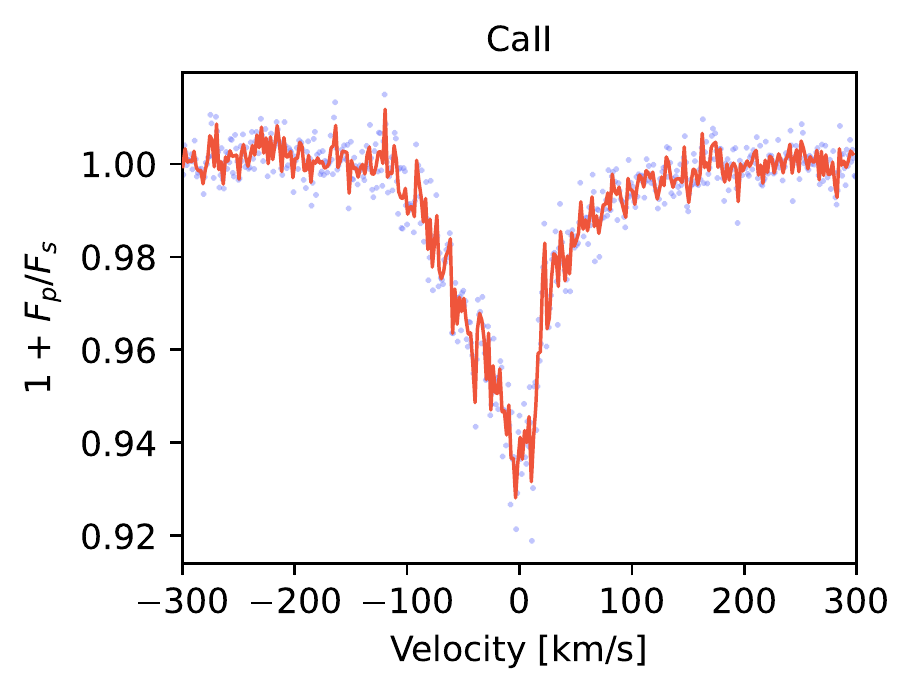}
         \label{fig:five over x}
     \end{subfigure}
     
     \begin{subfigure}[b]{0.33\textwidth}
         \centering
         \includegraphics[width=\textwidth]{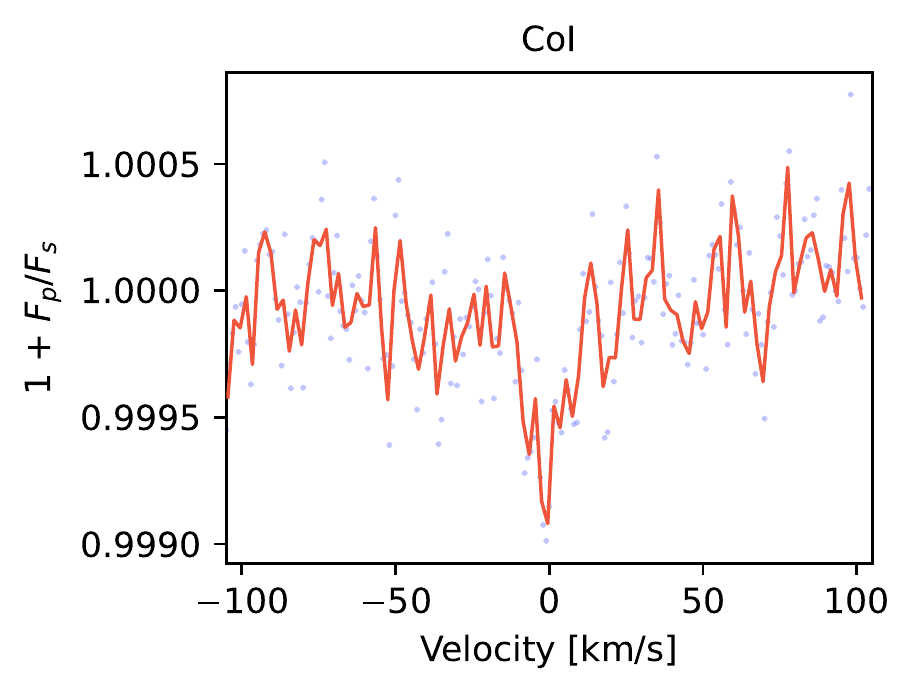}
         \label{fig:y equals x}
     \end{subfigure}
     \hfill
     \begin{subfigure}[b]{0.33\textwidth}
         \centering
         \includegraphics[width=\textwidth]{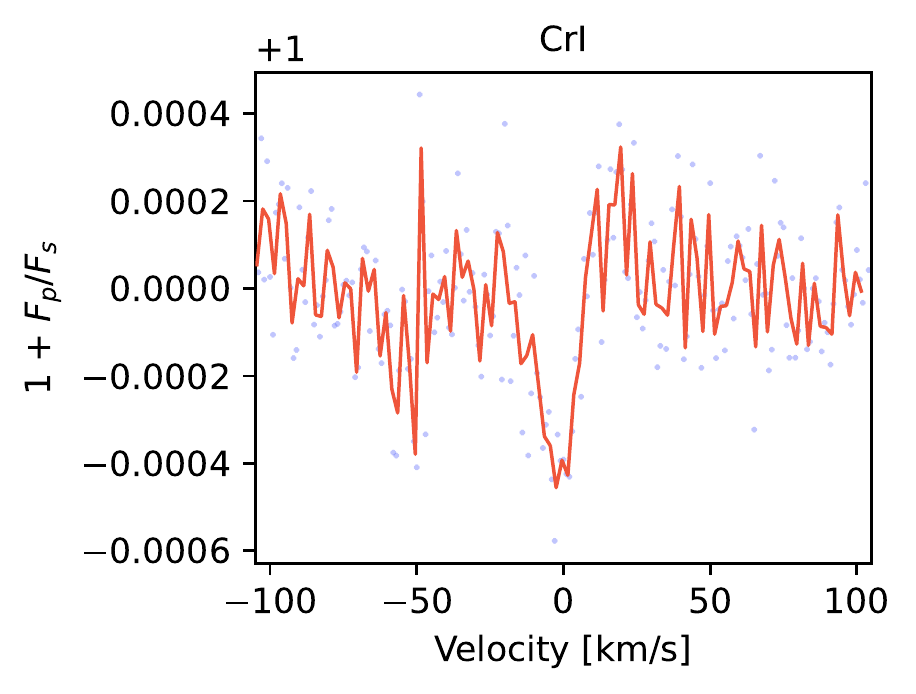}
         \label{fig:three sin x}
     \end{subfigure}
     \hfill
     \begin{subfigure}[b]{0.33\textwidth}
         \centering
         \includegraphics[width=\textwidth]{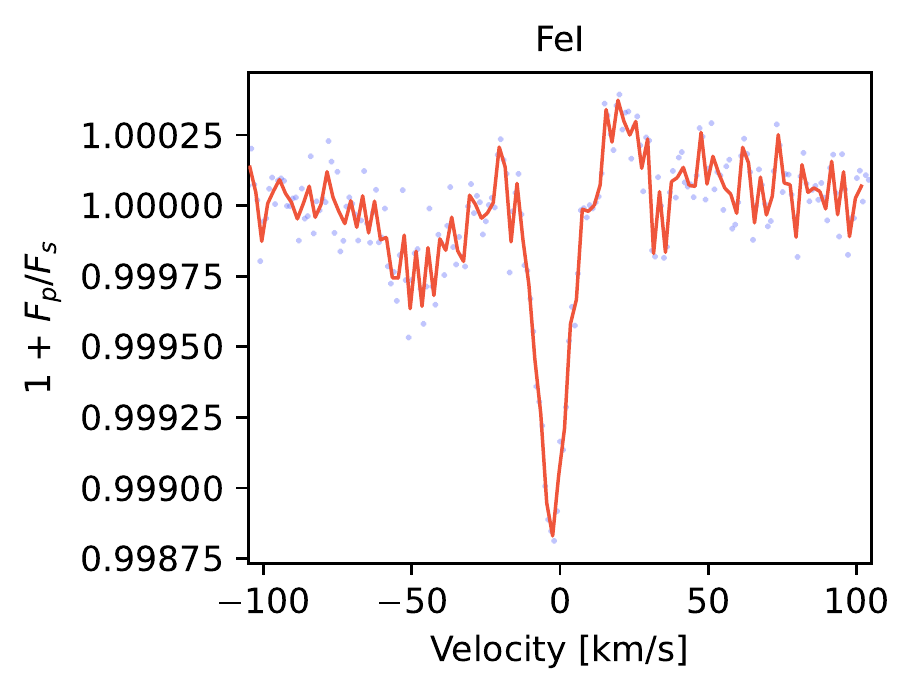}
         \label{fig:five over x}
     \end{subfigure}
        
     \begin{subfigure}[b]{0.33\textwidth}
         \centering
         \includegraphics[width=\textwidth]{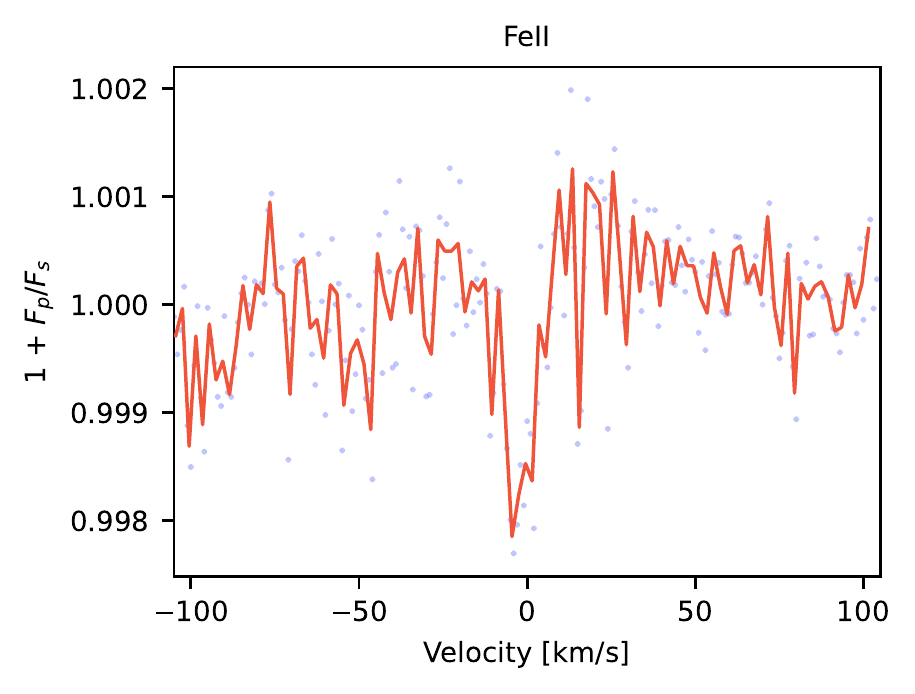}
         \label{fig:y equals x}
     \end{subfigure}
     \hfill
     \begin{subfigure}[b]{0.33\textwidth}
         \centering
         \includegraphics[width=\textwidth]{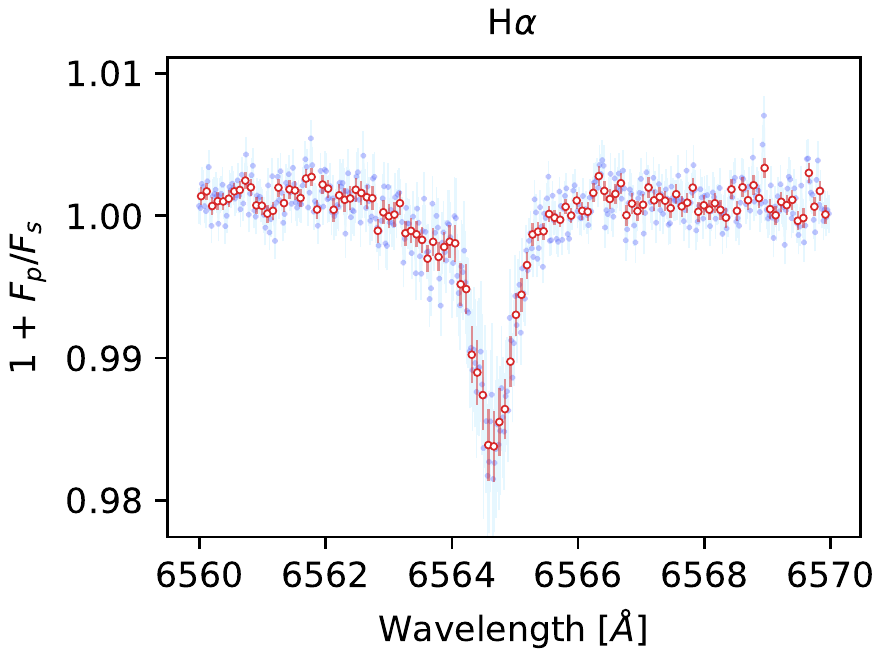}
         \label{fig:three sin x}
     \end{subfigure}
     \hfill
     \begin{subfigure}[b]{0.33\textwidth}
         \centering
         \includegraphics[width=\textwidth]{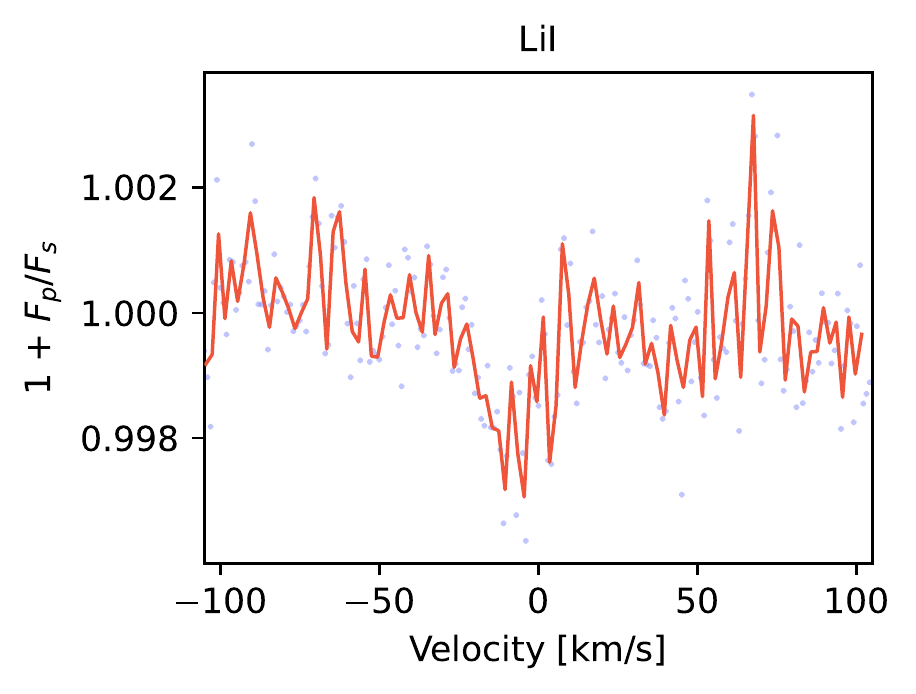}
         \label{fig:five over x}
     \end{subfigure}
        
     \begin{subfigure}[b]{0.33\textwidth}
         \centering
         \includegraphics[width=\textwidth]{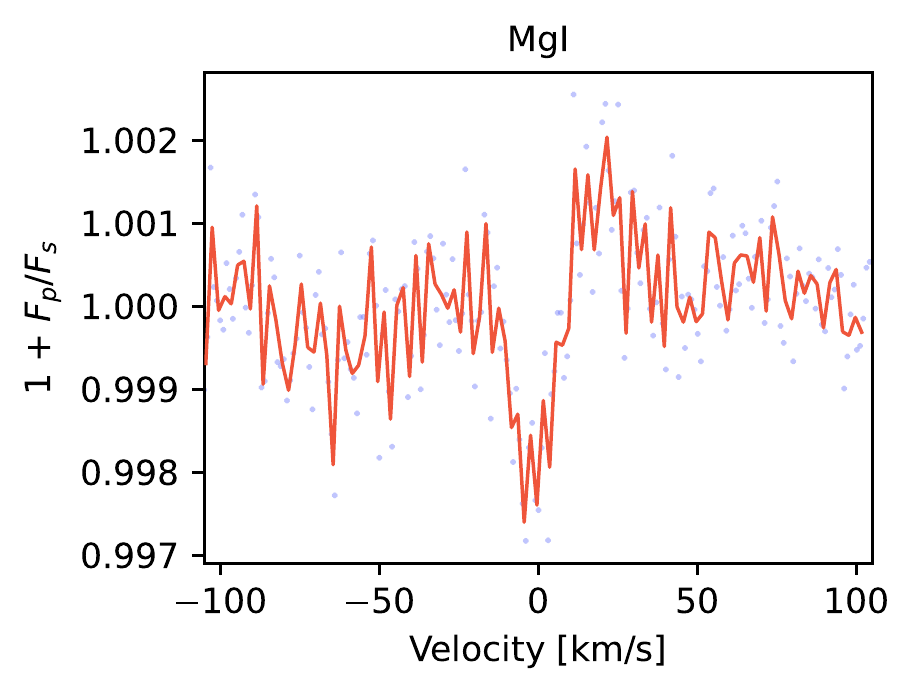}
         \label{fig:y equals x}
     \end{subfigure}
     \hfill
     \begin{subfigure}[b]{0.33\textwidth}
         \centering
         \includegraphics[width=\textwidth]{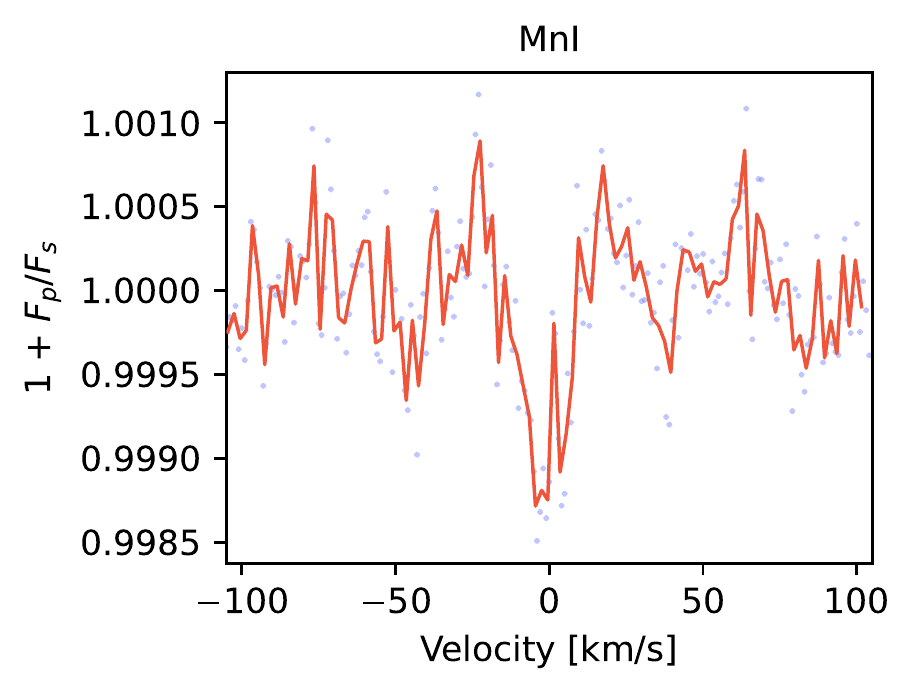}
         \label{fig:three sin x}
     \end{subfigure}
     \hfill
     \begin{subfigure}[b]{0.33\textwidth}
         \centering
         \includegraphics[width=\textwidth]{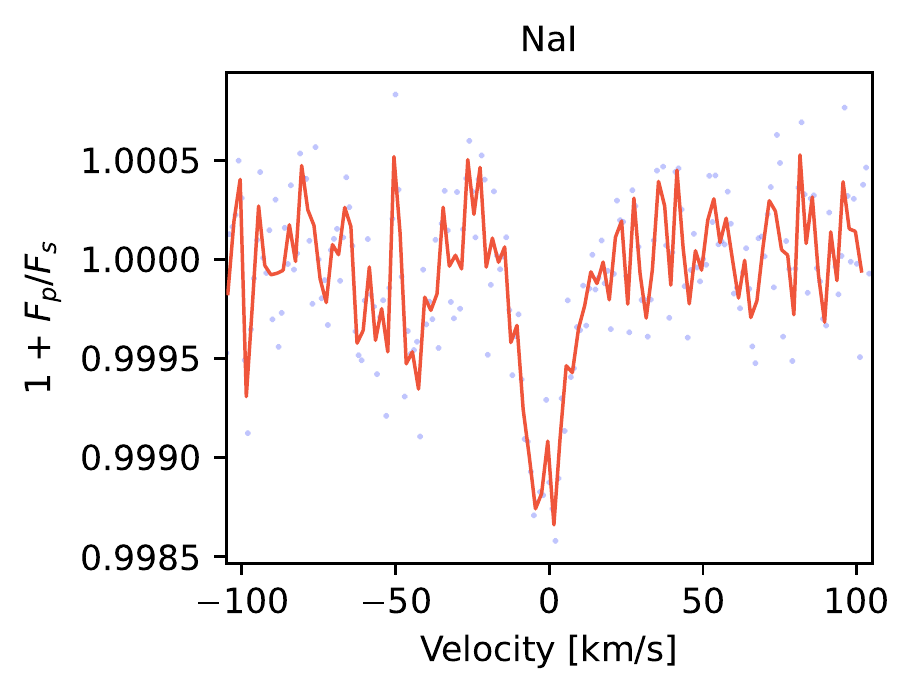}
         \label{fig:five over x}
     \end{subfigure}
     
    \begin{subfigure}[b]{0.33\textwidth}
         \centering
         \includegraphics[width=\textwidth]{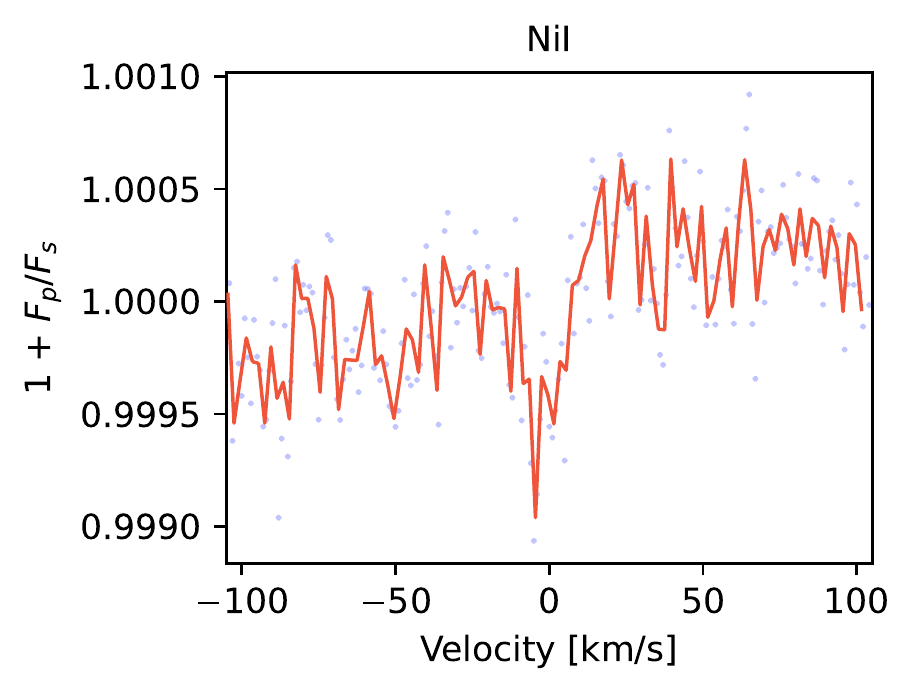}
         \label{fig:y equals x}
     \end{subfigure}
     \hfill
     \begin{subfigure}[b]{0.33\textwidth}
         \centering
         \includegraphics[width=\textwidth]{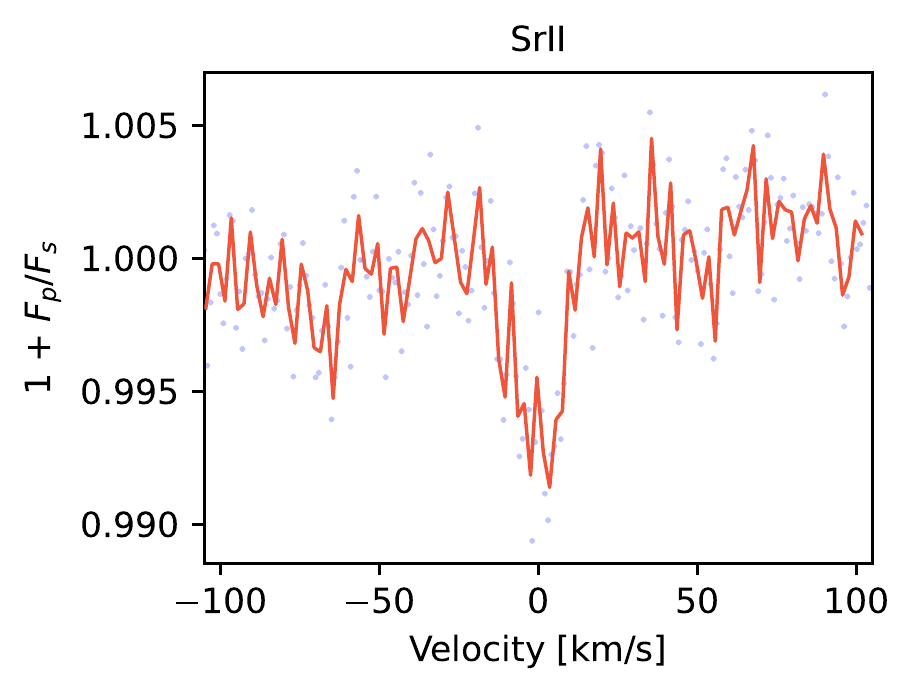}
         \label{fig:three sin x}
     \end{subfigure}
     \hfill
     \begin{subfigure}[b]{0.33\textwidth}
         \centering
         \includegraphics[width=\textwidth]{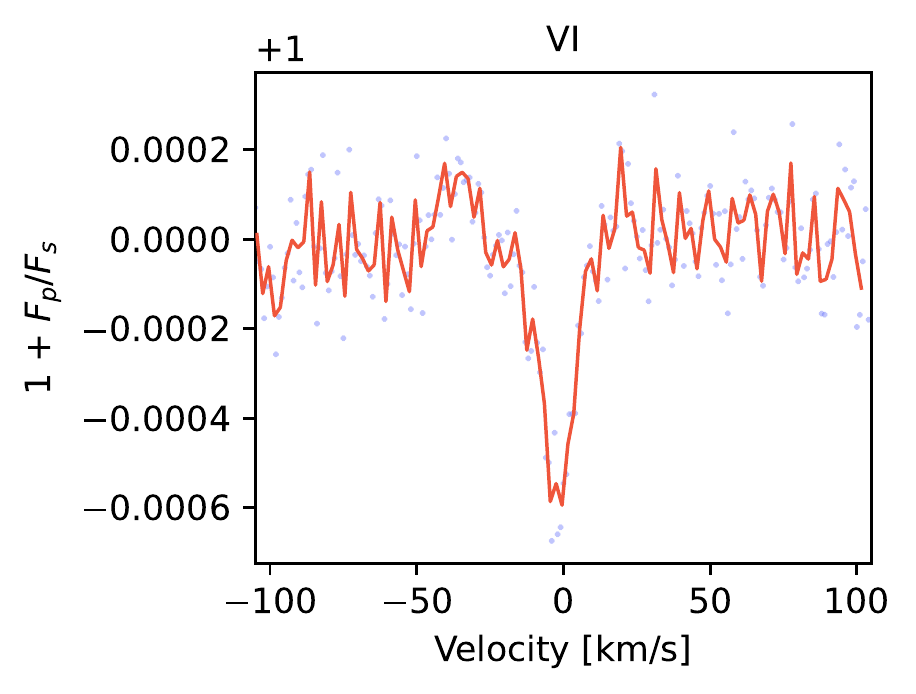}
         \label{fig:five over x}
     \end{subfigure}
     
        \caption{Same as Fig. \ref{fig:W76n1ccf} for WASP-121b, night 2 (4UT's - 2019 January 6). }
        \label{fig:W121n2ccf}
\end{figure*}

\begin{figure*}
     \centering
     \begin{subfigure}[b]{0.33\textwidth}
         \centering
         \includegraphics[width=\textwidth]{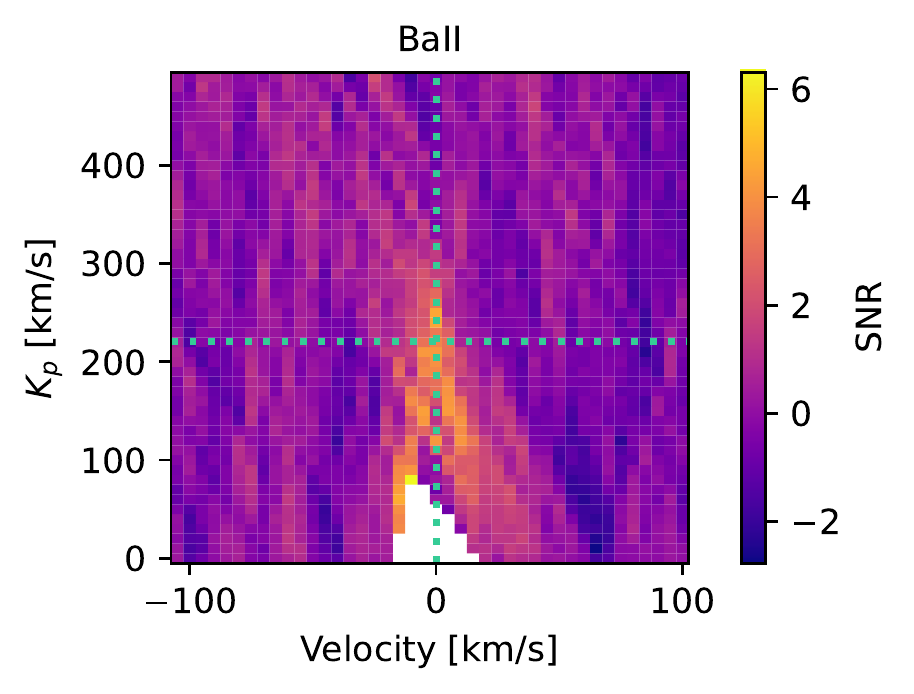}
         \label{fig:y equals x}
     \end{subfigure}
     \hfill
     \begin{subfigure}[b]{0.33\textwidth}
         \centering
         \includegraphics[width=\textwidth]{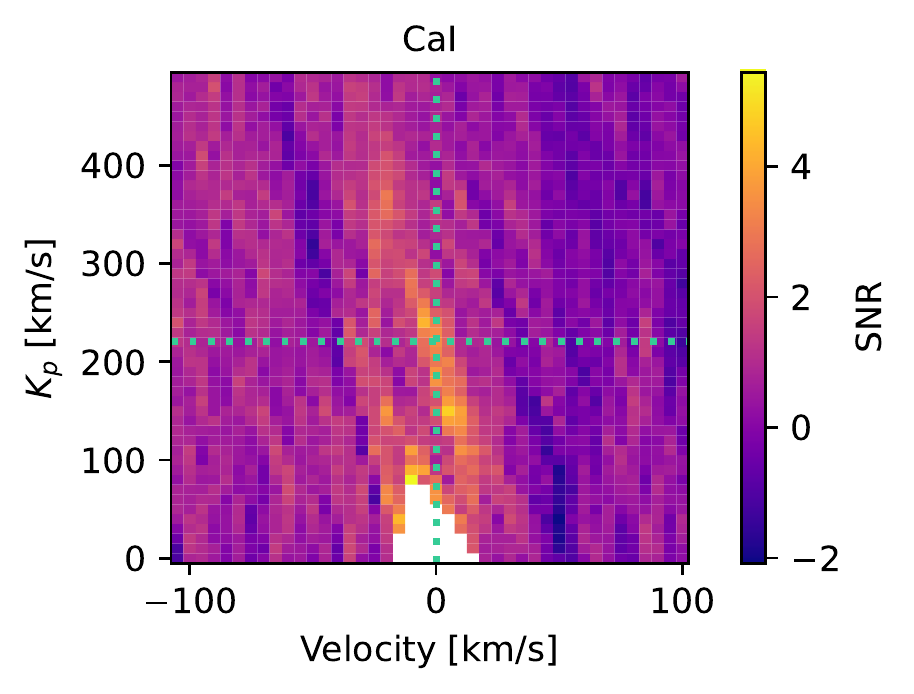}
         \label{fig:three sin x}
     \end{subfigure}
     \hfill
     \begin{subfigure}[b]{0.33\textwidth}
         \centering
         \includegraphics[width=\textwidth]{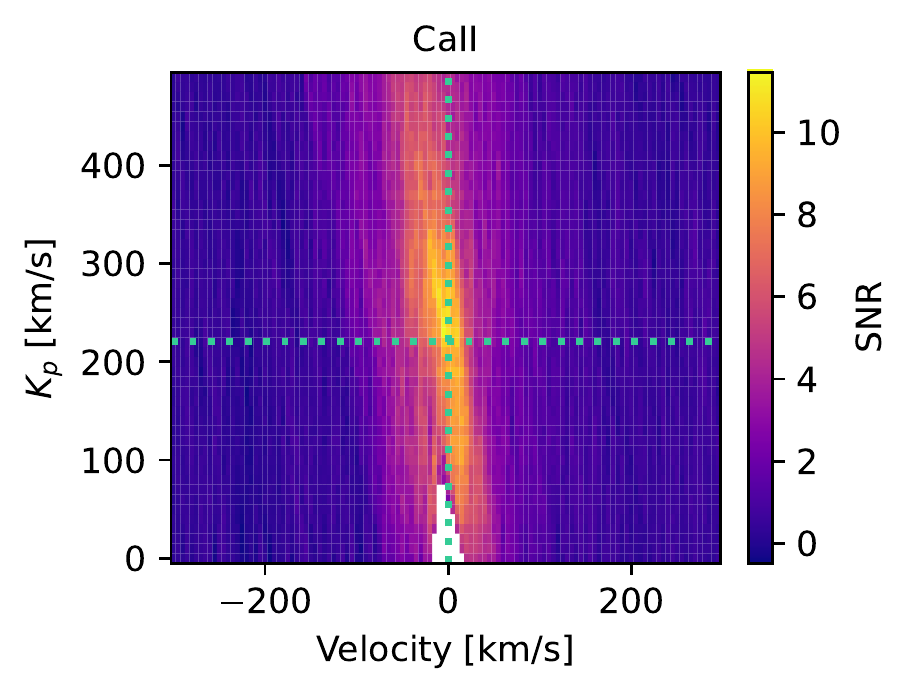}
         \label{fig:five over x}
     \end{subfigure}
     
     \begin{subfigure}[b]{0.33\textwidth}
         \centering
         \includegraphics[width=\textwidth]{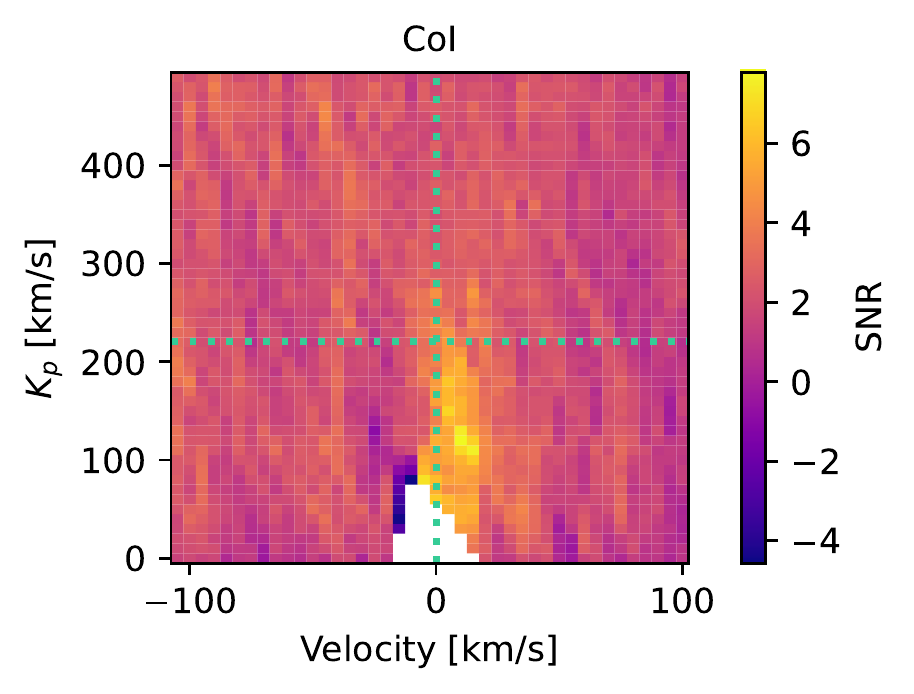}
         \label{fig:y equals x}
     \end{subfigure}
     \hfill
     \begin{subfigure}[b]{0.33\textwidth}
         \centering
         \includegraphics[width=\textwidth]{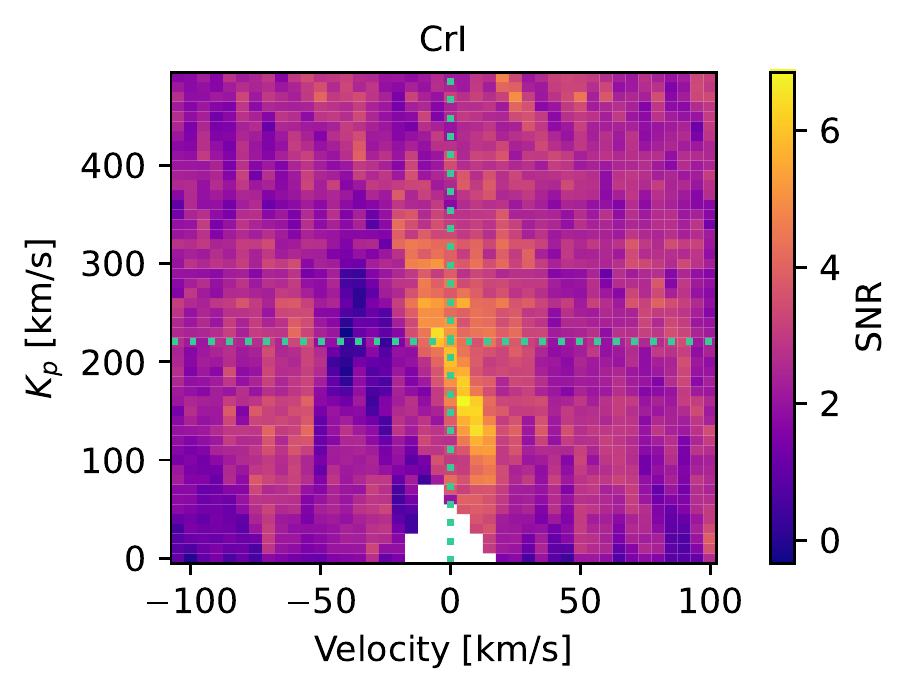}
         \label{fig:three sin x}
     \end{subfigure}
     \hfill
     \begin{subfigure}[b]{0.33\textwidth}
         \centering
         \includegraphics[width=\textwidth]{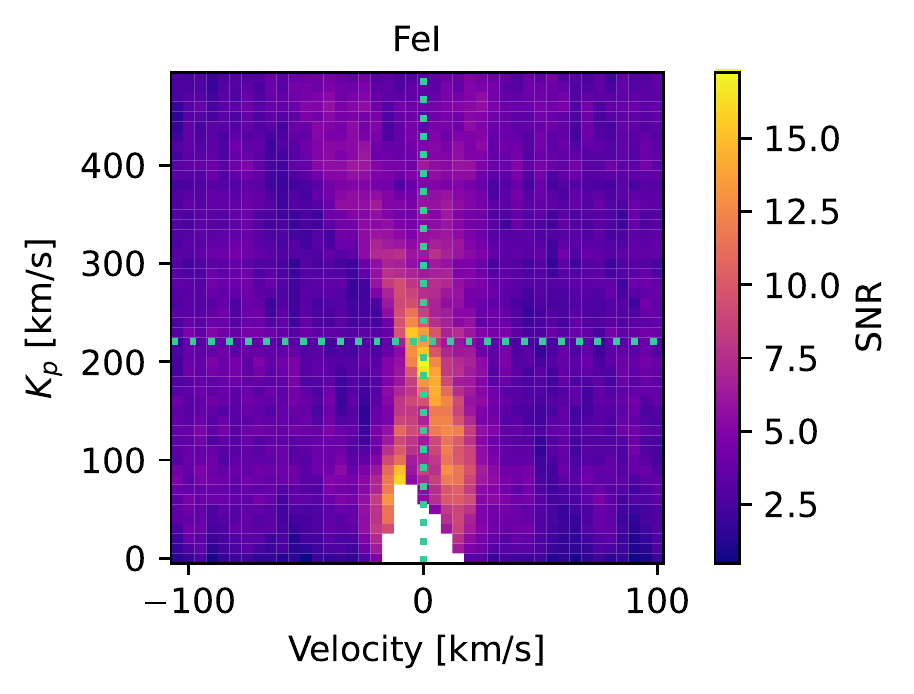}
         \label{fig:five over x}
     \end{subfigure}
     
     \begin{subfigure}[b]{0.33\textwidth}
         \centering
         \includegraphics[width=\textwidth]{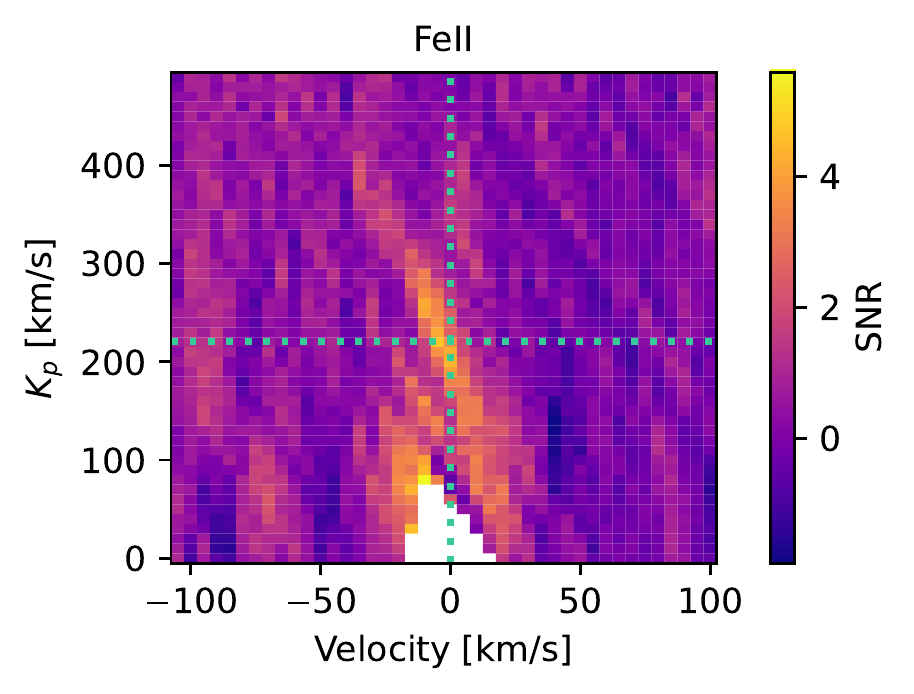}
         \label{fig:y equals x}
     \end{subfigure}   
     \hfill
     \begin{subfigure}[b]{0.33\textwidth}
         \centering
         \includegraphics[width=\textwidth]{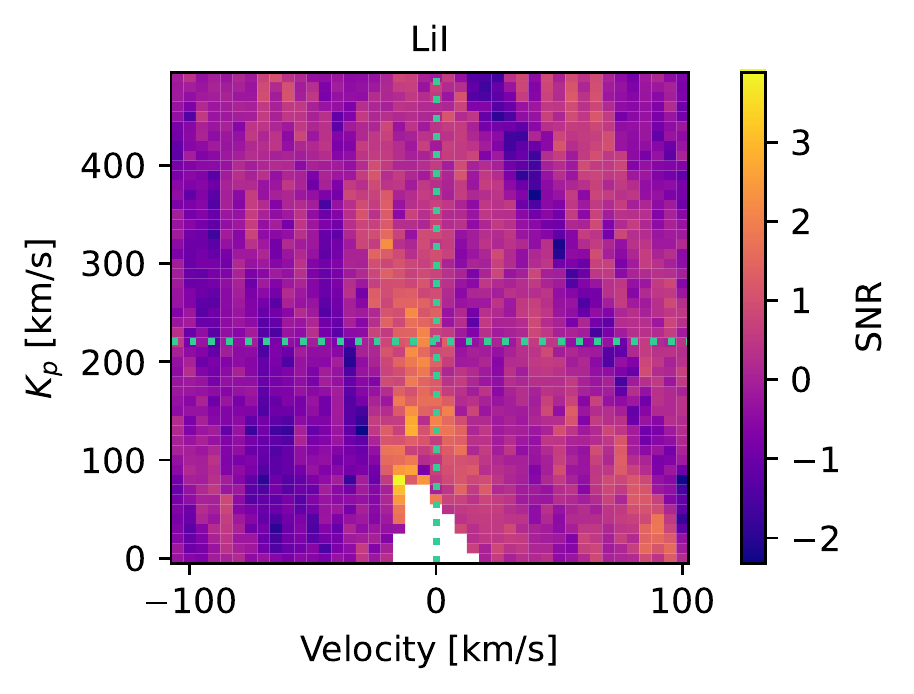}
         \label{fig:three sin x}
     \end{subfigure}
     \hfill
     \begin{subfigure}[b]{0.33\textwidth}
         \centering
         \includegraphics[width=\textwidth]{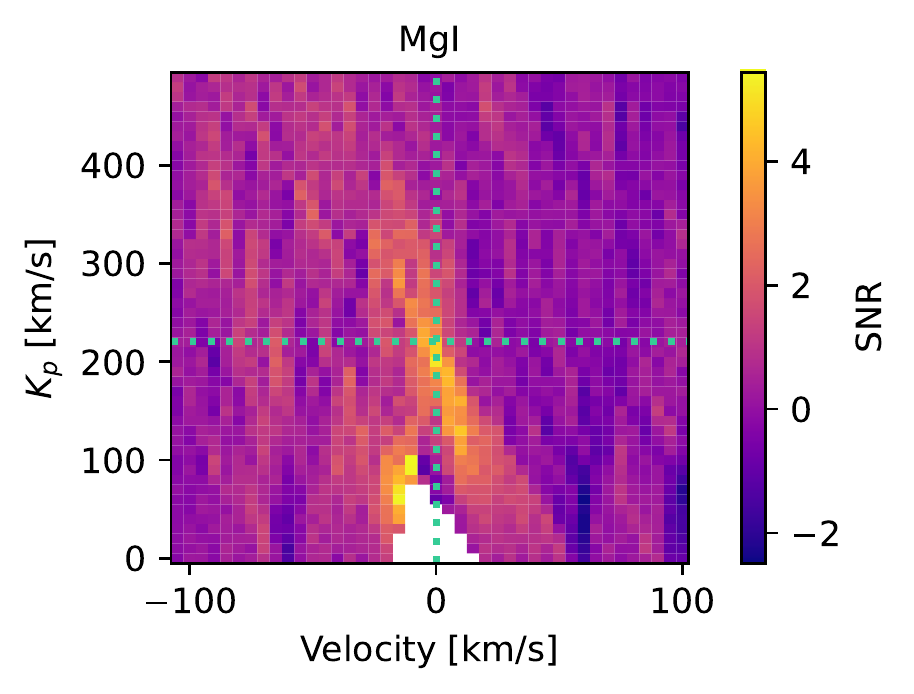}
         \label{fig:five over x}
     \end{subfigure}
        
     \begin{subfigure}[b]{0.33\textwidth}
         \centering
         \includegraphics[width=\textwidth]{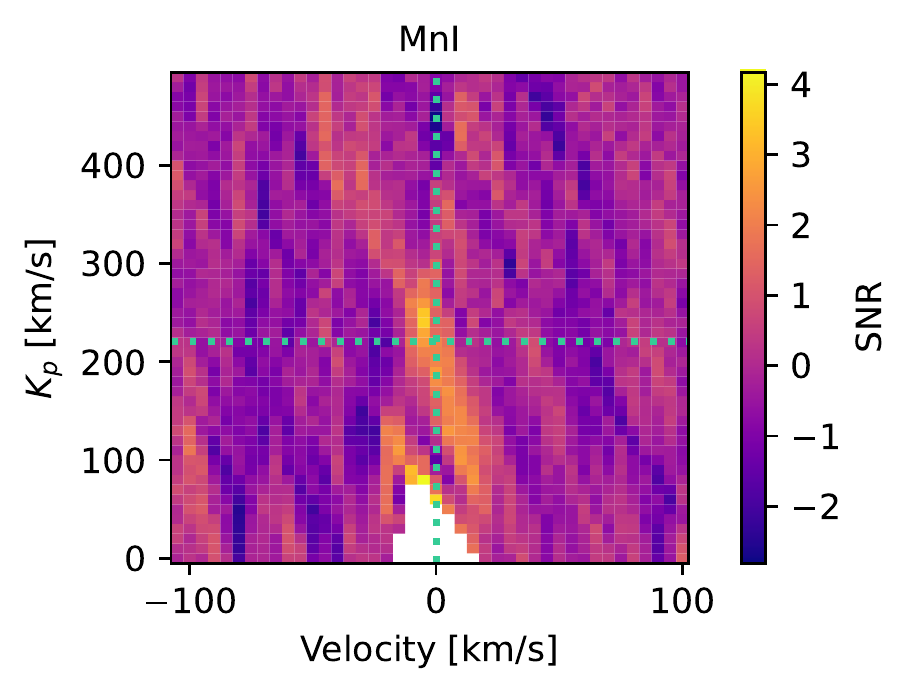}
         \label{fig:y equals x}
     \end{subfigure}
     \hfill
     \begin{subfigure}[b]{0.33\textwidth}
         \centering
         \includegraphics[width=\textwidth]{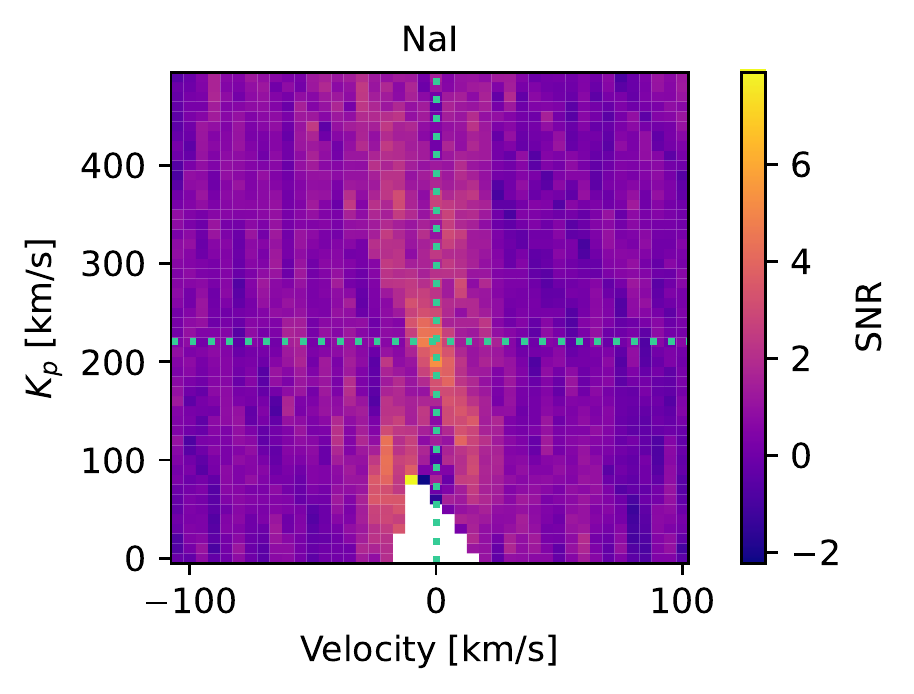}
         \label{fig:three sin x}
     \end{subfigure}
     \hfill
     \begin{subfigure}[b]{0.33\textwidth}
         \centering
         \includegraphics[width=\textwidth]{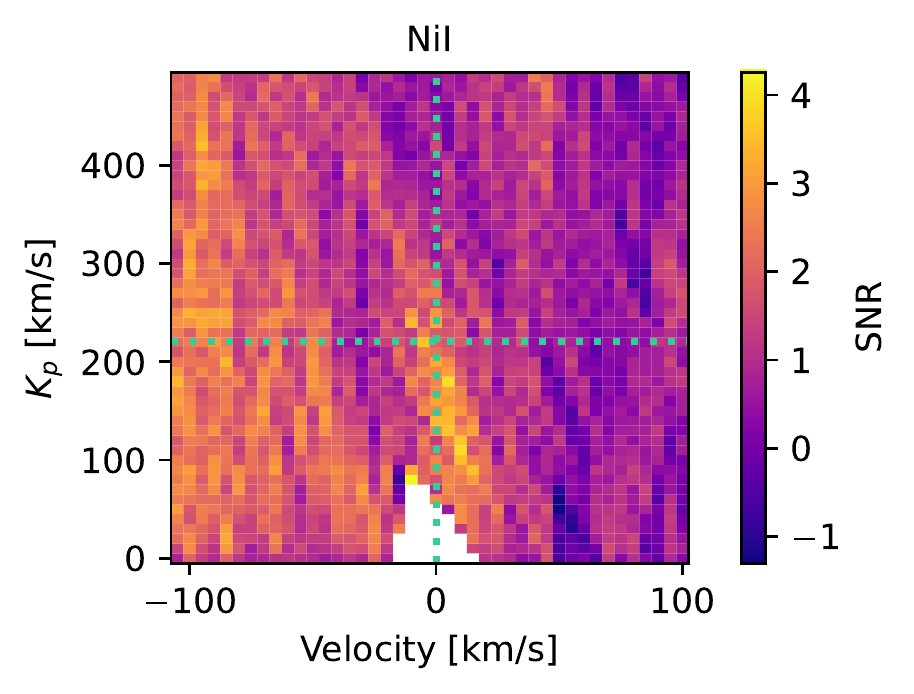}
         \label{fig:five over x}
     \end{subfigure}
     
          \begin{subfigure}[b]{0.33\textwidth}
         \centering
         \includegraphics[width=\textwidth]{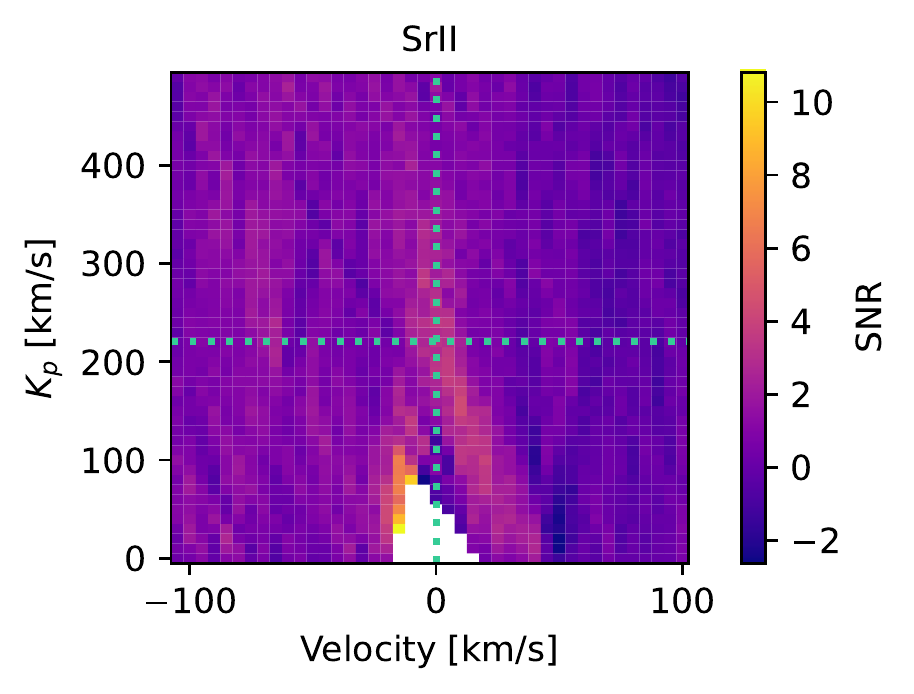}
         \label{fig:y equals x}
     \end{subfigure}
     \hfill
     \begin{subfigure}[b]{0.33\textwidth}
         \centering
         \includegraphics[width=\textwidth]{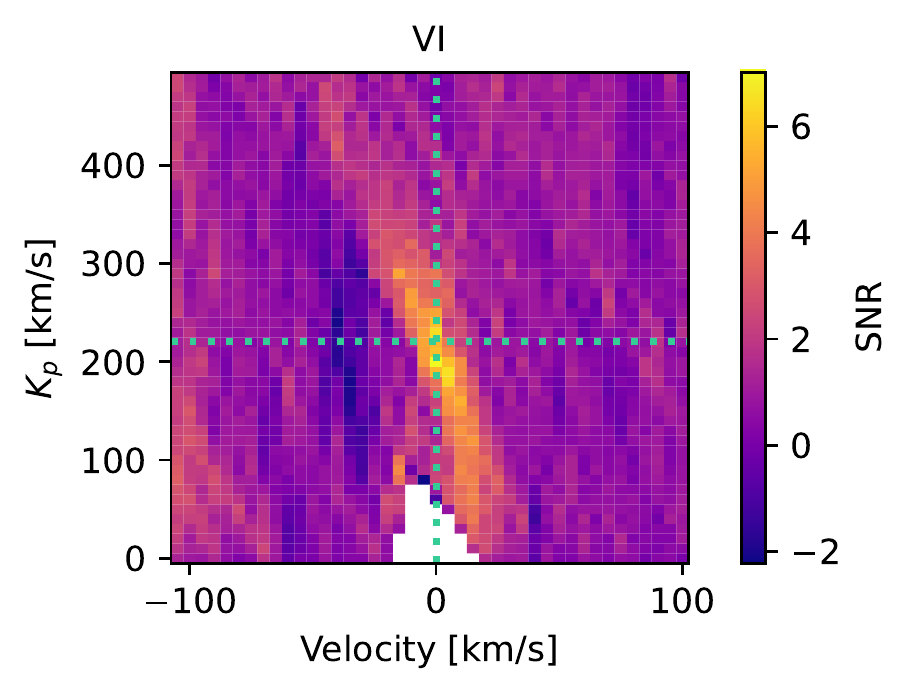}
         \label{fig:three sin x}
     \end{subfigure}
     \hfill
     \begin{subfigure}[b]{0.33\textwidth}
         \centering
         \includegraphics[width=\textwidth]{nothing.png}
         \label{fig:five over x}
     \end{subfigure}
     
        \caption{Same as Fig. \ref{fig:W76n1kp} for WASP-121b, night 2 (4UT's - 2019 January 6)}
        \label{fig:W121n2kp}
\end{figure*}

\FloatBarrier

\section{Individual barium II lines}
\label{Appendix:Individual_Barium}

\begin{figure*}[th]
     \centering
     \begin{subfigure}[b]{0.33\textwidth}
         \centering
         \includegraphics[width=\textwidth]{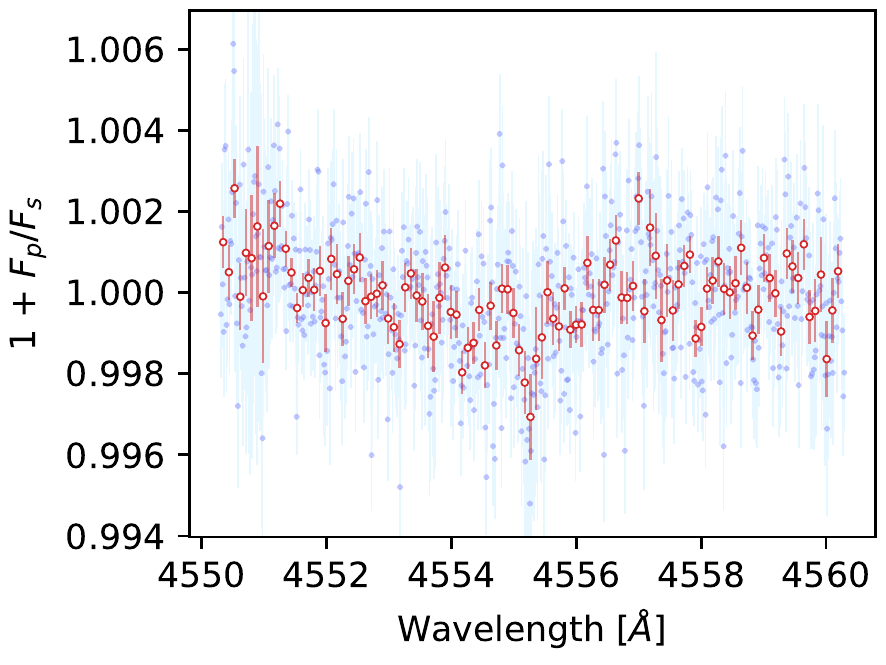}
         \label{fig:y equals x}
     \end{subfigure}
     \hfill
     \begin{subfigure}[b]{0.33\textwidth}
         \centering
         \includegraphics[width=\textwidth]{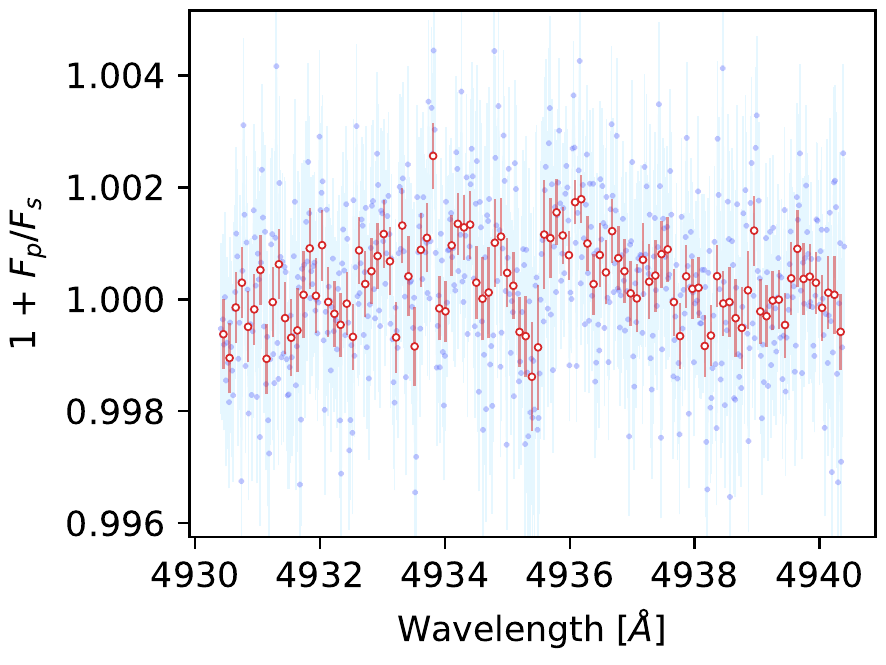}
         \label{fig:three sin x}
     \end{subfigure}
     \hfill
     \begin{subfigure}[b]{0.33\textwidth}
         \centering
         \includegraphics[width=\textwidth]{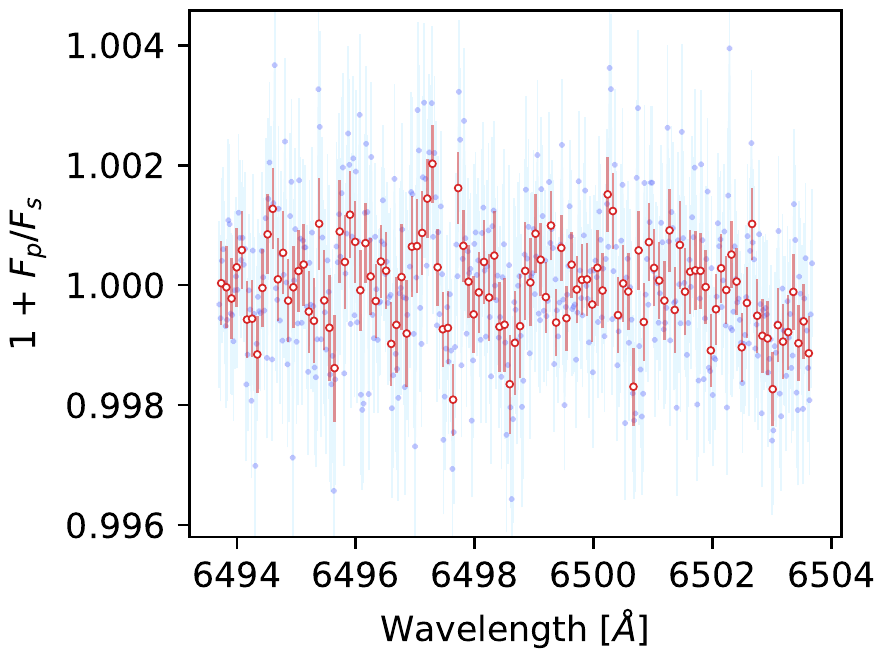}
         \label{fig:five over x}
     \end{subfigure}
        
        \caption{Planetary transmission spectrum for the WASP-121b, night 2 (4UTs, 2019 January 6) dataset, centered at the three strongest lines in the barium mask.   }
        \label{fig:ind_bar}
\end{figure*}

\end{appendix}

\end{document}